\documentclass[%
 reprint,
 superscriptaddress,
 preprintnumbers,
 amsmath,amssymb,
 prb,
]{revtex4-2}

\usepackage{graphicx}
\usepackage{dcolumn}
\usepackage{bm}
\usepackage{hyperref}
\usepackage{color}
\usepackage{braket}
\usepackage{mathtools}
\usepackage{subcaption}
\usepackage[maxfloats=256]{morefloats}
\newcommand{\w}{w}

\begin{document}

\preprint{}

\title{Electron-phonon interaction and phonons in 2d doped semiconductors}

\author{Francesco Macheda}
\affiliation{Istituto Italiano di Tecnologia, Graphene Labs, Via Morego 30, I-16163 Genova, Italy}
\author{Thibault Sohier}
\affiliation{Laboratoire Charles Coulomb (L2C), Université de Montpellier, CNRS, Montpellier, France}
\author{Paolo Barone}%
\affiliation{CNR-SPIN, Area della Ricerca di Tor Vergata, Via del Fosso del Cavaliere 100,
I-00133 Rome, Italy}%
\author{Francesco Mauri}
\affiliation{Dipartimento di Fisica, Università di Roma La Sapienza, Roma, Italy }%
\affiliation{Istituto Italiano di Tecnologia, Graphene Labs, Via Morego 30, I-16163 Genova, Italy}%
\begin{abstract}
Electron-phonon interaction and phonon frequencies of doped polar semiconductors are sensitive to long-range Coulomb forces and can be strongly affected by screening effects of free carriers, the latter changing significantly when approaching the two-dimensional limit. We tackle this problem within a linear-response dielectric-matrix formalism, where screening effects can be properly taken into account by generalized effective charge functions and the inverse scalar dielectric function, allowing for controlled approximations in relevant limits. We propose complementary computational methods to evaluate from first principles both effective charges -- encompassing all multipolar components beyond dynamical dipoles and quadrupoles -- and the static dielectric function of doped two-dimensional semiconductors, and provide analytical expressions for the long-range part of the dynamical matrix and the electron-phonon interaction in the long-wavelength limit. As a representative example, we apply our approach to study the impact of doping in disproportionated graphene, showing that optical Fr\"ohlich and acoustic piezoelectric couplings, as well as the slope of optical longitudinal modes, are strongly reduced, with a potential impact on the electronic/intrinsic scattering rates and related transport properties.
\end{abstract}

\maketitle

\section{Introduction}
The electron-phonon interaction (EPI) is one of the most thoroughly studied topics in solid state physics \cite{ziman2001electrons,ziman2001electrons,Grimvall1981TheEI,schrieffer1999theory,mahan1990many,RevModPhys.89.015003} due to the fundamental role it plays in the determination of a variety of physical properties. The prediction and interpretation of e.g., transport experiments \cite{PhysRevB.98.201201,PhysRevB.102.094308,Macheda2020,PhysRevB.100.085204,PhysRevResearch.3.043022,Lee2020,Ponce2020,PhysRevB.94.201201,macheda2021ab,PhysRevB.97.045201,PhysRevResearch.2.033055}, excited carriers relaxation \cite{PhysRevB.84.075449,Harb2006,Betz2013} and superconductivity \cite{PhysRev.108.1175,schrieffer1999theory}, relies on the accurate calculation of the EPI from first-principles,
which has become possible in recent years thanks to the development of density functional theory (DFT) \cite{PhysRev.136.B864,PhysRev.140.A1133}, density functional perturbation theory (DFPT) \cite{RevModPhys.73.515} and Wannier interpolation technique \cite{RevModPhys.84.1419,PhysRevB.76.165108,PhysRevB.82.165111}, as well as swift progress of computational infrastructures \cite{doi:10.1063/5.0005082,Pizzi2020,PONCE2016116}. 

In insulators and undoped semiconductors, long-range Coulomb interactions arise from the charge polarization and lead to non-analytic contributions to phonons and EPIs, including the well-known splitting of longitudinal (LO) and transverse (TO) optical modes \cite{doi:10.1119/1.1934059} as well as the Fröhlich \cite{doi:10.1080/00018735400101213}  and piezoelectric electron-phonon interactions \cite{PhysRevB.13.694}. In the absence of free carriers and in the long-wavelength limit, the form of such non-analytic contributions is known exactly, enabling a precise evaluation of related effects \cite{RevModPhys.73.515,PhysRevB.13.694,PhysRevLett.115.176401,PhysRevB.92.054307,PhysRevLett.125.136601}. Due to the strong screening provided by free carriers in partially filled bands, those long-ranged Coulomb interactions are expected to vanish in metals.
In doped semiconductors, on the contrary, Coulomb-mediated interactions are only partially screened by the small fraction of added charge carriers. In this intermediate situation between insulator and metal, the long-wavelength behaviour of phonons and EPI may be significantly altered, as recently shown for 3d doped semiconductors \cite{PhysRevLett.129.185902}. 
Going beyond semi-phenomenological corrections of the screened quantities \cite{Ren2020},
the formalism developed in \cite{PhysRevLett.129.185902} proposes a clean separation between bare quantities and their screening, thus enabling the evaluation of doping and temperature effects on both of them. 
In regimes where doping and temperature effects on the bare quantities are negligible, 
this allows for an efficient yet precise evaluation and interpolation of phonons and EPIs at a given temperature and doping via the modification of screening only. 

Given that electrostatically-doped 2D materials are at the heart of the quest for high-efficiency electronic devices \cite{Wang2012,Chhowalla2016}, the same rigorous treatment of screened Coulomb interactions in quasi-2d systems would be highly beneficial. However, it is now well established that dimensionality alters the dielectric screening behaviour in 2d undoped semiconductors and insulators, and consequently any phonon or EPI involving long-range Coulomb interactions \cite{PhysRevX.11.041027}. The implementation of DFPT in 2d boundary conditions \cite{PhysRevB.96.075448} has allowed to study those effects; most notably, the LO-TO splitting breaks down at zone center and increases linearly with momentum \cite{Verberck_prb2011,Sohier2017}, while the Fröhlich EPI stays finite \cite{PhysRevB.94.085415} instead of diverging as the inverse of momentum, as in 3D \cite{PhysRevB.13.694}. Yet, as for 3d materials, the doping effects on the 2d long-range Coulomb interaction are virtually always neglected \cite{doi:10.1063/5.0015102,PhysRevB.85.115317,Li2019,PhysRevB.103.075410,PhysRevB.104.235424,https://doi.org/10.48550/arxiv.2207.10187,https://doi.org/10.48550/arxiv.2207.10190}, and occasionally included via approximate models \cite{PhysRevB.87.235312}. Nonetheless, a direct computation of both phonons and EPIs in the presence of electrostatic doping, as recently implemented in DFPT \cite{PhysRevB.96.075448}, showed that metallic screening causes a vanishing of the linear-in-momentum LO-TO splitting specific of 2d semiconductors \cite{Sohier2017}, a sizeable effect that might be detected by momentum-resolved electron energy loss spectroscopy in a transmission electron microscope (TEM-EELS) \cite{Senga2019}. The intrinsic dimensionality reduction combined with careful Brillouin-zone sampling allowed to compute transport properties from the Boltzmann equation formalism in highly-doped 2d semiconductors without resorting to Wannier interpolation methods \cite{PhysRevMaterials.2.114010,Sohier2020}.
However, such a procedure is intrinsically costly and limited due to the fine sampling of electronic states needed to account for a small Fermi surfaces, especially at small dopings. Thus, as typical DFPT calculations scale like the cube of the number of atoms in the simulation cell, small dopings, large systems or multiple doping and temperature conditions of the same system remain out of reach. This latter shortfall in particular prevents any systematic study of a material transport properties, which is of the outmost importance for the development of electronic devices. Recently, a procedure based on the formalism of Ref. \cite{PhysRevX.11.041027} has been proposed to deal with some of the above shortfalls \cite{PhysRevB.106.115423}, but neglecting the effects of doping and temperature on bare interactions and on the local-fields components of the response. 

Motivated by the above considerations, we extend the theoretical framework recently outlined for 3d systems \cite{PhysRevLett.129.185902} in order to deal with quasi-2d doped semiconductors. Using a static linear-response dielectric matrix formulation, we introduce screened and unscreened effective charges. Along with the inverse scalar dielectric function, these are used to derive general expressions for the long-range Coulomb contributions to the dynamical matrix and EPI. At vanishing doping, they reduce to well-established formulas, both for 3d and 2d systems. This allows for controlled approximations of screening effects in appropriate doping and temperature regimes, whose range of validity can be assessed within the general theoretical framework. At variance with the 3d case, the presence of a non-periodic direction complicates the electrostatic problem and enforces the dependence of all the response functions on the out-of-plane component $z$. 
Nonetheless, a great deal of simplification comes when the wavelength of the interaction is much larger or much smaller than the typical out-of-plane thickness of the material $t$. We denote these regimes as the thin and thick limits, characterized respectively by a wavevector dependence of the in-plane Coulomb kernel being of the form $q^{-1}$ and $q^{-2}$. 
In such regimes, for materials displaying in-plane mirror symmetry, one can integrate out the out-of-plane variable via layer-averages to a good approximation. The resulting long-range components (LRC) of both the dynamical matrix and EPI can then be connected, as in the 3D case, to a properly modified version of the well-known phenomenological theory of Born and Huang \cite{doi:10.1119/1.1934059}, involving in-plane effective charges and the inverse scalar dielectric function $\epsilon^{-1}$. 
We further discuss their range of validity for the two-dimensional case. The layer-averaging procedure is naturally appropriate for those single layers materials where the physical observables are mainly determined by in-plane electrostatics. Nonetheless, even when out-of-plane perturbations are important (e.g. when considering remote couplings with surrounding materials), our approach allows for a simple extraction of phonon perturbations that are macroscopically unscreened both from the in-plane dielectric response and from the presence of periodically repeated images. This makes it suitable to be integrated in frameworks that account for remote screening in heterostructures such as the one proposed in Ref. \cite{PhysRevMaterials.5.024004}, which proposes a treatment of the out-of-plane electrostatics a-posteriori. 

Operatively, we propose a fast and precise technique based on first-principles calculations and Wannier interpolation that is grounded in our general theoretical formulation and supported by the computation of both screened and unscreened charge responses. Crucially, in controlled regimes, accurate quantities can be obtained from \textit{ab-initio} calculations performed only for the undoped setup, while still accounting for the most relevant doping and temperature effects beyond the state-of-the-art, and with a great reduction of the computational workload with respect to brute-force methods. We validate our findings in disproportionated graphene,
i.e., a particular realization of gapped graphene
which can be found in the presence of substrates causing a symmetry-breaking modulation of potential \cite{Zhou2007,Novoselov2007,PhysRevLett.115.136802} and that has been recently proposed to host strong polar responses \cite{Bistoni2019}. Our approach shows that there exists a small doping regime where the LRCs can be described using the effective charges value of the undoped setup at zero temperature and an RPA expression for the screening.
The existence of this regime is expected for any two-dimensional material.
In the strong doping regime, instead, this simplification doesn't occur and the effective charges are mostly determined by the appearance of intraband terms in the electronic polarizability; one shall then resort to the \textit{ab-initio} calculation of the macroscopically screened and unscreened effective charge functions in the doped setup.
%
As a practical example of the implications of our developments, we show that electronic lifetimes can be strongly affected by the presence of free carriers, with likewise critical implications for physical observables. In particular, the reduction of the electronic lifetimes with doping opens the way to the engineering of new upper limits to the carriers mobility in 2d materials via a fine-tuned choice of the doping and temperature regimes, with important consequences for the design of efficient field-effect transistors.

The paper is organized as follows: in Sec.  \ref{sec:II} we present the general theoretical framework of our approach, starting from the introduction of the response functions and of the effective charges for two-dimensional materials to arrive to the expressions of the LRCs of the dynamical matrix and of the EPI; in Sec. \ref{sec:compappr} we discuss the implementation of our theoretical considerations into an operative computational approach; in Sec. \ref{sec:III} we apply our developments to the case of disproportionated graphene; finally, in Sec. \ref{sec:IV} we draw our conclusions; the appendices are devoted instead to the treatment of technical details which integrate the derivations of the main text.
\label{sec:I}

\section{Theory}
\label{sec:II}
\subsection{Framework and main formulae}
\label{sec:frammainform}
We are interested in the effect of free-carriers on the dynamical matrix $C$ and electron-phonon interaction $g$ as a function of the chemical potential and of the temperature $T$. 
To reconcile with the doped semiconductor literature, we prefer to use the free carrier concentration $n$ as a variable, which evaluates to zero in the case of an undoped semiconductor, rather than the chemical potential which is more appropriate for metals.

Our aim is to evaluate the expressions of $C$ and $g$ on fine grids in reciprocal space at any given $(n,T)$, having at disposal with a reasonable effort only their value for the undoped setup at zero Kelvin. Following the same logic as Ref. \cite{PhysRevLett.129.185902}, we exploit the usual separation of short and long-range components of $C$ and $g$
\begin{align}
C = C^{\textrm{S}}+C^{\textrm{L}}, \quad g = g^{\textrm{S}} + g^{\textrm{L}},
\label{eq:separation}
\end{align}
and focus on the description of the long-range components since, as we will show, they are the only ones strongly affected by $(n,T)$. Their description, within well-defined and controlled regimes, requires the knowledge of two main ingredients: i) macroscopically unscreened effective charges and ii) the macroscopic inverse dielectric function. In such regimes, doping and temperature act on the long-range components mostly through the macroscopic inverse dielectric function, which can be approximately expressed as a function of quantities computed in absence of free-carriers. As a result, Eq. \ref{eq:separation} may be described using \textit{ab-initio} techniques only for $(n=0,T=0)$. Outside these regimes, we can still compute long-range components at reduced cost on a few selected line, exploiting crystal symmetries.

The theoretical approach that we use to deduce the various components of Eq. \ref{eq:separation} is based on the static dielectric matrix formulation of the linear response problem for quasi-2d materials. 
Despite the generality of the method and the possibility to obtain exact results, we find that employing the static RPA approximation \cite{PhysRev.82.625} entails a vast simplification of the derivations at a formal level, with at the same time the possibility to (carefully) generalize the conclusions even to the presence of exchange and correlation terms. 
As done in Ref. \cite{PhysRevLett.129.185902}, we therefore employ the RPA approximation in order to derive the main theoretical results.
When computing numerical results, we will reintroduce exchange-correlation effects and defer their theoretical discussion to appendices. In order to simplify the theoretical treatment, we also restrict our arguments to the in-plane electrostatic of quasi-2d systems with in-plane mirror symmetry. 
Such systems do not mix in-plane and out-of-plane responses at the first order in the expansion of the EPI, while higher orders (such as the piezoelectric coupling) may retain information regarding the out-of-plane components \cite{https://doi.org/10.48550/arxiv.2207.10190}. Nonetheless, the influence of such terms on e.g. the mobility comes mostly from the coupling with acoustic modes \cite{https://doi.org/10.48550/arxiv.2207.10190},  
which may be screened statically even at low doping since the plasma and phonon frequencies are comparable. We therefore prefer to keep only the leading order (yet accurate) description of the EPI and discuss its modification is presence of doping.

As anticipated, within these approximations we can obtain relations for the in-plane long-range components of the dynamical matrix and of the EPI involving effective charge functions and the two-dimensional macroscopic inverse dielectric function. In particular, we define the macroscopically unscreened/screened effective charges $\bar Z_{s,\alpha}(\mathbf{ q})$/$Z_{s,\alpha}(\mathbf{ q})$ starting from the expression of the total charge change arising in a crystal following a collective displacement of the atoms of type $s$ along the Cartesian direction $\alpha$ modulated via an in-plane wavevector $\mathbf{ q}$
\begin{align}
\delta \rho^{\textrm{tot}}_{s,\alpha}(\mathbf{ q},n,T) \coloneqq -i\frac{e  q}{At} Z_{s,\alpha}(\mathbf{ q},n,T), \label{eq:0}\\
Z_{s,\alpha}(\mathbf{ q},n,T) \coloneqq \epsilon^{-1}(\mathbf{ q},n,T)\bar Z_{s,\alpha}(\mathbf{ q},n,T), \label{eq:ZZbar}
\end{align}
where $e$ is the electric charge, $A$ the area of the primitive cell of the crystal, $\epsilon^{-1}$ is the two-dimensional macroscopic dielectric function which we will introduce in the next section and $t$ is the typical scale of the electronic response along the out-of-plane direction. The above expressions can be viewed as a generalization of static effective charges tensors to the case of materials with non integer electronic statistical occupations. The long-range components of the dynamical matrix and of the EPI may then be expressed as a function of the effective charges as
\begin{align}
C^{\textrm{L}}_{ss',\alpha\beta}(\mathbf{ q},n,T)=\frac{e^2q^2}{A}v(\mathbf{q}) \bar Z^{\textrm{c.c.}}_{s,\alpha}(\mathbf{ q},n,T) Z_{s',\beta}(\mathbf{ q},n,T)
\label{eq:31}
\end{align}
where $v(\mathbf{q})$ is the two-dimensional Fourier transform of the Coulomb potential, and as
\begin{align}
g^{\textrm{L}}_{\mathbf{ q}\nu, mm'}(\mathbf{ k})= \frac{ie^2q}{A}v(\mathbf{q}) \braket{u_{m\mathbf{ k+ q}}|u_{m'\mathbf{ k}}}  \times \nonumber \\
\sum_{s\alpha} Z_{s,\alpha}(\mathbf{ q},n,T) e^{\nu}_{s,\alpha}(\mathbf{ q})l_{\mathbf{ q}\nu}\left(\frac{M_0}{M_s}\right)^{1/2},
\label{eq:gasymp}
\end{align}
where  $e^{\nu}_{s,\alpha}$ is the eigendisplacement of the vibrational mode $\nu$ and $l_{\mathbf{ q}\nu}$ its zero point motion amplitude, while $u(\mathbf{ r},z)$ are the $\mathbf{ r}$-periodic part of the Bloch function and $M_0$ and $M_s$ are, respectively, a reference atomic mass and the atomic mass of the atom $s$. 

Beside the technical interest regarding an improved accuracy in the description of the phononic properties, at a deeper level our framework allows a precise understanding of screening mechanisms at a static level. The consequent extrapolation of the bare unscreened couplings at any given doping and temperature represents a necessary prerequisite to the inclusion of dynamical effects in the description of the screened interactions, which may be of crucial relevance, e.g., in the vicinity of the plasmon resonances. 

\subsection{2d electrostatic and response functions}
\label{sec:IIsubA}
Quasi-2d systems are periodic and infinite systems in two dimensions with a finite extension in the third spatial direction---in the case of monolayers this is on the order of the atomic scale, whereas for larger thin films it can reach up to the order of micron. Such systems present a natural distinction between in-plane and out-of-plane properties.
Indeed, a periodical quasi-2d system is naturally described by reciprocal space variables in the plane of the material, and a real space variable in the out-of-plane direction.
Without loss of generality, the out-of-plane direction is aligned to the $\hat z$ direction of a Cartesian coordinate system. 2d quasi-momenta are noted $\mathbf{k}$ for electrons and $\mathbf{q}$ for phonons.
To simplify the notation, we will not distinguish between 2d and 3d vectors, whose nature can be inferred from the context. 
The quasi-2d nature of the problem is reflected in the form of the Bloch theorem for periodic systems
\begin{equation}
\psi_{m\mathbf{k}}(\mathbf{ r},z)=\frac{e^{i\mathbf{ k} \cdot \mathbf{ r}}}{\sqrt{N}} u_{m\mathbf{ k}}(\mathbf{ r},z),
\label{eq:1}    
\end{equation}
where $m$ is the band index, $u_{m\mathbf{ k}}(\mathbf{ r},z)$ is the $\mathbf{ r}$-periodic part of the Bloch function and $N$ is the number of cells in the Born-von Karman supercell. The  electrostatics of quasi-2d systems is then formulated as a function of $(\mathbf{ q},z)$. In these variables, the Coulomb kernel reads as \cite{PhysRevB.96.075448,PhysRevX.11.041027} (see App. \ref{app:A1} for the transform conventions and notations)
\begin{equation}
v(\mathbf{ q},z-z')=4\pi\int_{-\infty}^{\infty} \frac{dq_z}{2\pi} \frac{e^{iq_z(z-z')}}{ q^2+q_z^2} = 2\pi \frac{e^{- q|z-z'|}}{ q}.
\label{eq:2}
\end{equation}
The above kernel is involved in the expression of the dielectric response, which in turn determines the LRCs of the dynamical matrix and EPI, as shown in detail in Secs.  \ref{sec:IIsubB} and \ref{sec:IIsubC}. Therefore, all the response functions related to the dielectric one (whose definitions are given in App. \ref{app:A3}) need to be expressed as a function of $(\mathbf{ q},z)$. In this spirit, the independent particle polarizability (IPP) of the Khon-Sham system---i.e. the  density-density response function of an independent particle system--- is written as \cite{giuliani2005quantum}
\begin{align}
& \chi^0(\mathbf{ q}+\mathbf{ G},\mathbf{ q}+\mathbf{ G'},z,z')=\frac{2e^2}{A}\sum_{mm'\mathbf{ k}}\frac{f_{m\mathbf{ k}}-f_{m'\mathbf{ k}+\mathbf{ q}}}{\epsilon_{m\mathbf{ k}}-\epsilon_{m'\mathbf{ k}+\mathbf{ q}}} \int d\mathbf{ r} \nonumber\\
& u_{m'\mathbf{ k}+\mathbf{ q}+\mathbf{ G}}(\mathbf{ r},z) u^{\textrm{c.c.}}_{m\mathbf{ k}}(\mathbf{ r},z) \int d\mathbf{ r'} u_{m\mathbf{ k}}(\mathbf{ r'},z')u^{\textrm{c.c.}}_{m'\mathbf{ k}+\mathbf{ q}+\mathbf{ G'}}(\mathbf{ r'},z'),
\label{eq:3}
\end{align}
where $f_{m\mathbf{ k}}$ is the Fermi-Dirac occupation of the states with energy $\epsilon_{m\mathbf{ k}}$, the factor 2 takes in account spin degeneracy, the spatial integration runs over the unit cell and we have used that
\begin{equation}
u_{m'\mathbf{ k}+\mathbf{ q}+\mathbf{ G}}(\mathbf{ r'},z')=e^{-i\mathbf{ G}\cdot\mathbf{ r'}}u_{m'\mathbf{ k}+\mathbf{ q}}(\mathbf{ r'},z'),
\label{eq:4}
\end{equation}
where the $\mathbf{G}$ are reciprocal lattice vectors. The dependence of Eq. \ref{eq:3} on the carrier concentration $n$ and temperature $T$, which is present both in the Fermi-Dirac distributions and in the periodic part of the Bloch wavefunctions (we disregard the latter in this work), has been left implicit not to overburden notation. We will follow the same rule in the rest of this work when possible.

Knowing the expressions for $\chi^0$ and for the Coulomb kernel, we can express the electronic dielectric response matrix, in the RPA for the Khon-Sham ground-state \cite{giuliani2005quantum}, as
\begin{align}
\epsilon(\mathbf{ q}+\mathbf{ G},\mathbf{ q}+\mathbf{ G'} & ,z,z')=\delta(z-z')\delta_{\mathbf{ G} \, \mathbf{ G'}} \label{eq:5} \\
-2\pi \int dz'' & \frac{e^{-|\mathbf{ q}+\mathbf{ G}||z-z''|}}{|\mathbf{ q}+\mathbf{ G}|}  \chi^{0}(\mathbf{ q}+\mathbf{ G},\mathbf{ q}+\mathbf{ G'},z'',z'). \nonumber
\end{align}
The form of Eqs. \ref{eq:3} and \ref{eq:5} allows for effective approximations of the $z,z'$ dependence of the response functions.
We first assume that the periodic part of the Bloch's wave functions can be approximated as
\begin{equation}
u_{m\mathbf{ k}}(\mathbf{ r},z)=u_{m\mathbf{ k}}(\mathbf{ r}) \frac{1}{\sqrt{t}}\theta(\frac{t}{2}-|z|),
\label{eq:6}
\end{equation}
where $\theta$ is the Heaviside function and $t$ is defined as the finite layer thickness outside which the electronic cloud vanishes completely. This approximation corresponds to consider the quasi-2d material as an electronically compact homogeneous layer along the out-of-plane direction. One could choose more accurate forms for the $z$ dependence of the wavefunction, but the asymptotic long range expansions of in-plane quantities, that are the focus of this work, do not depend on such choice, as elucidated in App. \ref{app:B}.

With the approximation of Eq. \ref{eq:6}, Eq. \ref{eq:3} becomes
\begin{align}
\chi^0(\mathbf{ q}+\mathbf{ G},\mathbf{ q}+\mathbf{ G'} ,z,z') = \frac{1}{t^2}\theta(\frac{t}{2}-|z|) \theta(\frac{t}{2}-|z'|) \times \nonumber \\
\chi^0(\mathbf{ q}+\mathbf{ G},\mathbf{ q}+\mathbf{ G'}),
\label{eq:7}
\end{align}
where $\chi^0(\mathbf{ q}+\mathbf{ G},\mathbf{ q}+\mathbf{ G'})$ corresponds exactly to the IPP of a two dimensional system \cite{PhysRevB.91.165428,PhysRevB.75.205418} (see App. \ref{app:A5}). 
In other words, the approximation of Eq. \ref{eq:6} implies that the IPP of a quasi-2d material can be modelled for the 2d case and then extended uniformly along the $z$ direction inside the layer thickness, while the presence of the $1/t^2$ pre-factor assures the correct dimensionality of the response. 

Next, the out-of-plane variables are integrated out of the response functions, via an average along the out-of-plane direction. We define the layer averaged dielectric matrix as (see App. \ref{app:A3}) 
\begin{align}
\tilde \epsilon(\mathbf{ q}+\mathbf{ G},\mathbf{ q}+\mathbf{ G'})= \int_{-\frac{t}{2}}^{\frac{t}{2}} dz \frac{dz'} {t} \epsilon(\mathbf{ q}+\mathbf{ G},\mathbf{ q}+\mathbf{ G'},z,z').
\label{eq:10}
\end{align}
This quantity relates to the 2D IPP as follows (see App. \ref{app:A5}):
\begin{align}
\tilde \epsilon(\mathbf{ q}+\mathbf{ G},\mathbf{ q}+\mathbf{ G'}) =& \delta_{\mathbf{ G} \, \mathbf{ G'}} - \nonumber \\
& \tilde{v}(\mathbf{ q}+\mathbf{ G})  \chi^{0}(\mathbf{ q}+\mathbf{ G},\mathbf{ q}+\mathbf{ G'}), \label{eq:13}
\end{align}
where
\begin{align}
\tilde{v}(\mathbf{ q}+\mathbf{ G}) & =\frac{4 \pi}{|\mathbf{ q}+\mathbf{ G}|^2t}\times \label{eq:12} \\
& \left( 1-\frac{2}{|\mathbf{ q}+\mathbf{ G}|t}e^{-|\mathbf{ q}+\mathbf{ G}|\frac{t}{2}}\sinh(|\mathbf{ q}+\mathbf{ G}|\frac{t}{2})\right). \nonumber
\end{align}
Eq. \ref{eq:13} and \ref{eq:12} are particularly pleasant because we can deduce the asymptotic behaviour of the dielectric response function in relevant limits where the Coulomb kernel assumes the simple expression
\begin{align}
\tilde{v}(\mathbf{ q}+\mathbf{ G})=
\begin{cases}
\frac{2\pi}{|\mathbf{ q}+\mathbf{ G}|} \quad  |\mathbf{ q}+\mathbf{ G}|t \ll 1\\
\frac{4\pi}{|\mathbf{ q}+\mathbf{ G}|^2t} \quad |\mathbf{ q}+\mathbf{ G}|t \gg 1
\end{cases}.
\label{eq:S2DLS3DL}
\end{align}
The above limits for a quasi-2d material are the aforementioned thin limit---$|\mathbf{ q}+\mathbf{ G}|t \ll 1$---and the thick limit---$|\mathbf{ q}+\mathbf{ G}|t \gg 1$, where $\mathbf{ q}$ is still intended to be small in order to allow Taylor expansions. In these limits Eq. \ref{eq:S2DLS3DL} shows that the dependence of the Coulomb kernel upon in-plane components of the wavevector assumes the formal expression typical of, respectively, a 2d and a 3d system.

This observation is in line with what was already noted in Ref. \cite{PhysRevB.96.075448}, i.e. there exists a scale that discriminates between the two and three dimensional character of the response functions.
The same can be easily shown also for the interacting polarizability $\chi$ by considering its Dyson equation. Of course, for a realistic material there will be a crossover between the two regimes as a function of $|\mathbf{ q}|$ or $|\mathbf{ G}|$. For example, if we consider systems where $t$ is of the order of the lattice parameter, as done in this work, then in the long wavelength limit $\epsilon(\mathbf{q},\mathbf{q})\coloneqq \epsilon(\mathbf{q})$---called the `head' of the dielectric matrix---will behave as for the case of a 2d material while $\epsilon(\mathbf{q+G},\mathbf{q+G'})$---the `body'---will mostly follow the 3d behaviour (see also App. \ref{app:D} for terminology). If instead the lattice parameter is very large, 
we can expect the full matrix to show 2d behaviours. Conversely, if we consider a multilayer structure with a small in-plane unit cell, then all the elements of the response are expected to become substantially 3d with the increase of the number of layers. 
When all the relevant elements of the response are in the thin and/or thick limit, then our layer averaging procedure is well justified, as explained in App. \ref{app:B}. 
Intermediate crossover regimes, existing in ranges that 
depend on material-dependant internal parameters, 
are far more complicated to treat since general asymptotic formulae cannot be deduced. Brute-force first-principles methods then have to be used to evaluate the response functions. 

From now on, unless otherwise stated, we drop the tilde notation for the layer-averaged quantities, which can be easily recognized from the context.

\subsection{Effective charges}
\label{sec:effchargeth}
The electrostatic response of materials to external perturbating potentials can be effectively described in terms of the effective charges, which are macroscopic quantities in the sense that they do not depend on $\mathbf{G}$ vectors explicitly. They do contain, however, all the information regarding the microscopic response of the material, which instead explicitly depends on the $\mathbf{G}$ vectors components (the so-called local-fields). 
To obtain an expression for the effective charges defined in Sec. \ref{sec:frammainform}, we first introduce the reciprocal space expression of the screened Coulomb potential \footnote{notice that the screened Coulomb potential depends on two reciprocal space index, as typical for the case of non-homogeneous electron gas; the homogeneous part of the interaction is indeed represented by the diagonal components.}
\begin{align}
\w(\mathbf{ q+G},\mathbf{ q}+\mathbf{ G'})=
\epsilon^{-1}(\mathbf{ q+G},\mathbf{ q}+\mathbf{ G'})v(\mathbf{ q+ G'}).
\label{eq:xiepsvc}
\end{align}
Explicit formulae relating the matrix elements of $\w$ and its inverse $\w^{-1}$ are given in App. \ref{app:D}; the long wavelength expansion of $\w$ is given in \ref{app:A6subI}, and the limits for its macroscopic components are
\begin{align}
\w(\mathbf{q})=\epsilon^{-1}(\mathbf{ q })v(\mathbf{q})=
\begin{cases}
2\pi\frac{1}{ q+\mathbf{ q}\cdot B' \cdot \mathbf{ q}} \quad \textrm{thin}\\
\frac{ 4\pi}{t}
\frac{1}{\mathbf{ q} \cdot B'' \cdot \mathbf{ q}} \quad \textrm{thick}
\end{cases},
\end{align}
were $B'$ and $B''$ are defined from the asymptotic form of Eq. \ref{eq:32}. They are, respectively, the tensorial generalization of the effective dielectric screening length $r_{eff}$ as defined in Ref. \cite{PhysRevB.96.075448} for a material with $\epsilon^0_{eff}\approx 1$, and of the electronic dielectric constant $\epsilon^{\infty}$.

Then, we consider a collective displacement of the atoms of type $s$ along the Cartesian direction $\alpha$ modulated via a wavevector $\mathbf{ q}$.
From the electronic point of view, this may be regarded as an external charge density perturbation, which can be expressed for in-plane displacements as (see App. \ref{app:A2} and Ref. \cite{PhysRevB.88.174106})
\begin{align}
t\delta \rho^{\textrm{ext}}_{s,\alpha}(\mathbf{ q}+\mathbf{ G})=-i\frac{\mathcal{Z}_s e}{A}\left[( q_{\alpha}+ G_{\alpha})e^{-i\mathbf{ G}\cdot\boldsymbol{ \tau}_{s}}\right],
\label{eq:28}
\end{align}
where $\mathcal{Z}_s e$ is the atomic charge and $\boldsymbol{ \tau}_{s}$ indicates the position of the atom in the unit cell.
Its relation to the total electrostatic potential $\delta V^{\textrm{tot}}_{s,\beta}(\mathbf{ q})$, obtained as the sum of the induced and the external potentials, in the thin and thick limits can simply be expressed in terms of the $\w$ tensor (see App. \ref{app:E}) as
\begin{align}
\delta V^{\textrm{tot}}_{s,\alpha}(\mathbf{ q})=\sum_{\mathbf{ G'}} \w(\mathbf{ q},\mathbf{ q}+\mathbf{ G'})t\delta \rho^{\textrm{ext}}_{s,\alpha}(\mathbf{ q}+\mathbf{ G'}). \label{eq:29}
\end{align}
As shown in App. \ref{app:expansion} in the RPA, we can define the macroscopically unscreened density response $\delta \bar \rho^{\textrm{tot}}$:
\begin{align}
\delta \rho^{\textrm{tot}}_{s,\alpha}(\mathbf{ q})=\epsilon^{-1}(\mathbf{ q})\delta \bar \rho^{\textrm{tot}}_{s,\alpha}(\mathbf{ q}),
\quad \delta \bar \rho^{\textrm{tot}}_{s,\alpha}(\mathbf{ q})=\frac{\delta V^{\textrm{tot}}_{s,\alpha}(\mathbf{ q})}{\w(\mathbf{ q})}
\label{eq:2inline}
\end{align}
and conveniently write
\begin{align}
\frac{\delta V^{\textrm{tot}}_{s,\alpha}(\mathbf{ q})}{\w(\mathbf{ q})} = -i\frac{e  q}{A} \bar Z_{s,\alpha}(\mathbf{ q}).\label{eq:30}
\end{align}
We obtain at last
\begin{align}
\bar Z_{s,\alpha}(\mathbf{ q})=i\frac{Aq}{e }\sum_{\mathbf{ G'}} \frac{\epsilon^{-1}(\mathbf{ q},\mathbf{ q}+\mathbf{ G'})}{\epsilon^{-1}(\mathbf{q})}\delta V^{\textrm{ext}}_{s,\alpha}(\mathbf{ q}+\mathbf{ G'}).
\end{align}
From the above equation it is evident that both $\bar Z_{s,\alpha}(\mathbf{ q})$ and $Z_{s,\alpha}(\mathbf{ q})$ include the local-fields components of the response that contribute to the variation of the macroscopic potential. They differ in the inclusion of the long-range, macroscopic components of the screening response through Eq. \ref{eq:ZZbar}. As anticipated, the macroscopic character of Eq. \ref{eq:30} is evident from the absence of any explicit reference to the lattice vectors $\mathbf{G}$ (see also App. \ref{app:E}). 
As shown in App. \ref{app:expansion}, $\delta \bar \rho^{\textrm{tot}}_{s,\alpha}(\mathbf{ q})$ is analytical and allows for a Taylor expansion. Therefore, $\bar Z_{s,\alpha}(\mathbf{ q})$ can be written as (expliciting the dependence on the doping level $n$ and $T$ dependence) 
\begin{align}
\bar Z_{s,\alpha}(\mathbf{ q},n,T)= \frac{i}{q}M_{s,\alpha}(n,T)+\frac{ q_{\beta}}{ q}Z^*_{s,\alpha\beta}(n,T)+ \nonumber \\
-\frac{i}{2} \frac{ q_{\beta}}{ q} q_{\gamma}Q_{s,\alpha\beta\gamma}(n,T)+\ldots \,. \label{eq:expansion}
\end{align}
Since the charge density change (and the effective charge functions) is a real quantity in direct space, Eq. \ref{eq:expansion} comprises alternating imaginary and real terms in the reciprocal space expansion.
The coefficients of the expansion are site-dependent tensorial quantities (with rank proportional to the order in $q$ of the expansion), and as such they comply with the site symmetries, transforming as the totally symmetric irreducible representation. 
The rank-1 tensor $M_{s,\alpha}(n,T)$ - a polar vector, transforming as a force - is strictly zero in insulators/semiconductors where $n=0$, as it arises from intraband terms.
In the presence of free carriers, it is allowed only for those atoms whose Wyckoff positions are not fixed by symmetry, i.e., that can be subject to forces that do not lower the crystallographic symmetries, and it is indeed related to the presence of a Fermi energy shift (see Eq. (79) of Ref. \cite{RevModPhys.73.515} and the discussion in Ref. \cite{PhysRevLett.129.185902}). The second term corresponds to Born effective charge tensors, the third to dynamical effective quadrupole tensors and so on (sums over repeated indexes are intended). In Eq. \ref{eq:expansion} we highlighted the doping and temperature dependence that mostly comes from intraband contribution in the IPP for $n\neq0$, as shown in App. \ref{app:A5}. 

\subsection{Long range components}
\subsubsection{Dynamical matrix}
\label{sec:IIsubB}
We are now ready to provide the asymptotic form for the LRC of the dynamical matrix of a generic 2d material, valid for insulators, semiconductors (doped or not) and metals, and which becomes exact in the thin and thick limits. We start from the component of the force constants matrix that gives rise to LRCs, expressed as a function of the inverse dielectric screening in an all-electron formalism \cite{PhysRevB.1.910}
\begin{align}
C_{ss',\alpha\beta}(\mathbf{ R}_p,\mathbf{ R}_{p'})= \frac{\partial^2}{\partial (\mathbf{ R}_p+\boldsymbol{\tau}_{s})_{\alpha} \partial (\mathbf{ R}_{p'}+\boldsymbol{\tau}_{s'})_{\beta}} \nonumber \\ 
\int d\mathbf{ r} \int_{-\infty}^{\infty} dz \epsilon^{-1}(\mathbf{ R}_{p}+\boldsymbol{ \tau}_s,\mathbf{ r},\tau_{sz},z)\frac{\mathcal{Z}_s\mathcal{Z}_{s'}e^2}{|\mathbf{r}-\mathbf{ R}_{p'}-\boldsymbol{\tau}_{s'}|},
\label{eq:20}
\end{align}
where $s,s'$ are atomic indexes, $\alpha,\beta$ are Cartesian indexes and the spatial integration runs over the whole crystal, while $\mathbf{ R}_p$ and $\boldsymbol{\tau}_s$ indicate the position of the atom $s$ in the cell $p$, as detailed in App. \ref{app:A1}. 
The above expression is amenable for in-plane derivatives (i.e. for $\alpha,\beta=x,y$) while the out-of-plane direction is more cumbersome and model-dependent. We restrict to the study of in-plane derivatives, i.e. to in-plane modes. The LRCs of the out-of-plane modes are generally less important on the final dispersion \cite{PhysRevX.11.041027}; if needed, a more refined approach based on Ref. \cite{PhysRevX.11.041027} should be developed. 
Further restricting to single layer materials with mirror symmetry \footnote{In the case of more layers, one can model the response of the material as constant along the out-of-plane direction within the slab of the material, or one can introduce more refined models where each layer is treated separately, and remote couplings are considered.}, we can then fix the out-of-plane atomic coordinate $\tau_{sz}=0$ and can rewrite the LRC, as shown in App. \ref{app:A6}, as
\begin{align}
C^{\textrm{L}}_{ss',\alpha\beta}(\mathbf{ q})=\frac{\mathcal{Z}_s\mathcal{Z}_{s'}e^2}{A} \sum_{\mathbf{ G} \, \mathbf{ G'}} \left( q_{\alpha}+  G_{\alpha}\right)\left( q_{\beta}+  G'_{\beta}\right) \times \nonumber \\
\w(\mathbf{ q},\mathbf{ q}+\mathbf{ G'})\frac{{\w}^{\textrm{c.c.}}(\mathbf{ q},\mathbf{ q}+\mathbf{ G})}{\w(\mathbf{ q})} e^{i\mathbf{ G}\cdot \boldsymbol{ \tau}_{s}-i\mathbf{ G'}\cdot\boldsymbol{ \tau}_{s'}}. \label{eq:27}
\end{align}
If we now recast Eq. \ref{eq:27} as a function of macroscopic physical quantities alone, i.e. that do not depend on $\mathbf{G}$, as firstly done by Born and Huang starting from a phenomenological theory valid only for the case of undoped semiconductors \cite{Born1954DynamicalTO}, we end up with Eq. \ref{eq:31}. One may also find useful to rewrite Eq. \ref{eq:31} in a more symmetric form, i.e. as a function of the screened Coulomb potential as
\begin{align}
C^{\textrm{L}}_{ss',\alpha\beta}(\mathbf{ q},n,T)=\frac{e^2q^2}{A}\w(\mathbf{q}) \bar Z^{\textrm{c.c.}}_{s,\alpha}(\mathbf{ q},n,T) \bar Z_{s',\beta}(\mathbf{ q},n,T).
\end{align}
Eq. \ref{eq:31} is non-analytical for semiconductors and insulators due to the asymptotic form of the dielectric screening \cite{PhysRevLett.129.185902}. The non-analyticity is cured by the presence of free-carriers for metals and doped semiconductors, for which $\lim_{\mathbf{q}\rightarrow 0}\epsilon^{-1}(\mathbf{q})= 0$. For the latter, however, this happens in a vanishingly small region around $\Gamma$ in the limit of zero doping.

To connect to well known formulae, we rewrite Eq. \ref{eq:31} at the leading order in the $\mathbf{ q}$ expansion for the case of an insulator or an undoped semiconductor
\begin{align}
C^{\textrm{L}}_{ss',\alpha\beta}(\mathbf{ q})= \frac{e^2}{A}\w(\mathbf{q})\sum_{\gamma}{ q}_{\gamma}Z^*_{s,\gamma\alpha}\sum_{\gamma'}{ q}_{\gamma'}Z^*_{s',\gamma'\beta};
\label{eq:36}
\end{align}
one can see that the thin limit of Eq. \ref{eq:36} is equivalent to the in-plane component of Eqs. 45-46 of Ref. \cite{PhysRevX.11.041027} or to Eq. 4 of Ref. \cite{Sohier2017}, while the thick limit is the standard textbook version of the LRC of the dynamical matrix given by Born and Huang \cite{doi:10.1119/1.1934059} (Eq. 18 of \cite{PhysRevB.43.7231}). The crossover between the thin and the thick limit of Eq. \ref{eq:36} can be observed not just as a function of $\mathbf{ q}$, where the crossover scale is simply given by the effective dielectric screening length $r_{eff}$, but also as a function of $t$ for those layered materials that can be brought continuously from the one layer setup to the bulk form (as for example increasing the number of layers of h-BN in the AA stacking \cite{Verberck_prb2011,Sohier2017,Paolohbn}). As argued in App. \ref{app:B}, in this case the crossover scale given by $t$ is to be more appropriately intended as the dielectric thickness of the material.
\subsubsection{EPI}
\label{sec:IIsubC}
\begin{figure}
\includegraphics[width=\columnwidth]{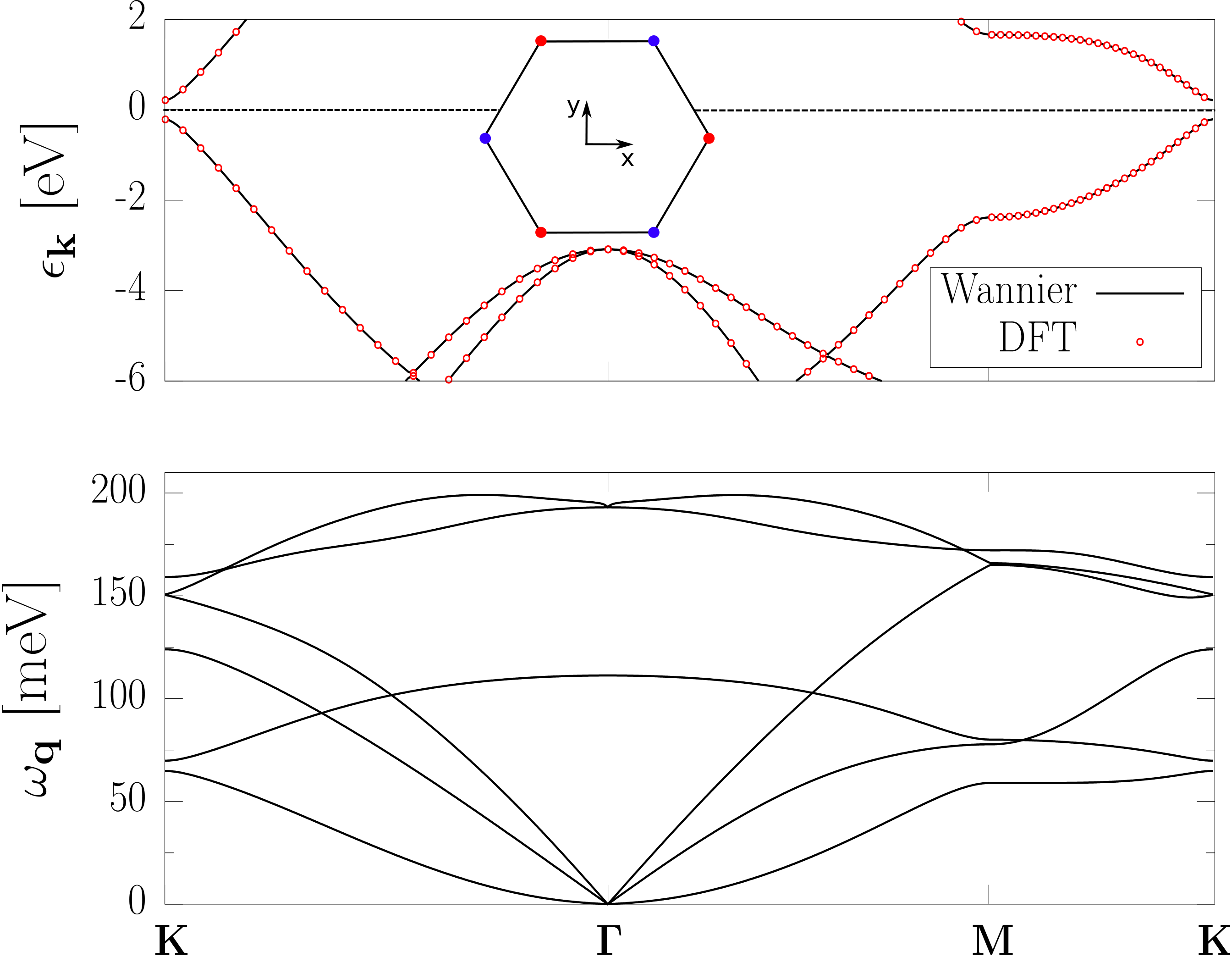}
\caption{\small Upper) \textit{ab-initio} band structure and its Wannier interpolation in the vicinity of the Fermi level, taken as energy reference at 0 eV; inset: schematics of disproportionated graphene, with the cartesian reference systems used in this work. Lower) Phonon band structure as obtained from Fourier interpolation.}
\label{fig:1}
\end{figure}
After the derivation of the LRC of the dynamical matrix in Sec. \ref{sec:IIsubB}, we can proceed along the same line and derive the LRC of the EPI. In this case, we start from the all-electron expression \cite{RevModPhys.89.015003}
\begin{align}
g_{\mathbf{ q}\nu}(\mathbf{ r},z)=\int d\mathbf{ r'}dz' \epsilon^{-1}(\mathbf{ r},\mathbf{ r'},z,z')g^{\textrm{b}}_{\mathbf{ q}\nu}(\mathbf{ r'},z'),
\label{eq:38}
\end{align}
where $g^{\textrm{b}}$ is the bare electron-phonon coupling induced by a unit displacement of wavevector $\mathbf{ q}$ of the atoms along a mode $\nu$
\begin{align}
g^{\textrm{b}}_{\mathbf{ q}\nu}(\mathbf{ r},z)=e^{-i\mathbf{q}\cdot\mathbf{r}}\Delta_{\mathbf{ q}\nu} V^{\textrm{e-n}}(\mathbf{r},\{\mathbf{R}_p+\boldsymbol{\tau_s}\}), \label{eq:39}
\end{align}
where 
\begin{align}
V^{\textrm{e-n}}(\mathbf{r},\{\mathbf{R}_p+\boldsymbol{\tau_s}\})=-\sum_{ps} \frac{\mathcal{Z}_se^2}{|\mathbf{r}-\mathbf{R}_p-\boldsymbol{\tau}_{s}|}.
\end{align}
$V^{\textrm{e-n}}$ is therefore the cell-periodic electron-nuclei interaction and $\Delta_{\mathbf{ q}\nu}$ indicates the variation following a phonon displacement, as more precisely defined in App. \ref{app:A4}. We bracket Eq. \ref{eq:38} with the Bloch functions of Eq. \ref{eq:6}, leading to
\begin{align}
& \braket{\psi_{m\mathbf{ k+ q}}|g_{\mathbf{ q}\nu}(\mathbf{ r},z)|\psi_{m'\mathbf{ k}}}=\int_{-\frac{t}{2}}^{\frac{t}{2}} \frac{dz}{t} dz' \sum_{\mathbf{ G} \, \mathbf{ G'} \, \mathbf{ G''}} u^{\textrm{c.c.}}_{m\mathbf{ k+ q}}(\mathbf{ G}) \times \nonumber \\
& u_{m'\mathbf{ k}}(\mathbf{ G'})\epsilon^{-1}(\mathbf{ q}+\mathbf{ G'}-\mathbf{ G},\mathbf{ q}+\mathbf{ G''},z,z')g^b_{\mathbf{ q}\nu}(\mathbf{ G''},z'),\label{eq:40}
\end{align}
where
\begin{align}
g^{\textrm{b}}_{\mathbf{ q}\nu}(\mathbf{ G''},z') & =i\frac{2\pi e^2}{A}\sum_{s\alpha} \frac{\mathcal{Z}_se^{-|\mathbf{ q}+\mathbf{ G''}||z'|}}{|\mathbf{ q}+\mathbf{ G''}|} \times \label{eq:41}\\
& ( q_{\alpha}+ G''_{\alpha}) e^{-i\mathbf{ G''}\cdot \boldsymbol{ \tau}_s}e^{\nu}_{s,\alpha}(\mathbf{ q})l_{\mathbf{ q}\nu}\left(\frac{M_0}{M_s}\right)^{1/2}, \nonumber 
\end{align}
and $l_{\mathbf{ q}\nu}$ is the zero point motion amplitude (App. \ref{app:A4}). Following exactly the same procedures as in Sec. \ref{sec:IIsubB} we obtain the LRC as
\begin{align}
& g^{\textrm{L}}_{\mathbf{ q}\nu, mm'}(\mathbf{ k})= \frac{i 2\pi e^2|\mathbf{ q}|}{A}\sum_{\mathbf{ G} \, \mathbf{ G'}} u^{\textrm{c.c.}}_{m\mathbf{ k+  q}}(\mathbf{ G}) u_{m'\mathbf{ k}}(\mathbf{ G'}) \times \label{eq:42}\\
& \w^{\textrm{c.c.}}(\mathbf{ q},\mathbf{ q}+\mathbf{ G'}-\mathbf{ G})\sum_{s,\alpha}\bar Z_{s,\alpha}(\mathbf{ q}) e^{\nu}_{s,\alpha}(\mathbf{ q})l_{\mathbf{ q}\nu}\left(\frac{M_0}{M_s}\right)^{1/2} \nonumber.
\end{align}
Differently from the dynamical matrix case, we haven't yet isolated the head of the $\w$ tensor in the above expression. In general, we cannot set $\mathbf{ G'}=\mathbf{ G}$ blindly because we would lose corrections to the non-leading order expansion of the EPI due to the dependence of the wings of $\w$ on $\mathbf{ q}$. We will however set $\mathbf{ G'}=\mathbf{ G}$ and verify \textit{a posteriori} that this approximation is correct. We do expect the terms coming from $\mathbf{G}\neq\mathbf{G'}$ to be small because $\w(\mathbf{q},\mathbf{q}+\mathbf{G'}-\mathbf{G})$ contributes to a further power of $q$ with respect to the terms coming from $\w(\mathbf{q})$ (see also the discussion in Ref. \cite{PhysRevLett.125.136601}). Eq. \ref{eq:42} then becomes Eq. \ref{eq:gasymp}, which may be rewritten as a function of $\w(\mathbf{q})$ as
\begin{align}
g^{\textrm{L}}_{\mathbf{ q}\nu, mm'}(\mathbf{ k})= \frac{ie^2q}{A}\w(\mathbf{q}) \braket{u_{m\mathbf{ k+ q}}|u_{m'\mathbf{ k}}}  \times \nonumber \\
\sum_{s\alpha} \bar {Z}_{s,\alpha}(\mathbf{ q},n,T) e^{\nu}_{s,\alpha}(\mathbf{ q})l_{\mathbf{ q}\nu}\left(\frac{M_0}{M_s}\right)^{1/2},
\end{align}
where we have supposed that the unperturbed Bloch functions and the phonon polarizations do not change appreciably with $n$ and $T$. To connect with well known formulae, at the leading order for undoped semiconductors we find
\begin{align}
g^{\textrm{L}}_{\mathbf{ q}\nu, mm'}(\mathbf{ k})=i\w(\mathbf{q}) \braket{u_{m\mathbf{ k+ q}}|u_{m'\mathbf{ k}}}\times \nonumber \\
\sum_{s,\alpha\beta} q_{\beta}Z^*_{s,\alpha\beta}e^{\nu}_{s,\alpha}(\mathbf{ q}) l_{\mathbf{ q}\nu}\left(\frac{M_0}{M_s}\right)^{1/2}.
\label{eq:43}
\end{align}
One can see that Eq. \ref{eq:43} in the thin limit is equivalent to Eq. 8 of Ref. \cite{PhysRevB.94.085415}, while in the thick limit it is equivalent to Eq. 4 of \cite{PhysRevLett.115.176401}---considering the phase difference as explained App. \ref{app:A1}--- and to the long wavelength expansion of Eq. 9 of \cite{PhysRevB.92.054307} once the matrix element between the periodic part of the Bloch functions is approximated to $\delta_{mn}$.

\section{Computational approach}
\label{sec:compappr}
\subsection{Connection with theory and general strategy}
\label{sec:connection}
The theoretical framework developed in the previous sections requires the knowledge of the effective charge functions, whose expression is given in Eqs. \ref{eq:ZZbar} and \ref{eq:expansion}, in order to determine the LRCs of the dynamical matrix and the EPI. The central quantity is the layer-averaged total charge density change per unit surface, $t\delta \rho_{s,\alpha}^{\textrm{tot}}(\mathbf{ q},n,T)$, that in the RPA approximation is connected to the macroscopically unscreened and screened effective charges $\bar Z_{s,\alpha}(\mathbf{ q},n,T)$ and $Z_{s,\alpha}(\mathbf{ q},n,T)$. More precisely, in App. \ref{app:expansion} we show that in RPA it is possible to deduce the macroscopically unscreened density change $\delta \bar \rho^{\textrm{tot}}_{s,\alpha}(\mathbf{ q})$, connected to $\bar Z_{s,\alpha}(\mathbf{ q},n,T)$ by Eq. \ref{eq:rhototid}, by simply imposing that the macroscopic component of the layer-averaged electrostatic potential is zero. We also remark and stress that in RPA the connection between screened and unscreened quantities is simply attained through the macroscopic inverse dielectric function.

To connect to the theoretical derivations with a deeper insight, we can sum up the above observations saying that the layer-averaging procedure performed in the theoretical section and App. \ref{app:expansion} was engineered to separate short and long range components of the dynamical matrix and of the EPI, and to mathematically demonstrate the analiticity of the expansion Eq. \ref{eq:expansion}, which terms can be deduced from the computation of $\delta \bar \rho^{\textrm{tot}}_{s,\alpha}(\mathbf{ q})$. 
However, in \textit{ab-initio} calculations we can in principle retain the out-of-plane dependence in the solution of the response problem for those quantities that we demonstrated to be analytic functions of in-plane momentum. In fact, the reintroduction of the z-dependence here does not impact the analiticity of the expressions. 

From an intuitive point of view, this corresponds to treat our slab of material as a compact layer when it responds to the long-range macroscopic electrostatic, but \textit{not} disregarding its local-field dependence (even the out-of-plane) when looking at the analyitical unscreened charges and potentials. We detail in Sec. \ref{sec:compeffcharges} how this can be done in practice. 
We anticipate that within this treatment, for \textit{both} in-plane and out-of-plane perturbations, our computational approach can be used to deduce macroscopic charge densities and potentials with the correct out-of-plane dependence, which are macroscopically unscreened both from the in-plane dielectric response and from the presence of periodically repeated images.
This is of paramount importance if one wants to extract `bare' couplings to be inserted within formalisms to compute remote EPI and their screening in van der Waals heterostructures \cite{PhysRevMaterials.5.024004}. 

We conclude this section by noting that for realistic DFT calculations we will need to include exchange and correlation terms; in this case the connection between screened and unscreened quantities is not as straightforward as in the RPA case. The procedure of unscreening itself is not unique, as detailed in App. \ref{app:xc}. In the framework of this work, it suffices to say that a good approximation for the \textit{ab-initio} evaluation of the long range components is attained by computing the inverse dielectric function in the RPA+\textit{xc} approximation (i.e. RPA plus the response of the approximated DFT exchange-correlation functional \cite{RevModPhys.89.015003}) and computing the unscreened effective charge functions by setting to zero the total macroscopic electrostatic+\textit{xc} potential---more details are given in Sec. \ref{sec:compeffcharges} and App. \ref{app:xc}. 

\subsection{Evaluation of the macroscopic inverse dielectric function}
\label{sec:IIIsubBsub2}
A practical way to compute the layer-averaged inverse dielectric matrix for a given doping and temperature is to resort to \textit{ab-initio} calculations. This approach is based on the methodology developed in Ref. \cite{PhysRevB.91.165428}, and consists in evaluating the inverse dielectric screening as
\begin{align}
\tilde {\epsilon}^{-1}(\mathbf{ q+ G},\mathbf{ q+ G'})=\frac{\delta \tilde V^{\textrm{tot}}(\mathbf{ q+ G})}{\delta \tilde V^{\textrm{ext}}(\mathbf{ q+ G'})} \label{eq:48},
\end{align}
i.e. as the variation of the layer-averaged Khon-Sham potential $\delta \tilde V^{\textrm{tot}}$ as a consequence of the variation of the external potential $\delta \tilde V^{\textrm{ext}}$. In 2d-materials the above equation is valid provided that the Coulomb potential used in the DFT framework has been modified with the Coulomb cutoff technique \cite{PhysRevB.96.075448}. Indeed, Eq. \ref{eq:48} correctly takes in account the local-fields (LFs) corrections to $\epsilon^{-1}(\mathbf{ q},n,T)$ if $\delta V^{\textrm{tot}}(\mathbf{ q},n,T)$ is the total electronic potential change at the end of the DFPT cycle. 
If instead we stop the DFPT cycle after the first iteration, i.e. when the response of the system is still non-interacting, and use Eq. \ref{eq:48} with the corresponding potential change, we obtain $\epsilon^{-1}(\mathbf{ q},n,T)$ without LFs corrections. We remind that, typically, LFs corrections are on the order of $10\%$ on $\epsilon^{\infty}$ for 3d materials \cite{PhysRevB.23.6615,PhysRevB.33.7017,refId0} or on the inverse dielectric function for 2d materials \cite{PhysRevB.91.165428}; we will show in Sec. \ref{sec:invdielf} that such corrections are small also in the system under study in this work.

 At the RPA level, the macroscopic dielectric function can also be obtained (through Eq. \ref{eq:13}) from the evaluation of the IPP, which is amenable to simplifying approximations in the doped case at finite temperature. As done in Ref. \cite{PhysRevLett.129.185902}, one can split the IPP into two contributions as:
\begin{align}
& \chi^{0,\textrm{dop.}}(\mathbf{ q},n,T)= \chi^{0,\textrm{undop.}}(\mathbf{ q})+\delta \chi^{0}(\mathbf{ q},n,T) \label{eq:47} \\
& \chi^{0,\textrm{undop.}}(\mathbf{ q}) = \frac{2}{A}\sum_{mm'\mathbf{ k}}  \frac{\theta(\epsilon_{m'\mathbf{ k}})-\theta(\epsilon_{m\mathbf{ k}})}{\epsilon_{m\mathbf{ k}}-\epsilon_{m'\mathbf{ k+ q}}} |\braket{u_{m\mathbf{ k}}|u_{m'\mathbf{ k+ q}}}|^2 \nonumber \\
& \delta\chi^{0}(\mathbf{ q},n,T) = \frac{2}{A}\sum_{mm'\mathbf{ k}}  \frac{\delta f_{m\mathbf{ k}}-\delta f_{m'\mathbf{ k+ q}}}{\epsilon_{m\mathbf{ k}}-\epsilon_{m'\mathbf{ k+ q}}} |\braket{u_{m\mathbf{ k}}|u_{m'\mathbf{ k+ q}}}|^2. 
\nonumber
\end{align}
where $\delta f=f^{\textrm{dop.}}-f^{\textrm{undop.}}$ and $\epsilon_{m\mathbf{ k}},\epsilon_{m'\mathbf{ k}}$ are taken with respect to the chemical potential. 
We can then write
\begin{align}
\epsilon(\mathbf{ q},n,T)\approx \frac{1}{\epsilon^{-1,\textrm{undop.}}(\mathbf{ q})}-e^2 v(\mathbf{q})\delta \chi^0(\mathbf{ q},n,T),
\end{align}
and approximate $\epsilon^{-1}(\mathbf{ q},n,T)\approx 1/\epsilon(\mathbf{ q},n,T)$, which corresponds to neglecting LFs. 
We have also used that, within a rigid-band approximation, band energies and overlaps can be obtained from DFT calculations in the undoped setup. A practical simplification comes from neglecting the full $\mathbf{ q}$ dependence of the dielectric response and using the asymptotic expressions
\begin{align}
1/\epsilon^{-1,\textrm{undop.}}(\mathbf{ q})\approx 
\begin{cases}
1+r_{\textrm{eff}} q \quad \textrm{thin} \\
\epsilon^{\infty} \quad \textrm{thick}
\end{cases}.
\end{align}
Further neglecting the sum over valence/conduction states in $\delta \chi^0$ for electron/hole doping, and taking $\braket{u_{m\mathbf{ k}}|u_{m'\mathbf{ k}+\mathbf{ q}}}=\delta_{mm'}$ \cite{PhysRevB.91.165428} in $\delta \chi^0$, one finally gets for the macroscopic inverse dielectric function:
\begin{align}
& \epsilon^{-1}(\mathbf{ q},n,T)\approx
\begin{cases}
\left[ 1+r_{\textrm{eff}}  q-2\pi e^2/ q \delta \chi^0(\mathbf{ q},n,T)\right]^{-1} \quad \textrm{thin} \\
\left[ \epsilon^{\infty}-\frac{4\pi}{t} e^2/ q^2 \delta \chi^0(\mathbf{ q},n,T)\right]^{-1} \quad \textrm{thick}
\end{cases}, \label{eq:6.1} \\
& \delta \chi^0(\mathbf{ q},n,T) \approx \frac{2}{A}\sum_{m\mathbf{ k}}  \frac{\delta f_{m\mathbf{ k}}-\delta f_{m\mathbf{ k+ q}}}{\epsilon_{m\mathbf{ k}}-\epsilon_{m\mathbf{ k+ q}}}.
\nonumber
\end{align}
The above equation depends on quantities that can be computed directly in the undoped case ($r_{eff}$ and $\epsilon^{\infty}$) or that can be evaluated via Wannier interpolation ($\epsilon_{n\mathbf{k}}$). For this second case, restricting the sum over $m$ only to valence/conduction bands is particularly convenient and allows to consider only a subset of bands which are situated near the chemical potential. Given the above, the $n$ and $T$ dependence of Eq. \ref{eq:6.1} stems only from the occupation functions $f_{n\mathbf{k}}$, which makes it easy to implement.

\subsection{Computation of effective charges}
\label{sec:compeffcharges}
Within the Coulomb cutoff technique of Ref. \cite{PhysRevB.96.075448}, the 2D system is still replicated  periodically along the $\hat z$ direction, implying a discrete Fourier transform with reciprocal space vectors $G_z$. 
Standard DFPT calculations readily provide all the $\{\mathbf{ G},G_z\}$ components of screened $t\delta \rho^{\textrm{tot}}(\mathbf{ q}+\mathbf{ G},G_z,n,T)$, from which we select in particular the macroscopic and layer-averaged $\mathbf{ G}=0,G_z=0$ component. 
To obtain the macroscopically unscreened $\delta \bar \rho^{\textrm{tot}}(\mathbf{ q},n,T)$ and the related $\bar Z(\mathbf{ q},n,T)$, we perform the DFPT calculation while setting to zero the $\{\mathbf{ G}=0,G_z\}$ components of i) the change of the local part of the pseudopotential and ii) change of the Hartree and exchange-correlation potentials.
This corresponds to the condition $\delta V^{\textrm{tot}}(\mathbf{ q})=0$. In particular, setting \textit{all} the $G_z$ components to zero in the DFPT problem (as opposed to disregarding only the $G_z=0$ component) is coherent with the procedure of layer-averaging the Maxwell's equation, as presented in App. \ref{app:E}, and with the formal derivation of the effective charge expansion (App. \ref{app:expansion}). 
We observe that, even though the theory of Sec. \ref{sec:II} has been developed from an all-electron perspective, the implementation of pseudopotentials in the calculation does not spoil the conclusion regarding the long wavelength expansions presented in Sec. \ref{sec:II}. This is because the approximations introduced by pseudopotentials may in general fail to reproduce the all-electron results on the response function in the opposite limit, i.e. for $\mathbf{ q}\rightarrow \infty$ (when core contributions may become visible).
\\
We stress that we only modify the electrostatic problem for the $\{\mathbf{ G}=0,G_z\}$ component, and not for all the $\mathbf{G}$.
This means that for $\mathbf{G}\neq0$ we are still keeping the out-of-plane dependence of the response problem, which translates in retaining the correct out-of-plane behaviour of (unscreened) quantities that we demonstrated to be analytic functions of in-plane momenta. From a formal perspective, this may be thought as restoring the out-of-plane variable when evaluating expressions such as Eq. \ref{eq:indcharge}, an operation that does not spoil the analyticity of the expressions. Secondly, for in-plane perturbations our condition is asymptotically equivalent to the procedure used to deduce in-plane effective charge functions via DFPT at $\mathbf{q}=0$, as shown in the results sections. 

\subsection{Evaluation of the LRCs}
\label{sec:IIIsubBsub3}
We now discuss the general procedure that shall be used within our framework to practically implement an efficient interpolation of the long range components of the dynamical matrix and of the EPI, for general $n$ and $T$.\\

For insulators and undoped semiconductors Eqs. \ref{eq:31} and \ref{eq:gasymp} (usually approximated using $\langle u_{m\mathbf{k+q}}|u_{m'\mathbf{k}}\rangle \sim \delta_{mm'}$) are non-analytical asymptotic formulae in reciprocal space deduced in the $\mathbf{q}\rightarrow 0$ limit and can therefore be used only in the neighborhoods of $\Gamma$. 
Such non-analyticity implies that their expression is not prone to be transformed in real space (in order to be then back interpolated on the full BZ) using a finite set of Wannier functions or plane waves. To overcome this problem, within the state-of-the-art methodology one performs an \textit{ansatz} for the real space transform of Eq. \ref{eq:31} and \ref{eq:gasymp} that is able to recover the $\mathbf{ q} \rightarrow 0$ asymptotic behaviours of Eqs. \ref{eq:31} and \ref{eq:gasymp} (up to a certain order) once they are transformed back in reciprocal space on the full BZ \cite{PhysRevB.50.13035,PhysRevB.92.054307}. Notice that the \textit{ansatz} is not uniquely determined. Alternatively, one can perform the \textit{ansatz} directly in Fourier space by extending each order expansion of Eqs. \ref{eq:31} and \ref{eq:gasymp} at $\mathbf{q+G}$ vectors in such a way that periodicity and continuity at zone boundary is fulfilled \cite{PhysRevB.92.054307}. This extended expression for the LRCs on the full BZ (which we will refer to as xLRCs or xLs) is then used to perform the following interpolation scheme for a pair of points $(\mathbf{ k}_0,\mathbf{ q}_0)$: one first interpolates the well-behaved differences $C^{\textrm{S}}=C-C^{\textrm{xL}}$ and $g^{\textrm{S}}=g-g^{\textrm{xL}}$, where $C$ and $g$ are obtained \textit{ab-initio} on a coarse grid defined on the full BZ, via Fourier or Wannier interpolation \cite{PhysRevB.76.165108}; then, the xLRC are re-added evaluating their closed-form expression at $(\mathbf{ k}_0,\mathbf{ q}_0)$. 

For the case of doped semiconductors, as already discussed, Eq. \ref{eq:31} and \ref{eq:gasymp} are in principle analytical, even though the extension of the region where the non analyticity is cured depends on the magnitude of the carrier concentration. 
Practically, for small doping levels it is not guaranteed that Wannier interpolation can be performed without isolating the LRCs, but instead we should repeat the above described procedure for each $n$ and $T$.
The xLRCs would then be formally obtained from their expression in the undoped setup substituting $Z_{s,\alpha}^{\textrm{undop.}}(\mathbf{ q})\rightarrow Z_{s,\alpha}(\mathbf{ q},n,T)$ and $\bar Z_{s,\alpha}^{\textrm{undop.}}(\mathbf{ q})\rightarrow\bar Z_{s,\alpha}(\mathbf{ q},n,T)$. 
In practice, for the doping levels and temperatures studied in this work, we will assume (and verify \textit{a posteriori}) that the analytical, short-range components of the matrix elements are the same as the undoped setup at zero temperature \footnote{At a given (small) doping, this is certainly valid in the usual case that the initial grid is coarse enough that the xLRCs of the undoped and doped setups coincide on that grid. However, if the initial coarse grid is fine enough to sample points within the metallic region of the response, then the different forms of the xLRCs invalidates the identity of the SRCs between the undoped and the doped setups on the coarse grid. This is when we make an approximation.}, i.e. that $C=C^{\textrm{S},\textrm{undop.}}+C^{\textrm{xL},\textrm{dop.}}(n,T)$ and $g=g^{\textrm{S},\textrm{undop.}}+g^{\textrm{xL},\textrm{dop.}}(n,T)$. In this case we can perform only one \textit{ab-initio} calculation of $C$ and $g$ on the coarse grid for the undoped setup at zero temperature, interpolate $C^{\textrm{S},\textrm{undop.}}=C-C^{\textrm{xL},\textrm{undop.}}$ and $g^{\textrm{S},\textrm{undop.}}=g-g^{\textrm{xL},\textrm{undop.}}$ on fine meshes, and finally add $C^{\textrm{xL},\textrm{dop.}}(n,T)$ and $g^{\textrm{xL},\textrm{dop.}}(n,T)$ with the correct $n$ and $T$ dependencies on the fine grid. 
Notice that if we are interested in obtaining the correct expressions for $C$ and $g$ \textit{only} in a small region in the neighborhood of $\Gamma$, then the last step of the above procedure can be simplified since the xLRCs in reciprocal space can be evaluated using directly the asymptotic expansions Eqs. \ref{eq:31} and \ref{eq:gasymp}.

\section{Results}
\label{sec:III}
\subsection{Disproportionated graphene}
\label{sec:IIIsubA}
The sublattice symmetry of graphene can be broken via the interaction with substrates as, e.g., SiC \cite{Zhou2007,Novoselov2007,PhysRevLett.115.136802}. Here, every second carbon atom has a neighbor in the bottom layer so that a different potential is felt by atoms belonging to different sublattices, thus breaking the equivalence between carbon atoms, with a consequent reported band splitting up to 0.5 eV between the valence and conduction bands at the $\mathbf{K}$ point. In this setup, so-called disproportionated graphene is equivalent to monolayer h-BN from a symmetry point of view, and non-zero Born effective charge tensors and piezoelectric coefficients arise in the system \cite{Bistoni2019}. This system may be efficiently simulated creating an imbalance of valence charge between two otherwise equivalent neighbouring carbon atoms, more precisely by adding/subtracting a fractional value of valence electrons (and a compensating ion charge) to their pseudopotentials. The Born effective charge tensor, defined as
\begin{align}
Z^*_{s,\alpha\beta}=\frac{A}{e}\frac{\partial P_{\alpha}}{\partial \upsilon_{s,\beta}(\mathbf{0})}\Big|_{\mathbf{E=0}} \label{eq:35}
\end{align}
where $P_{\alpha}$ is a Cartesian component of the polarization vector per unit surface and $\upsilon_{s,\beta}(\mathbf{0})$ is the displacement of the atom $s$ along the cartesian direction $\beta$ (as already defined in Sec. \ref{sec:effchargeth} and App. \ref{app:A1}), is diagonal with two independent components $Z^*_{\parallel}$ and $Z^*_{\perp}$ describing in-plane and out-of-plane Born effective charges \cite{PhysRevB.84.180101}.
Since we focus on the in-plane components of the Born effective charges only, which are two order of magnitude bigger than the out-of-plane ones in disproportionated graphene, we will adopt the following notation $Z^*_{s,xx}=Z^*_{s,yy}\coloneqq Z^*_s \coloneqq (-1)^s Z^*$. 
This is consistent with the charge neutrality condition for the undoped system (for the doped system, as we will see, the sum over the atom of the Born effective charges may assume non-zero values). The piezoelectric tensor
\begin{align}
\bar e_{\alpha\beta\gamma}=\frac{\partial P_\alpha}{\partial \epsilon_{\beta\gamma}}\Big|_{\mathbf{E=0}}, \label{eq:45}
\end{align}
with $\epsilon_{\beta\gamma}$ being the strain tensor, has instead a single independent coefficient $\bar e_{yyy}=-\bar e_{yxx}=-\bar e_{xyx}=-\bar e_{xxy}$, where $\hat x$ and $\hat y$ are the Cartesian directions of the reference frame of the inset of Fig. \ref{fig:1}.
In the frozen-ion approximation, the piezoelectric tensor is expressed as a function of the dynamical quadrupole charges as \cite{PhysRevB.84.180101}
\begin{align}
\bar e_{\alpha\beta\gamma}^{\textrm{FI}}=-\frac{1}{2A}\sum_s\left(Q_{s,\beta\alpha\gamma}+Q_{s,\alpha\gamma\beta}-Q_{s,\gamma\alpha\beta}\right) = \label{eq:46} \\
-\frac{i}{A}\big[\frac{\partial ( q  \bar Z_{s,\beta})}{\partial  q_{\alpha} \partial  q_{\gamma}} +\frac{\partial ( q  \bar Z_{s,\alpha})}{\partial  q_{\beta} \partial  q_{\gamma}} -\frac{\partial ( q \bar Z_{s,\gamma})}{\partial  q_{\alpha} \partial  q_{\beta}}\big]_{\mathbf{ q}=0},\nonumber
\end{align}
which accordingly admit the non-zero components $Q_{s,yyy}=-Q_{s,yxx}=-Q_{s,xyx}=-Q_{s,xxy}\coloneqq Q_s$; notice that the definition of $\bar e^{\textrm{FI}}_{\alpha\beta\gamma}$ in terms of second derivatives of $\bar Z$ allows the generalization to the case of finite doping and temperature. We also have $\bar e^{\textrm{FI}}_{yyy}=-\bar e^{\textrm{FI}}_{yxx}=-\bar e^{\textrm{FI}}_{xyx}=-\bar e^{\textrm{FI}}_{xxy}\coloneqq \bar e^{\textrm{FI}}$.

In disproportionated graphene the values of the Born effective charges and of the quadrupoles are topological properties that, adopting a simple tight-binding model consisting in a graphene Dirac Hamiltonian plus an on-site energy diagonal term $\pm \Delta$, do not depend on the magnitude of the opened gap \cite{Bistoni2019} and may be expressed in terms of the Berry curvature of the band manifold. 
Nonetheless, it is found that along with the reduction of the on-site energy and the band gap, the region of $\mathbf{k}$-point where the Berry curvature is relevantly different from 0 is closer and closer to $\mathbf{K}$ \cite{Bistoni2019}. 
It follows that for this system, even for small dopings at finite temperature, important modifications to the values of $Z^*_s$ and $Q_s$ may appear, differently from what found for 3d 3C-SiC \cite{PhysRevLett.129.185902}.
For this reason, we will study the different qualitative behaviours of the dynamical matrix and EPI in a weak and strong doping regimes (WDR and SDR respectively), as well as intermediate regimes. Disproportionated graphene is a model candidate to benchmark the whole of our theoretical developments and compare exact results (within the DFT framework) with controlled approximations.
\\
As a conclusion to this section, we show in Fig. \ref{fig:1} a schematics of disproportionated graphene and the Cartesian reference frame that we adopt in this work, alongside with its \textit{ab-initio} electronic band structure (first-principles and Wannier interpolation obtained with the computational parameter explained in Sec. \ref{sec:IIIsubBsub1}) and with the Fourier interpolated phonon dispersion.

\subsection{Computational parameters}
\label{sec:IIIsubBsub1}
To perform the first-principles calculations we use a private version of the QUANTUM ESPRESSO (QE) code \cite{doi:10.1063/5.0005082}, firstly developed in Ref. \cite{Senga2019} to compute EELS cross sections via the extraction of the screened density response to external Cartesian perturbations. We use a  lattice parameter of $a=4.6487$ Bohr, PBE-GGA functionals \cite{PhysRevLett.77.3865} and norm-conserving pseudopotentials where the valence charge of the two carbon atoms has been altered, respectively by $\pm 0.1e$, in order to mimic the sublattice symmetry breaking of disproportionated graphene \cite{Bistoni2019}. The result is a gap opening at the $\mathbf{K}$-point of around $E_G\sim0.4$ eV, as shown in Fig. \ref{fig:1}. We sample the BZ with telescopic grids---following Ref. \cite{PhysRevB.103.134304}---that in the densest region are equivalent to Monkhorst-Pack grids of dimension $256^2$ in order to well describe the region around the chemical potential, and use a energy cutoff for the plane wave basis set of $90$ Ry.

To perform the \textit{ab-initio} calculations in the doped setup we simulate hole 
densities from to $n\sim9.5\times 10^{9}$cm$^{-2}$ to $n \sim 9.5 \times 10^{12}$cm$^{-2}$ at finite temperature.
We define an effective Fermi momentum $k_F$ as the wavevector where the hole occupation halves from its value at the top of the valence band. The doping is simulated in a double-gate setup where the carrier concentration introduced in the graphene layer is compensated by two equally distanced gates that are positioned $\pm 8.3704$ Bohrs away from the graphene plane along the $\hat z$ direction.
The Coulomb kernel is treated within the 2d Coulomb-cutoff technique as developed in Ref. \cite{PhysRevB.96.075448}, using an interlayer distance between graphene periodic images of $c=37.7941$ Bohr. 

As already anticipated and shown later, the methodology developed in this work requires to perform Wannier interpolation of electronic and vibrational quantities only for the undoped setup. In this case, the Wannier interpolation of the electronic properties is obtained using the SCDM method for the determination of the starting projections \cite{Damle2015,Vitale2020} as implemented in Wannier90 \cite{Pizzi_2020} and EPW \cite{PONCE2016116}, using 5 Wannier functions and a Gaussian entanglement with $\mu$ equal to the top of the valence band and $\sigma=10$ eV. Wannierization proves to be of fundamental importance when describing the electronic properties of disproportionated graphene.
In fact, simple tight-binding models cannot fully account for the complete orbital composition of the Bloch functions implying poor accuracy in the determination of, e.g., the angular dependency of the EPI. The Wannier interpolation of the electron-phonon matrix elements and of the dynamical matrix is performed using a private version of EPW, adapted for the study of the current work, using coarse meshes of $48\times 48\times 1$ for the electrons and $12\times 12\times 1$ for the phonons. The convergence of the electronic inverse scattering times (see Sec. \ref{sec:IIIsubE}) is obtained using fine grids of $\mathbf{ q}$ points of dimensions $700^2$ and a Gaussian smearing for the Dirac delta functions of $15$ meV. 

\subsection{Inverse dielectric function}
\label{sec:invdielf}
\begin{figure}
\includegraphics[width=\columnwidth]{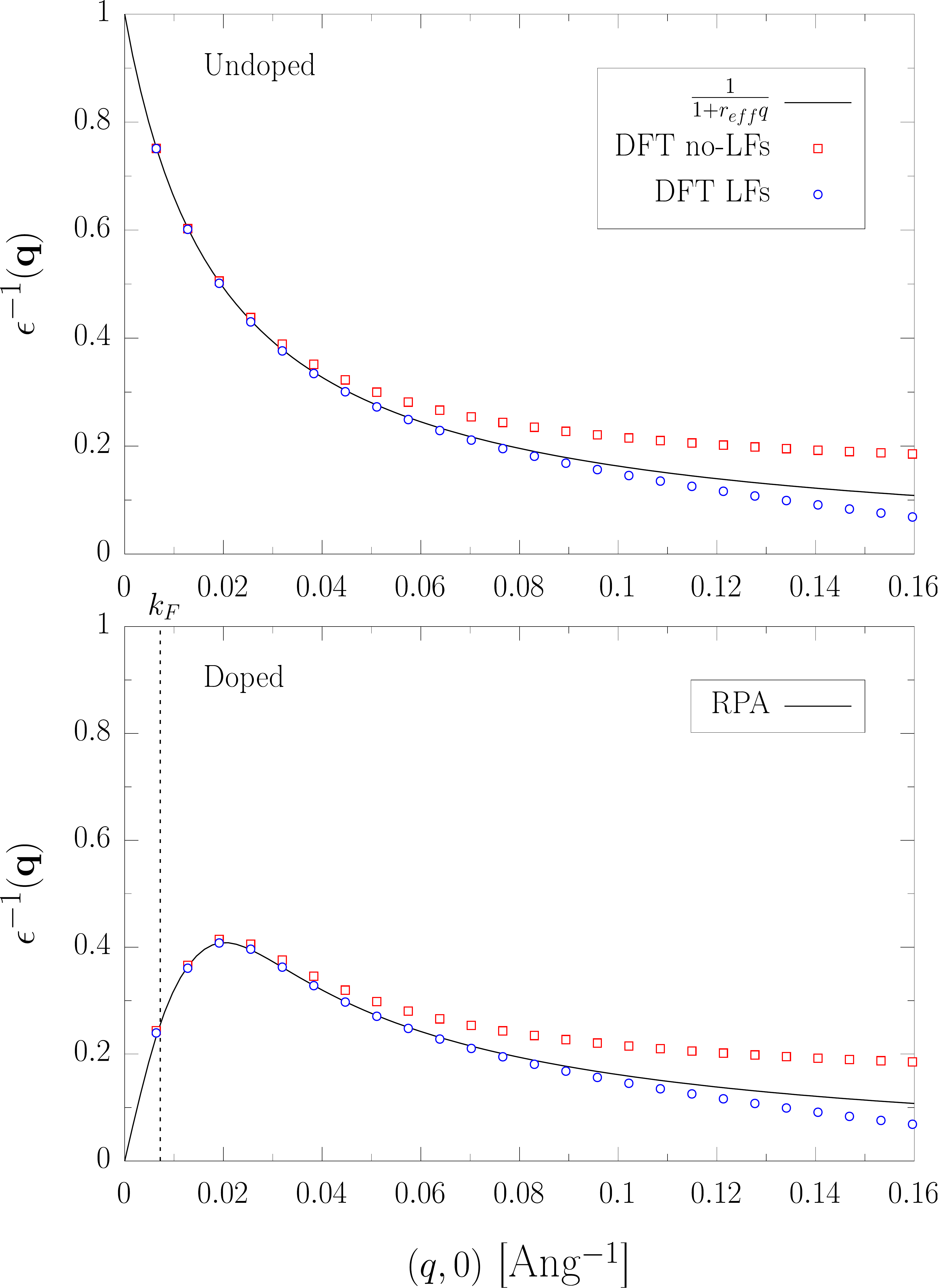}
\caption{\small Upper) Dielectric response, for disproportionate graphene in the undoped case, from the model of Eq. \ref{eq:TSscreening}, compared with the DFT calculation (RPA+\textit{xc}) obtained using Eq. \ref{eq:48} with or without the inclusion of local-fields (LFs) corrections, as explained in Sec. \ref{sec:compappr}, along the reciprocal space Cartesian line form $(q,0)$. The Cartesian reference frame used in this work is depicted in the inset of Fig. \ref{fig:1}. Lower) Same quantities for the case of $n\sim9.5\times 10^{9}$cm$^{-2}$ and a temperature of $4$ meV, compared with the RPA approximation of Eq. \ref{eq:6.1}. The effective Fermi wavevector $k_F$, defined in the text, is represented by a black vertical dashed line.}
\label{fig:2}
\end{figure}
The accurate description of the inverse dielectric function is a focal point of our approach. We first note that LFs have a quantitatively small effect on the asymptotic small $\mathbf{q}$ value of $\epsilon^{-1}(\mathbf{q})$, as shown in Fig. \ref{fig:2}. 
It is evident that in both doped and undoped cases, LFs have an impact on $\epsilon^{-1}$ of the order of some percent for wavevectors smaller than $\mathbf{ q}\sim 0.02$Bohr$^{-1}$. Thus, they may be neglected in this region, which is the most interesting in practice. 
Incidentally, this shows that the inclusion of \textit{xc} effects in the inverse dielectric response is of little importance to our aims. Importantly, we notice that the approximation of the inverse dielectric function via Eq. \ref{eq:6.1} works very well.

\subsection{Effective charges}
\begin{figure}
\includegraphics[width=\columnwidth]{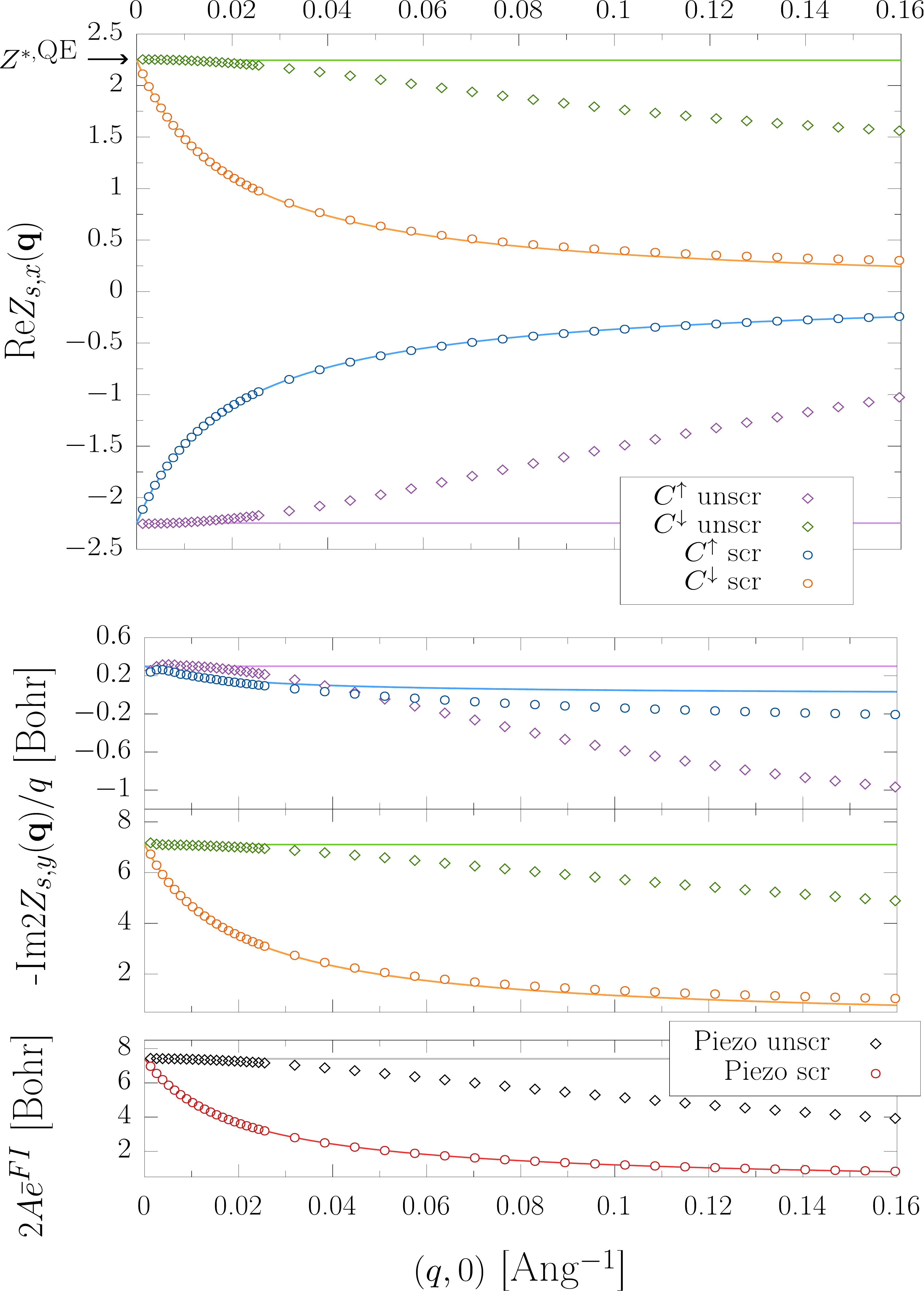}
\caption{\small Upper) Real part of the macroscopically unscreened (`unscr') and screened (`scr') effective charge for undoped disproportionated graphene at zero temperature, for the carbon atom with increased/decreased valence charge $C^\uparrow/C^\downarrow$. Lower) Same notation for the imaginary part divided by $ q$, summed over the atoms in the lowest panel to obtain $2A\bar e^{\textrm{FI}}$ (`piezo') as intercept with the vertical axis (see Eq. \ref{eq:46}). In the upper panel, the continuous horizontal lines correspond to $Z^{*,\textrm{QE}}_s$, while in the lower one to the fitted values of the quadrupoles $Q_s$ and of $2A\bar e^{\textrm{FI}}$. The lines passing through the screened charge data are, respectively, $\frac{Z^{*,\textrm{QE}}_s}{(1+r_{\textrm{eff}} q)}$, $\frac{Q_s}{(1+r_{\textrm{eff}} q)}$ and $\frac{2A\bar e^{\textrm{FI}}}{(1+r_{\textrm{eff}} q)}$ where $r_{\textrm{eff}}\sim97$Bohr is computed using Eq. \ref{eq:reff}.}
\label{fig:chargeneutral}
\end{figure}
\label{sec:IIIsubC}
\begin{table}[h]
\begin{center}
\renewcommand{\arraystretch}{0.90}
\begin{tabular}{| c | c | c |}
\hline
\multicolumn{1}{|c|}{$Z_{s,\alpha}$ \rule{0pt}{3ex}} & \multicolumn{2}{c|}{Reciprocal space direction}  \\[1ex]
\hline
\rule{0pt}{3ex} & $( q,0)$  & $(0, q)$ \\[1ex]
\hline
$Z_{s,x}$ \rule{0pt}{3ex} & $\frac{(-1)^sZ^*+\mathcal{O}( q^2)}{1+r_{\textrm{eff}} q+\mathcal{O}( q^2)}$ & $\mathcal{O}( q^2)$ \\[1.5ex]
\hline
$Z_{s,y}$ \rule{0pt}{3ex} & $\frac{1/2 Q_s  q+\mathcal{O}( q^2)}{1+r_{\textrm{eff}}  q+\mathcal{O}( q^2)}$ & $\frac{(-1)^sZ^*  q-1/2Q_s  q+\mathcal{O}( q^2)}{1+r_{\textrm{eff}}  q+\mathcal{O}( q^2)}$  \\[1.5ex]
\hline
\end{tabular}
\end{center}
\caption{\small Reciprocal space expansion of $ Z_{s,\alpha}(\mathbf{ q})$, for the case of undoped disproportionated graphene along different directions of the BZ; $\epsilon^{-1}(\mathbf{ q})$ is approximated as in Eq. \ref{eq:TSscreening}.}
\label{tab:2dcart}
\end{table}
Now, we can study the macroscopically screened and unscreened layer-averaged effective charges, $Z_{s,\alpha}(\mathbf{ q})$ and $\bar Z_{s,\alpha}(\mathbf{ q})$, for undoped disproportionated graphene along a specific direction in the reciprocal space. 
We focus on the in-plane components of the effective charges. Within this framework, $M_{s,\alpha}(n,T)=0$ by symmetry considerations.

The choice of the line is performed in order to  isolate the different terms of the expansion of Eq. \ref{eq:expansion}. In fact, the symmetry properties of the effective charges tensors enforce specific behaviours, as shown in Tab. \ref{tab:2dcart} for two possible directions. 
We choose the $( q,0)$ line since the leading order coefficients of $\bar Z_{s,x}(\mathbf{ q})$ and $\bar Z_{s,y}(\mathbf{ q})$ are respectively the Born effective charges and the dynamical quadrupole tensors. 
Since there are no lines where the leading order of the expansion of $\bar Z_{s,\alpha}(\mathbf{ q})$ is represented by the octupole term, we stop our expansion at the quadrupole level. 
We then plot in the upper panel of Fig. \ref{fig:chargeneutral} the real part of $\bar Z_{s,x}(\mathbf{ q})$ and $ Z_{s,x}(\mathbf{ q})$, alongside with the value of Born effective charges calculated from \textit{ab-initio} DFPT calculation at $\mathbf{q}=0$, which we denote by $Z^{*,\textrm{QE}}_s$. 
For small wavevectors, both $\textrm{Re}\bar Z_{s,x}(\mathbf{ q})$ and $\textrm{Re}Z_{s,x}(\mathbf{ q})$ tend to $Z^{*,\textrm{QE}}_s$.
For larger wavevectors ($ q\sim 0.02$ Bohr$^{-1}$), the unscreened charge departs from its asymptotic value. Indeed, the identification of $\bar Z_{s,x}(\mathbf{ q})$ with the effective charges starts to degrade because we are exiting from the thin limit---in fact, if we evaluate $t\sim 6$ Bohr, which is the typical interlayer distance in graphite, at $0.02$ Bohr$^{-1}$ we have $qt\sim0.1$.

As it regards $\textrm{Re}Z_{s,x}(\mathbf{ q})$ instead, we notice that it follows the simple model from Ref. \cite{PhysRevB.94.085415} 
\begin{align}
\textrm{Re}Z_{s,x}(\mathbf{q})=\epsilon^{-1}(\mathbf{q}) Z_s^{*,\textrm{QE}} \\
\epsilon^{-1}(\mathbf{ q})=\frac{1}{1+r_{\textrm{eff}}  q},
\label{eq:TSscreening} \\
r_{\textrm{eff}}=(\epsilon^{\infty,\textrm{QE}}-1)\frac{c}{2}
\label{eq:reff},
\end{align}
way beyond the expected regime of validity, where $c \gg t$ is the separation of the periodic images used in the simulation. 
Dividing the the Born effective charges with the simplest thin limit asymptotic expression of $\epsilon^{-1}(\mathbf{ q})$ thus works surprisingly well.

This seems to be a feature common to different materials \cite{PhysRevB.94.085415,PhysRevMaterials.5.024004,Sohier2017}, due to cancellation of errors between the wrong estimations of the effective charge by its asymptotic value and of the screening with respect to the local field inclusion at finite wavevectors. All the above conclusions can be drawn also analyzing the quadrupole and piezoelectric tensors, as done in the lower panel of Fig. \ref{fig:chargeneutral}. At difference with the Born effective charge case, the \textit{ab-initio} calculation of quadrupole from DFPT in QE is not implemented at the time of writing.
We therefore estimate the quadrupole values directly from the asymptotic values of $-\textrm{Im}2\bar Z_{s,y}(\mathbf{ q})/ q$. In the same way, we estimate also the value of the frozen ion piezoelectric tensor, as defined in Eq. \ref{eq:46}.
We find that $\bar e^{\textrm{FI}}=6.1\times10^{-10} \frac{\textrm{C}}{\textrm{m}}$, in perfect agreement with the values found in Ref. \cite{Bistoni2019} .

\begin{figure*}
\includegraphics[width=\textwidth]{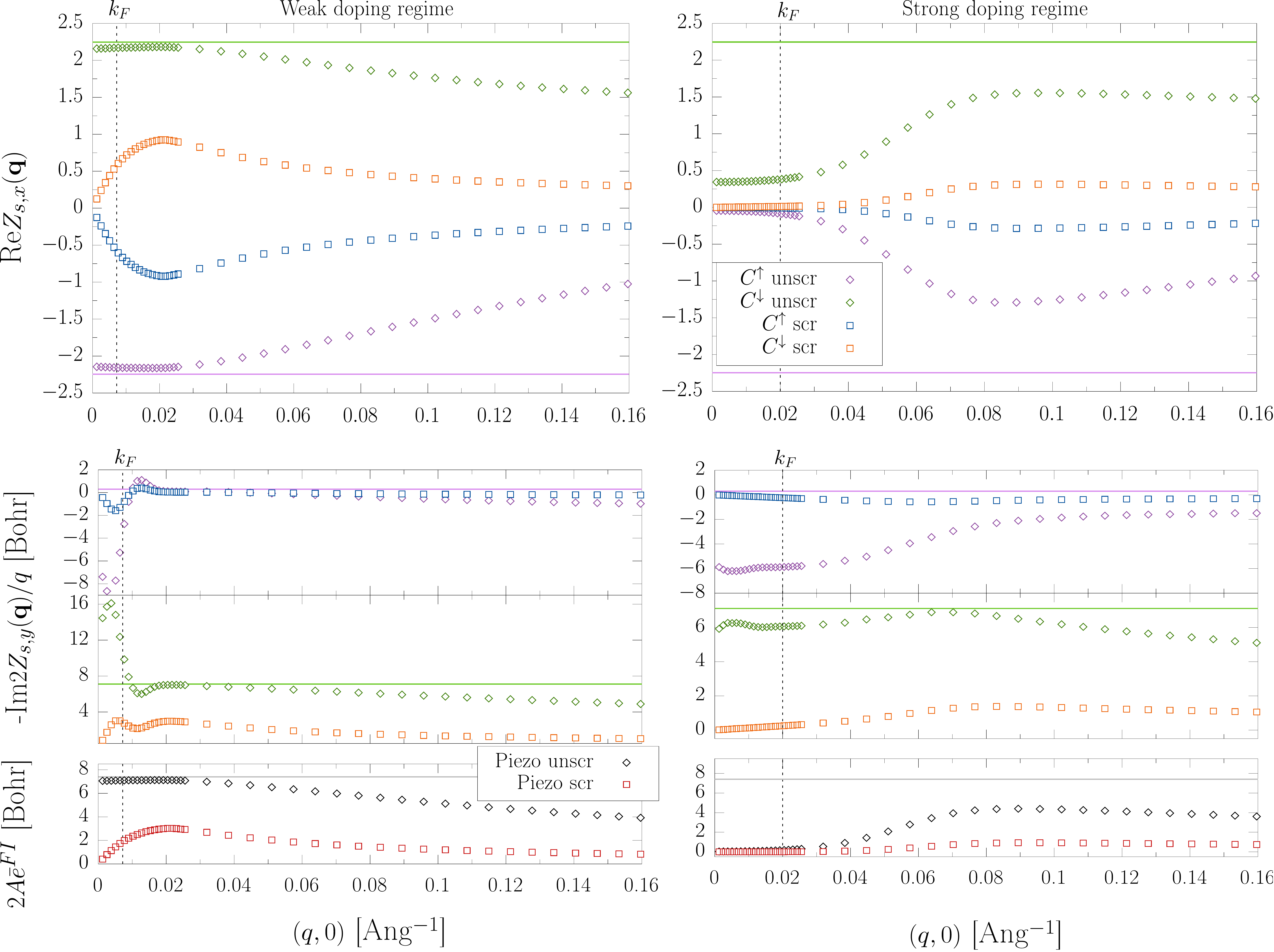}
\caption{\small Same quantities and same symbol-colour scheme as in Fig. \ref{fig:chargeneutral}, for the weak and strong doping regimes (left and right columns). The horizontal lines are the values of the asymptotic quantities deduced for the undoped setup.}
\label{fig:chargepanel1}
\end{figure*}

We now consider the same quantities but in the doped case,  starting with the WDR and SDR at, respectively, $n\sim9.5\times 10^{9}$cm$^{-2}$ and $n\sim1.9\times 10^{12}$cm$^{-2}$. In the first case, to mimic the zero temperature regime we use a Fermi-Dirac (FD) occupation with a temperature of $4$ meV, while in the second we use a Methfessel-Paxton (MP) distribution  \cite{PhysRevB.40.3616} with a smearing of $40$ meV---we will come back on elucidating this choice later. We plot in Fig. \ref{fig:chargepanel1} the same quantities as Fig. \ref{fig:chargeneutral}, respectively for the WDR on the left column and for the SDR on the right column. 
In the WDR regime we observe a reduction of the asymptotic value of $\textrm{Re}\bar Z_{s,x}(\mathbf{ q},n,T)$ with respect to the undoped case that only slightly impacts on the change of $\textrm{Re}Z_{s,x}(\mathbf{ q},n,T)$ at small wavevector, which is instead mostly determined by the influence of metallic screening. 
We also notice that in the regime where metallicity fades out, i.e. at large $\mathbf{ q}$, the charge response returns to the typical value for the semiconductor response. 

The same conclusions cannot seemingly be drawn for the quadrupole term, since peculiar behaviours arise at small $\mathbf{ q}$. This is not a serious issue since the relevant macroscopic parameter which enters the EPI in the long wavelength limit is the piezoelectric tensor, which is well-behaved. Indeed, in general at the quadrupole order we see that the added carriers substantially alter the unscreened part of the response through the $\mathcal{C}(\mathbf{ q},n,T)$ term of Eq. \ref{eq:Minterintra}, which originates mainly from the intraband contributions to the IPP coming from partially filled bands. In the charge expression, this means that $-\textrm{Im}2\bar Z_{s,x}(\mathbf{ q},n,T)/ q$ strongly departs from the behaviour shown in the undoped case. 
Nonetheless, if we look at the value of the piezoelectric tensor in the WDR as defined by the second derivative of the $\bar Z$ quantities (Eq. \ref{eq:46}), we notice that this is very similar to $\bar e^{\textrm{FI}}$, because the sum over atoms of the term coming from $\mathcal{C}(\mathbf{ q},n,T)$ results to cancel out almost completely.
It follows that $-\textrm{Im}2\sum_s Z_{s,x}(\mathbf{ q},n,T)/ q \approx 2A \epsilon^{-1}(\mathbf{ q},n,T) \bar e^{\textrm{FI}} $. The independence of the frozen ion piezoelectric tensor from doping in the WDR is interesting since the acoustic EPI is directly proportional to $\bar e^{\textrm{FI}}$ in the long wavelength limit.
This justifies neglecting the complicated doping and temperature dependence coming from the $\mathcal{C}(\mathbf{ q},n,T)$ term in the evaluation of the leading orders of the EPI.

In the SDR the situation is very different. The value of the Born effective charges is completely suppressed and so is the value of the dynamical quadruple tensors, due to the strong modifications of the wings of the dielectric response at high doping concentrations. Notice that now the sum rule for the Born effective charges $\sum_s Z^*_s=0$ is strongly broken.
This is allowed because the total dynamical matrix is expressed as a function of Eq. \ref{eq:22} as $ C^{\textrm{tot}}_{ss',\alpha\beta}(\mathbf{ q})=C_{ss',\alpha\beta}(\mathbf{ q})-\delta_{ss'}\sum_{s''} C_{ss'',\alpha\beta}(\mathbf{ 0})$, and the relevant translational invariance condition $\lim_{\mathbf{ q}\rightarrow 0}\sum_{ss'}C^{\textrm{tot}}_{ss',\alpha\beta}(\mathbf{ q})=C^{\textrm{tot}}_{ss',\alpha\beta}(\mathbf{ 0})$ \cite{PhysRevB.1.910} is trivially satisfied in the metallic regime of the dielectric response even if $\sum_s Z^*_s\neq0$.

\begin{figure}
\includegraphics[width=\columnwidth]{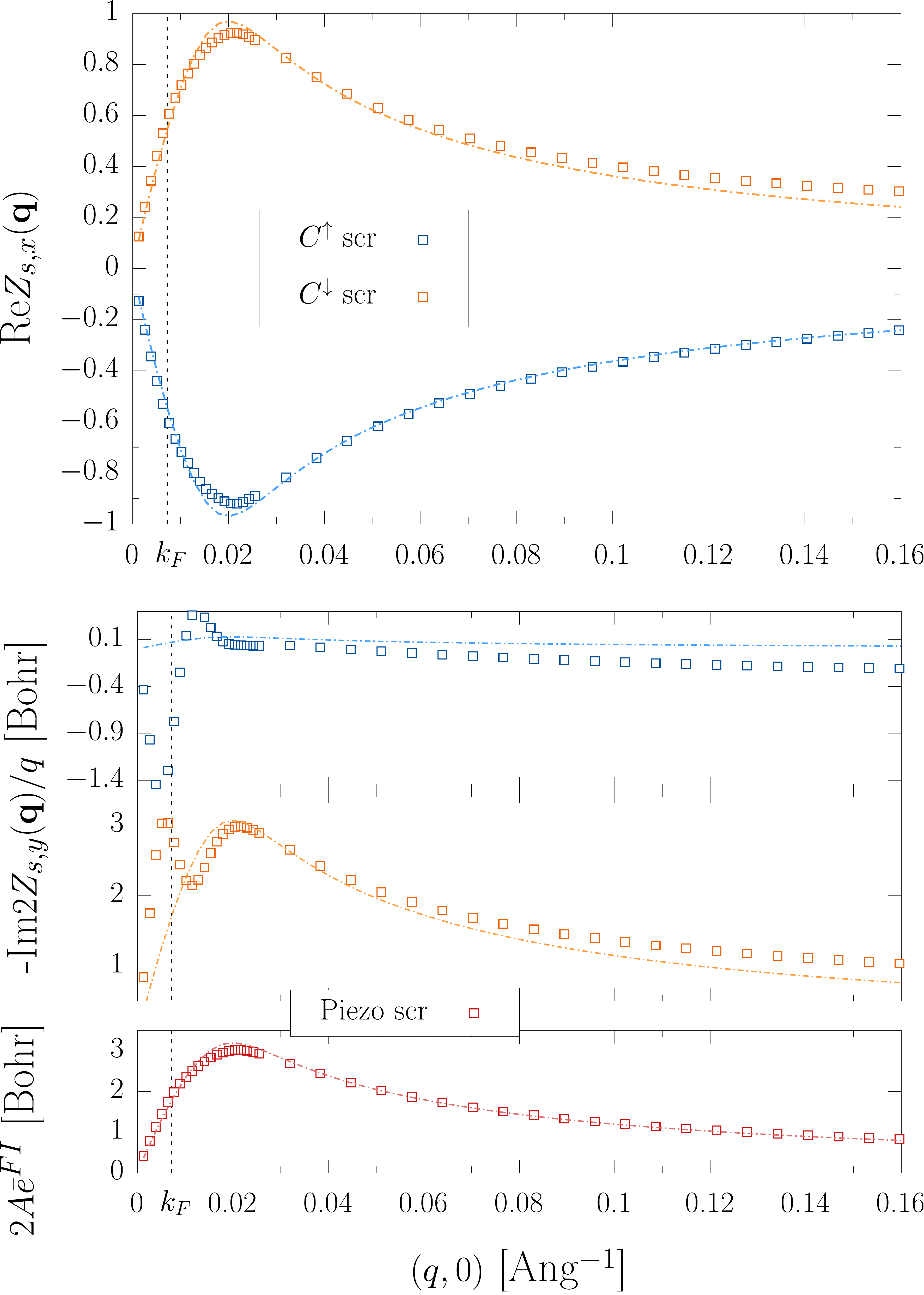}
\caption{\small Same quantities and same symbol-colour scheme as in Fig. \ref{fig:chargeneutral}, for the weak doping regime, compared to the results obtained screening $Z_s^{*,\textrm{QE}},Q_s$ and $\bar e^{FI}$ with the $\epsilon^{-1}(\mathbf{q},n,T)$ evaluated via Wannier interpolation of Eq. \ref{eq:6.1} (dashed lines).}
\label{fig:chargepanel2}
\end{figure}

\begin{figure}
\includegraphics[width=\columnwidth]{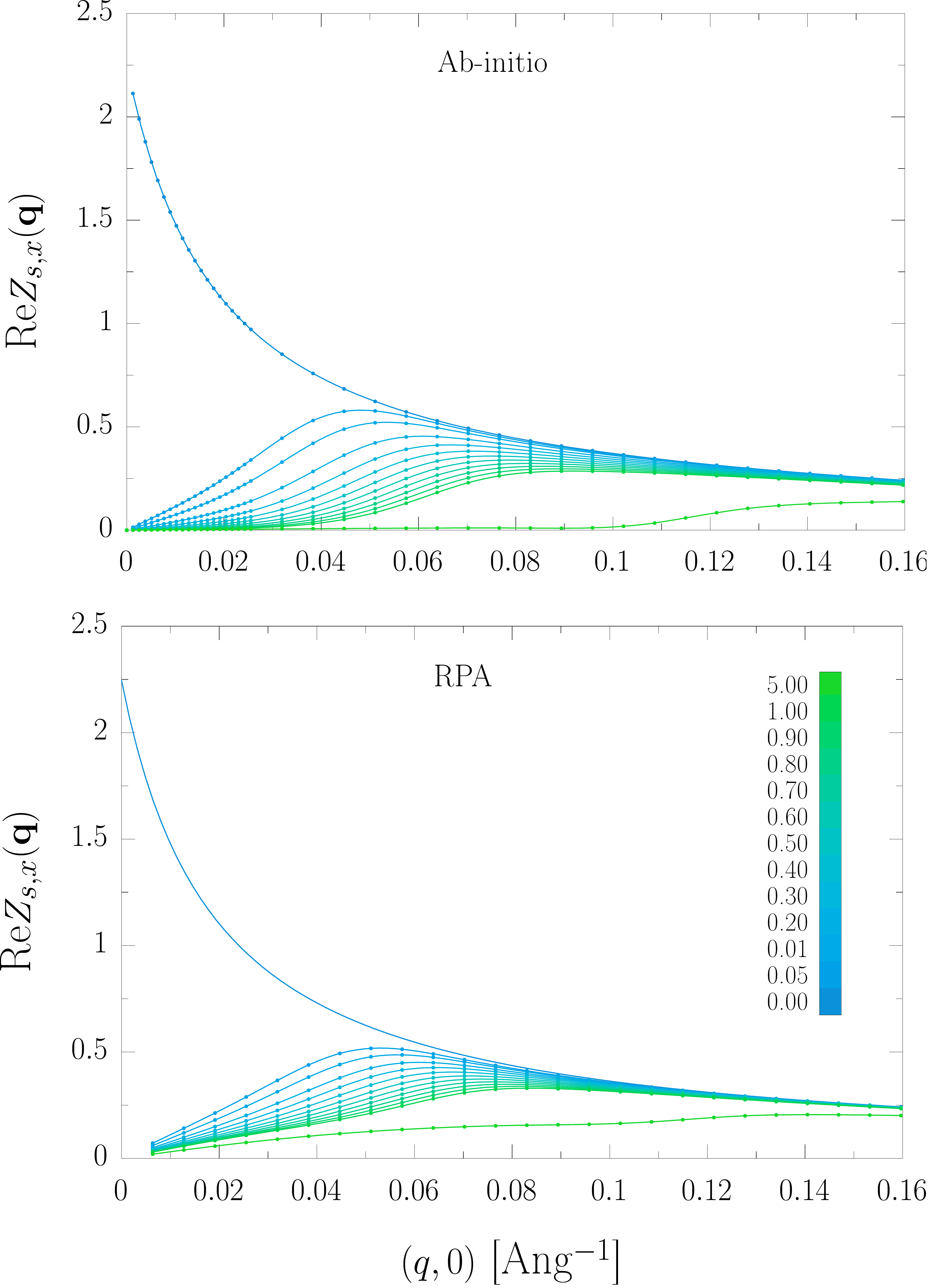}
\caption{\small Comparison of $\textrm{Re}Z_{s,x}(\mathbf{ q})$ obtained with \textit{ab-initio} calculations and the RPA approximation as explained in the text (upper and lower), for different values of hole concentration per unit cell, reported in units of $10^{-3}e$ units in the legend.}
\label{fig:charge}
\end{figure}

To conclude, the qualitative difference between the WDR and the SDR is that in the first regime the $n,T$ dependencies of the Born effective charges and of the piezoelectric tensor can be considered, to a good approximation, to enter only in the head of the inverse dielectric screening. 
This is highly important because it implies a simple operative procedure to determine the LRCs of the dynamical matrix and EPIs, disregarding the effects of doping on unscreened effective charges (the $\mathcal{C}$ term in Eq. \ref{eq:Minterintra}). To quantify this observation, Fig. \ref{fig:chargepanel2} shows the \textit{ab-initio} screened charges of Fig. \ref{fig:chargepanel1} together with the expressions $\epsilon^{-1}(\mathbf{ q},n,T)Z^{*\textrm{QE}}_s$, $ \epsilon^{-1}(\mathbf{ q},n,T)Q_s$ and $2A \epsilon^{-1}(\mathbf{ q},n,T)\bar e^{\textrm{FI}}$,  where $\epsilon^{-1}(\mathbf{ q},n,T)$ is evaluated via Wannier interpolation of Eq. \ref{eq:6.1}.
We also perform the same comparison in Fig. \ref{fig:charge} for $\textrm{Re} \bar Z(\mathbf{ q},n,T)$ for several different densities, using a MP smearing of $40$ meV to evaluate the statistical occupations. MP smearing has the drawback of losing the correspondence with a physical temperature, but it requires coarser samplings of the BZ in order to get converged results, which is of extreme importance in order to perform numerous calculations. In general, we find a good agreement between \textit{ab-initio} calculation and the RPA modelling but, as expected, this increasingly deteriorates while approaching the SDR.

\subsection{Frequencies and EPI}
Now that the behaviour of $\bar Z_{s,\alpha}(\mathbf{ q},n,T)$ and $Z_{s,\alpha}(\mathbf{ q},n,T)$ is understood, we turn to the interpolation of the dynamical matrix and EPI, using the procedure exposed in Sec. \ref{sec:IIIsubBsub3}. In principle, Eqs. \ref{eq:31} and \ref{eq:gasymp} are rigorously valid only in the thin and thick limits. 
Nonetheless, we can expect the validity of Eq. \ref{eq:gasymp} to be extended like that of $Z_{s,\alpha}(\mathbf{ q},n,T)$ in  Eq. \ref{eq:TSscreening}.
For Eq. \ref{eq:31} the considerations are a bit different since the LRC of the dynamical matrix depends both on the unscreened and the screened charge density changes. 
Here, the validity of Eq. \ref{eq:31} is extended using for $\bar Z_{s,\alpha}(\mathbf{q},n,T)$ the leading order asymptotic value obtained for the undoped setup, i.e. $Z^{*,QE}_s$.
This is coherent with the fact that at large $\mathbf{ q}$ the LRC perform a crossover to the value of the undoped setup. At sufficiently small wavevectors instead, the correct numerical value of $\bar Z_{s,\alpha}(\mathbf{ q},n,T)$ is not thoroughly needed since the whole LRC is suppressed via screening.

As discussed in the previous section, the evaluation of $Z_{s,\alpha}(\mathbf{ q},n,T)$ may be performed exploiting the RPA approximation of the head of the inverse dielectric screening. If this is not precise enough, as in the SDR,  $Z_{s,\alpha}(\mathbf{ q},n,T)$ may be evaluated \textit{ab-initio}.
In that case, we shall perform calculations on the minimum number of lines that are required by a symmetry analysis to describe all the relevant independent components of $Z_{s,\alpha}(\mathbf{ q},n,T)$, and then use symmetry relations to complete the description on the full BZ. Also, we can sample $Z_{s,\alpha}(\mathbf{ q},n,T)$ on each needed line on a relatively coarse set of points and then perform an inexpensive linear interpolation on finer sets of points. All the various steps that compose our strategy carefully avoid to resort to brute-force calculations of all the matrix elements of the dynamical matrix and EPI on the full BZ. Of course, for special regimes which fall very far from the thin or thick limits, our machinery is no more applicable.
\begin{figure*}
\includegraphics[width=\textwidth]{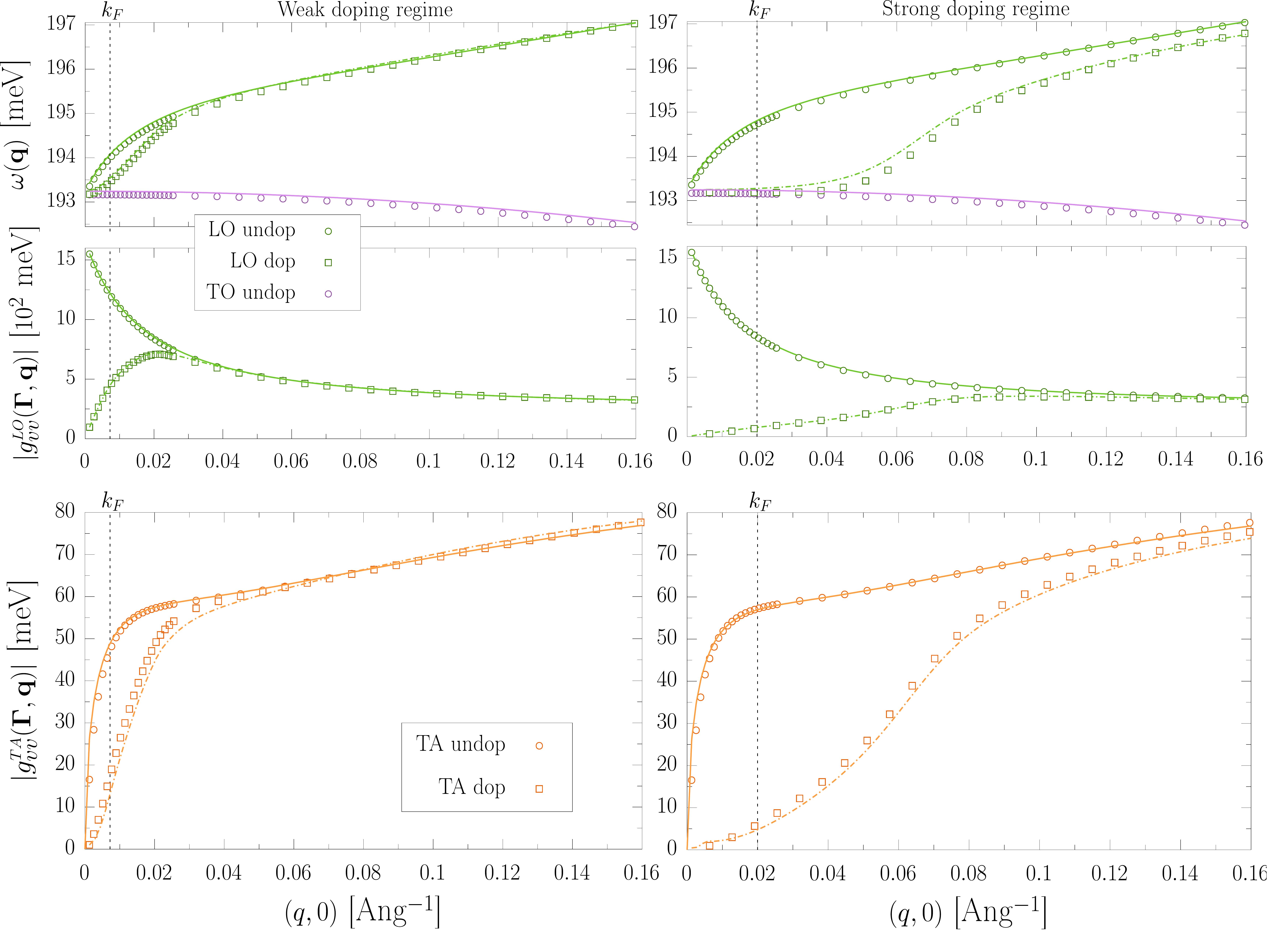}
\caption{\small Values for the TO and LO phonon frequencies (upper row) and of the EPI (central and lower rows) for the undoped (circles) and doped (squares) setups, in the WDR (left column) and in the SDR (right column). The central row shows the Fr\"olich coupling with the LO mode (LO), while the lower one is relative to the coupling between the TA mode and the dynamical effective quadrupoles. The continuous and dashed lines represent the Wannier interpolation of, respectively, the undoped and the doped setups. The TO modes and their EPI are not showed for the doped case as they are substantially left invariant by doping.}
\label{fig:interp}
\end{figure*}

\begin{figure*}
\includegraphics[width=\textwidth]{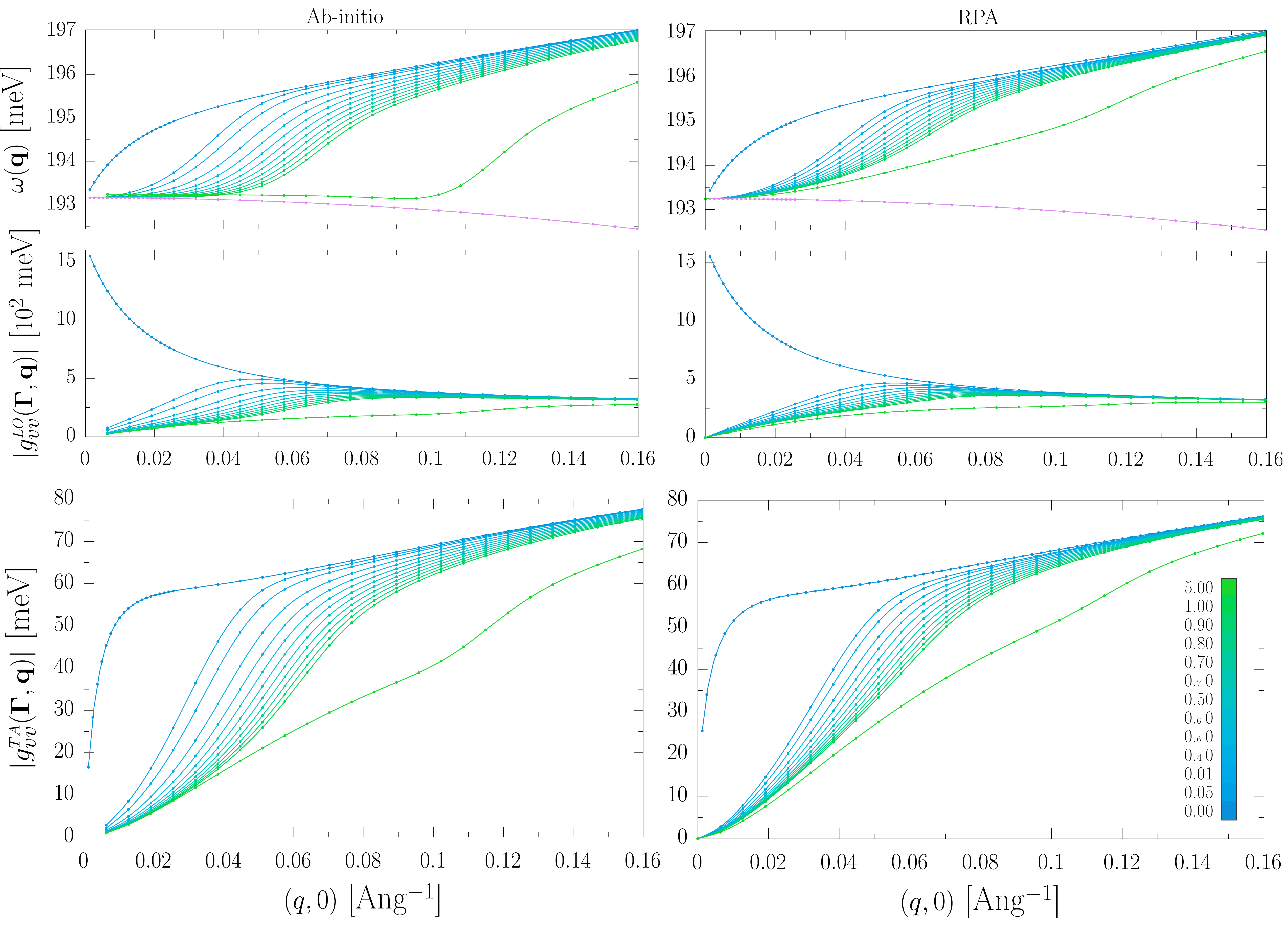}
\caption{\small Same quantities as in Fig. \ref{fig:interp}, compared between \textit{ab-initio} calculations (left column) and the RPA approach as described in the text (right column).}
\label{fig:EPCFREQdoping}
\end{figure*}

The previous observations are quantified in Fig. \ref{fig:interp}, interpolating on the line $( q,0)$ for the dynamical matrix and EPI for the WDR and SDR (left and right columns). 
In both cases the LRCs are evaluated using directly the $Z_{s,\alpha}(\mathbf{ q},n,T)$ coming from \textit{ab-initio} calculations. The long-range features of the coupling and of the phonon dispersion are strongly suppressed in the region near $k_F$, and an excellent agreement between \textit{ab-initio} calculations and interpolated quantities is obtained. Finally, as for effective charges, frequencies and EPI values for different dopings are plotted in Fig. \ref{fig:EPCFREQdoping}. \textit{Ab-initio} results are compared with those obtained computing the LRCs with the RPA approximation to $\epsilon^{-1}(\mathbf{ q},n,T)$. As for the charges, we find that the general trends agree in the WDR while they are increasingly different while approaching the SDR.
\label{sec:IIIsubD}
\subsection{Lifetimes}
\label{sec:IIIsubE}
\label{sec:5}
We can finally quantify the impact that the correct descriptions of the EPI and phonon frequencies have on physical observables; in particular, we compute the electronic inverse lifetimes evaluated in the Self Energy Relaxation Time Approximation (SERTA) \cite{Ponce2020} as
\begin{align}
\tau^{-1}_{m\mathbf{ k}}&=\frac{2\pi}{\hbar} \sum_{m'\nu} \int\frac{d\mathbf{ q}}{A_{\textrm{BZ}}} |g_{\mathbf{ q}\nu,m'm}(\mathbf{ k})|^2 \times \nonumber \\
&\big[(1-f_{m'\mathbf{ k}+\mathbf{ q}}+n_{\nu\mathbf{ q}})\delta(\epsilon_{m\mathbf{ k}}-\epsilon_{m'\mathbf{ k}+\mathbf{ q}}-\hbar \omega_{\nu\mathbf{ q}}) + \nonumber \\
&(f_{m'\mathbf{ k}+\mathbf{ q}}+n_{\nu\mathbf{ q}})\delta(\epsilon_{m\mathbf{ k}}-\epsilon_{m'\mathbf{ k}+\mathbf{ q}}+\hbar \omega_{\nu\mathbf{ q}}) \big], 
\label{eq:SERTA}
\end{align}
where $A_{\textrm{BZ}}$ is the area of the BZ, and $n_{\nu\mathbf{ q}}$ is the Bose-Einstein occupation factor of the phonon of frequency $\omega_{\nu\mathbf{ q}}$. 

\begin{figure*}
\includegraphics[width=2\columnwidth]{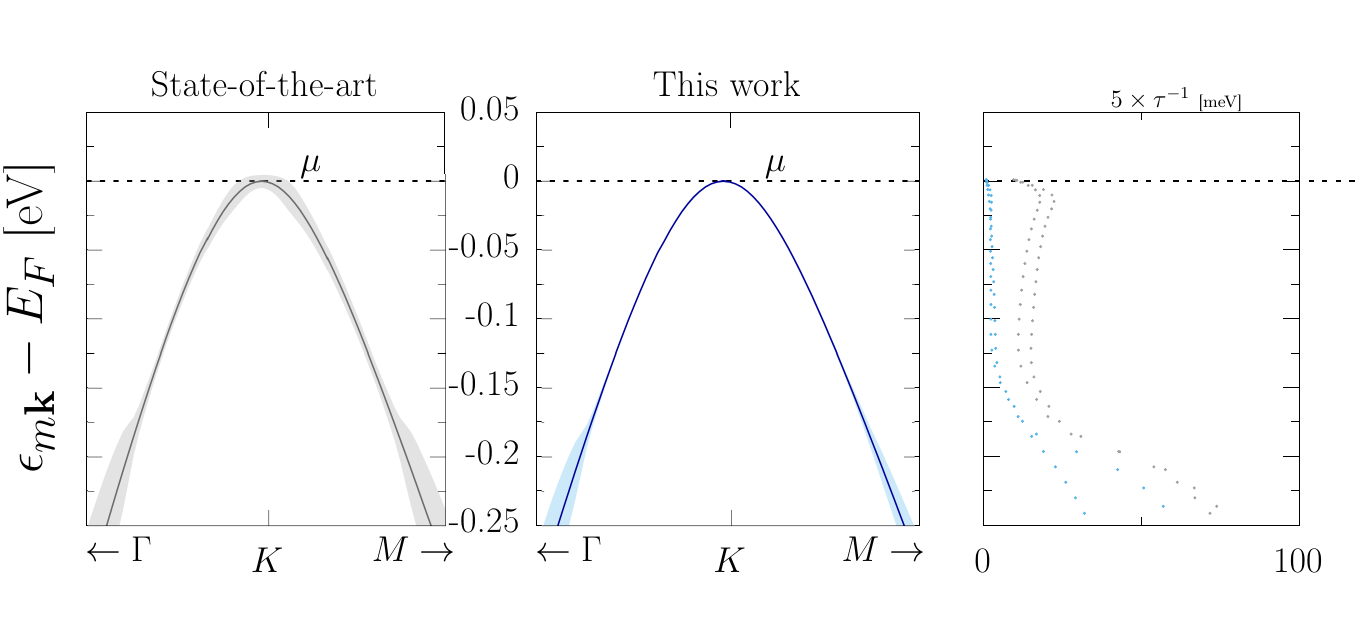}
\caption{\small Comparison of inverse lifetime scattering computed along high symmetry lines using Eq. (\ref{eq:SERTA}) where the LRCs are computed using Eqs. (\ref{eq:31})-(\ref{eq:gasymp}) with and without $n$ and $T$ dependence: the inverse lifetimes are given in absolute magnitude in the right panel, and as linewidths in the left and central ones. The light blue dots and the central panel are computed with the method proposed in this work, while grey dots and the left panel are obtained using state-of-the-art methods; the black dotted line indicates the position of the chemical potential.}
\label{fig:invtau}
\end{figure*}

We compare the cases where the asymptotic expressions of the LRCs are evaluated using the effective charges of the undoped setup screened by the undoped dielectric screening or with correct screening dependence on $n$ and $T$, within the RPA approximation to $\epsilon^{-1}(\mathbf{ q},n,T)$. 
We choose a setup tailored to highlight the differences between the two approaches. 
We fix the chemical potential at the top of valence band and set $T=300~$K, so that the numerical difference in the approaches is only due to screening. 
We do not take into account the change in the effective charges values due to $n$ and $T$ since this is highly material dependent and the current aim is to compare the present treatment of screening with state-of-the-art methods, more than a refined calculation of the electronic lifetimes. As shown in Fig. \ref{fig:invtau},  discarding the doping dependence of the screening, as currently done in most state-of-the-art first-principles calculations within the rigid-band approximation, implies a very strong overestimation of the inverse scattering times in the region near the top of the valence band.
This is mostly attributed to a wrong estimation of the piezoelectric coupling between electrons and TA phonons.
For electronic states that are below  $~200$ meV from the top of the valence band, we also observe an overestimation of the scattering times which are mostly determined by the $\Gamma$ optical phonons, due to the overestimation of the Fr\"olich coupling. As mentioned before, in the case of disproportionated graphene even stronger reductions of the electronic inverse lifetimes may come from the change of the effective charge tensors with doping and temperature, which are routinely not taken in account in state-of-the-art calculations. Notice that the value of the scattering times does not change further relevantly with increasing the number of carriers.
This means that our lifetimes are one typical of the high doping regime where dynamical effects are negligible. Also, such dynamical effects may not prevent the screening of piezoelectric coupling since plasma and acoustic frequencies may be of the same order of magnitude.

\label{sec:6}
\section{Conclusions}
\label{sec:IV}
We analyse the dependence on doping and temperature of effective charges, EPI and phonon frequencies in quasi-two-dimensional doped semiconductors, within a linear-response dielectric-matrix formulation that allows for controlled approximations of the effect of electronic screening. We further propose a fast and accurate interpolation method based on Wannier functions that enables a quantitative analysis at a feasible computational cost. 
We show that neglecting free-carrier screening on the piezoelectric and Fr\"ohlich interactions, as done in state-of-the-art computational approaches, leads to a substantial overestimation of scattering rates in specific doping-temperature regimes, which may have a strong impact on the determination of transport properties. 
However, our general formulation is not limited to those couplings, but applies to any EPI that are accessible within DFPT. The proposed approach for dealing with electronic screening lays the foundation for further extensions tackling other less conventional types of EPI, such as the vector coupling or the electron two-phonon scattering, that may play an important role in polar metals and doped quantum paraelectrics \cite{GASTIASORO2020168107,PhysRevLett.126.076601,PhysRevB.105.125142,PhysRevB.94.224515}. As a concluding remark, we mention that the extension to finite-frequencies dependence within a time-dependent approach can also give access to nonadiabatic effects on effective charges and, hence, on lattice dielectric properties \cite{PhysRevB.103.134304, PhysRevLett.128.095901} which contributes to the shapes of the dynamical structure factor probed by EELS \cite{Senga2019} and other inelastic scattering experiments \cite{Sinha2001}.

\textit{Acknowledgements---}We acknowledge the European Union’s Horizon 2020 research and innovation program under grant agreements no. 881603-Graphene Core3 and the MORE-TEM ERC-SYN project, grant agreement No 951215. We acknowledge that the results of this research have been achieved using the DECI resource \textit{Mahti CSC} based in Finland at https://research.csc.fi/-/mahti with support from the PRACE aisbl; we also acknowledge PRACE for awarding us access to Joliot-Curie Rome at TGCC, France. \textbf{This paper is dedicated to the living memory of Dr. Nicola Bonini}. 

\appendix 
\section{Conventions, definitions and derivations}
\label{app:A}
\subsection{Fourier transforms}
\label{app:A1}
We assume Born-von Karman cyclic boundary conditions and consider a supercell made up of $N$ primitive cells, whose lattice vectors are $\mathbf{ T}_1=N_1\mathbf{ t}_1$, $\mathbf{ T}_2=N_2\mathbf{ t}_2$ with $N_1N_2=N$, where $\mathbf{ t}_1,\mathbf{ t}_2$ are the direct lattice vectors (and $\mathbf{ g}_1,\mathbf{ g}_2$ the reciprocal ones). We define the Fourier transform in the Born-von Karman supercell as
\begin{align}
f(\mathbf{ q}+\mathbf{ G})= \int f(\mathbf{ r}) e^{-i(\mathbf{ q}+\mathbf{ G})\cdot\mathbf{ r}} d\mathbf{ r}, \\
f(\mathbf{ r})= \frac{1}{NA}\sum_{\mathbf{ q} \, \mathbf{ G}} f(\mathbf{ q}+\mathbf{ G}) e^{i(\mathbf{ q}+\mathbf{ G})\cdot\mathbf{ r}},
\label{eq:A0}
\end{align}
where $A$ is the area of the unit cell, the integration is intended to run over the Born-von Karman supercell and $\mathbf{ q}=m_1/N_1 \mathbf{ g}_1 + m_2/N_2 \mathbf{ g}_2$ where $m_1,m_2$ are integer numbers, while $\mathbf{ G}$ are defined as reciprocal lattice vectors. In the case of quantities which are cell-periodic, the above expression is evaluated at $\mathbf{ q}=0$, and the domain of integration is reduced to the primitive cell while taking $N=1$. Analogously, the Fourier transform of quantities dependent on two real space indexes is written as 
\begin{align}
f(\mathbf{ q}+\mathbf{ G},\mathbf{ q}+\mathbf{ G'})= \frac{1}{NA}\int d\mathbf{ r}d\mathbf{ r'} f(\mathbf{ r},\mathbf{ r'}) \times \nonumber \\
e^{-i\left[(\mathbf{ q}+\mathbf{ G})\cdot\mathbf{ r}-(\mathbf{ q}+\mathbf{ G'})\cdot\mathbf{ r'}\right]}, \\
f(\mathbf{ r},\mathbf{ r'})=\frac{1}{NA} \sum_{\mathbf{ q} \, \mathbf{ G} \, \mathbf{ G'}} f(\mathbf{ q}+\mathbf{ G},\mathbf{ q}+\mathbf{ G'}) \times \nonumber \\
e^{i\left[(\mathbf{ q}+\mathbf{ G})\cdot\mathbf{ r}-(\mathbf{ q}+\mathbf{ G'})\cdot\mathbf{ r'}\right]},
\label{eq:A2}
\end{align}
where we have used that $\mathbf{ q}$ is the same in both reciprocal space arguments as a consequence of $f(\mathbf{ r},\mathbf{ r'})=f(\mathbf{ r}+\mathbf{ R},\mathbf{ r'}+\mathbf{ R})$ for any $\mathbf{ R}$ belonging to the direct lattice---when $\mathbf{G}=\mathbf{G'}=0$ we will often shorten $f(\mathbf{ q},\mathbf{ q}) \coloneqq f(\mathbf{ q})$.
Since we need to transform also along the out-of-plane direction to obtain the expression for the Coulomb kernel of Eq. \ref{eq:1}, we define
\begin{align}
f(\mathbf{ q}+\mathbf{ G},q_z)= \int f(\mathbf{ r},z) e^{-i(\mathbf{ q}+\mathbf{ G})\cdot{\mathbf{ r}}}e^{-iq_z z} d\mathbf{ r}dz,\\
f(\mathbf{ r},z)= \frac{1}{2\pi NA} \int \sum_{\mathbf{ q} \, \mathbf{ G}} f(\mathbf{ q}+\mathbf{ G},q_z) e^{i(\mathbf{ q}+\mathbf{ G})\cdot{\mathbf{ r}}}e^{iq_z z} dq_z.
\label{eq:A1}
\end{align}
With these definitions, the 3d transform of the Coulomb kernel is
\begin{align}
v(\mathbf{ q}+\mathbf{ G},\mathbf{ q}+\mathbf{ G'},q_z)=\frac{4\pi e^2}{|\mathbf{ q}+\mathbf{ G}|^2+q_z^2}\delta_{\mathbf{ G} \, \mathbf{ G'}},
\end{align}
from which Eq. \ref{eq:1} follows using the residue theorem on the antitransform to the $z$ variable. 

For the transform of the force constants matrix, we define 
\begin{align}
C_{ss',\alpha\beta}(\mathbf{ q})= \sum_{p} C_{ss',\alpha\beta}(\mathbf{ R}_p)e^{i\mathbf{ q}\cdot(\mathbf{ R}_p+\boldsymbol{ \tau}_s-\boldsymbol{ \tau}_{s'})}, \nonumber \\
C_{ss',\alpha\beta}(\mathbf{ R}_p)=\frac{1}{N} \sum_{\mathbf{ q}} C_{ss',\alpha\beta}(\mathbf{ q})e^{-i\mathbf{ q}\cdot(\mathbf{ R}_p+\boldsymbol{ \tau}_s-\boldsymbol{ \tau}_{s'})},
\end{align}
where $\mathbf{ R}_p$ is the vector that identifies the cell $p$ and $\boldsymbol{ \tau}_s,\boldsymbol{ \tau}_{s'}$ are the basis vectors of the atoms $s,s'$ in the unit cell. Accordingly, a phonon of polarization $e^{\nu}_{s,\alpha}(\mathbf{ q})$ induces a displacement $\mathbf{\upsilon}$ of the atom $s$ along the direction $\alpha$ in the cell $p$ that can be written as
\begin{align}
\upsilon^{\nu}_{sp,\alpha}(\mathbf{ q}) = e^{\nu}_{s,\alpha}(\mathbf{ q}) e^{i\mathbf{ q}\cdot(\mathbf{ R}_p+\boldsymbol{ \tau}_s)}.
\end{align}
The above definition differentiates from the one of Refs. \cite{PhysRevB.55.10337,PhysRevB.55.10355,PhysRevB.43.7231,PhysRevB.76.165108} for the explicit presence of $\boldsymbol{ \tau}_s$ at the exponential, while it is the same of Refs. \cite{PhysRevB.1.910,PhysRevB.88.174106}---the advantage of this definition is that the response to such monochromatic perturbation behaves as a scalar under a reference frame change. We also define
\begin{align}
\upsilon^{\nu}_{s,\alpha}(\mathbf{ q}) = e^{\nu}_{s,\alpha}(\mathbf{ q}) \sum_{p} e^{i\mathbf{ q}\cdot(\mathbf{ R}_p+\boldsymbol{ \tau}_s)}.
\label{eq:phaseconv}
\end{align}
\subsection{Definition of charge density change}
\label{app:A2}
In general, for the monochromatic linear response problem we may write the external perturbation on the atoms as \cite{PhysRevB.88.174106} $\upsilon_{s,\alpha}(\mathbf{ q})=\sum_p \lambda_{\alpha} e^{i\mathbf{ q}\cdot(\mathbf{ R}_p+\boldsymbol{ \tau}_s)}$ where $\boldsymbol{\lambda}$ is the dimensional amplitude of the atomic displacement. For a generic charge density or potential $f$ that depends on the set of atomic positions $\{\mathbf{ R}_p+\boldsymbol{\tau}_s\}$, we define its variation in the linear response regime with respect to the external perturbation $\upsilon_{s,\alpha}(\mathbf{ q})$ as
\begin{align}
\frac{\partial f(\mathbf{r},\{ \mathbf{ R}_p+\boldsymbol{\tau}_s\})}{\partial \lambda_{\alpha}}\Big |_{\lambda_{\alpha}=0}= \nonumber \\ \sum_{\mathbf{ G}}\delta f_{s,\alpha}(\mathbf{ q}+\mathbf{ G},z,\{ \mathbf{ R}_p+\boldsymbol{\tau}_s\}) e^{i(\mathbf{ q}+\mathbf{ G})\cdot \mathbf{ r}};
\end{align}
$\sum_{\mathbf{ G}}\delta f_{s,\alpha}(\mathbf{ q}+\mathbf{ G},z,\{ \mathbf{ R}_p+\boldsymbol{\tau}_s\}) e^{i\mathbf{ G}\cdot \mathbf{ r}}$ is clearly a cell-periodic function and, using a notation similar to the one of Ref. \cite{RevModPhys.73.515}, may be as well indicated with $f^{\mathbf{ q}}(\mathbf{r})$;
throughout the text we shorten the notation by defining $\delta f_{s,\alpha}(\mathbf{ q}+\mathbf{ G},z)\coloneqq\delta f_{s,\alpha}(\mathbf{ q}+\mathbf{ G},z,\{ \mathbf{ R}_p+\boldsymbol{\tau}_s\})$.
As an example, the ionic charge density change induced by $\upsilon_{s,\alpha}(\mathbf{ q})$ is written, given the unperturbed expression for the ionic charge density $\rho^{\textrm{ion}}(\mathbf{r})=\sum_{ps} \mathcal{Z}_s e \delta(\mathbf{r}-\mathbf{ R}_p-\boldsymbol{\tau}_s)$, as
\begin{align}
\frac{\partial \rho^{\textrm{ion}}(\mathbf{r},\{ \mathbf{ R}_p+\boldsymbol{\tau}_s\})}{\partial \lambda_{\alpha}}\Big |_{\lambda_{\alpha}=0}=\sum_{\mathbf{ G}}\delta \rho^{\textrm{ion}}_{s,\alpha}(\mathbf{ q}+\mathbf{ G},z) e^{i(\mathbf{ q}+\mathbf{ G})\cdot \mathbf{ r}}.
\end{align}
Since $\delta \rho^{\textrm{ion}}_{s,\alpha}$ is itself the perturbing external charge density of the electrostatic problem, we rename it $\delta \rho^{\textrm{ext}}_{s,\alpha}$ for clarity. For a 2d material where all the atoms are disposed on one single layer and are subjected only to in-plane perturbations of the atoms, which is the matter of study of this work, we obtain 
\begin{align}
\delta \rho^{\textrm{ext}}_{s,\alpha}(\mathbf{ q}+\mathbf{ G},z)=-i\frac{\mathcal{Z}_s e}{A}\left[( q_{\alpha}+ G_{\alpha})e^{-i\mathbf{ G}\cdot\boldsymbol{ \tau}_{s}}\right]\delta(z),
\label{eq:rhoextz}
\end{align}
which, when layer-averaged as defined in Eq. \ref{eq:layeraveraging}, gives Eq. \ref{eq:28}. The same definitions may be extended to the variation of potentials. Notice that in case of the use of pseudopotentials, following a collective atomic displacement the local part of the pseudopotential produces an external charge in a form similar Eq. \ref{eq:rhoextz}, but which includes a form factor coming from the shape of the pseudopotential. The form factor is though relevantly different from zero for $q\gg \frac{1}{R_c}$, where $R_c$ is the core radius of the pseudopotential; similarly, the non-local part of the pseudopotential is irrelevant in the $q\ll \frac{1}{R_c}$ limit. 
\subsection{Response functions}
\label{app:A3}
We define the following linear response functions in real space:
\begin{align}
\Delta \rho^{\textrm{ind}}(\mathbf{r}) = \int d\mathbf{r'} \chi(\mathbf{r},\mathbf{r'}) \Delta V^{\textrm{ext}}(\mathbf{r'}),\label{eq:firsteq}\\
\Delta \rho^{\textrm{ind,H}}(\mathbf{r}) = \int d\mathbf{r'} \Pi(\mathbf{r},\mathbf{r'}) \Delta V^{\textrm{tot,H}}(\mathbf{r'}), \label{eq:A14}\\
\Delta \rho^{\textrm{ind}}(\mathbf{r}) = \int d\mathbf{r'} \chi^{\textrm{0}}(\mathbf{r},\mathbf{r'}) \Delta V^{\textrm{tot}}(\mathbf{r'}),\label{eq:chiir}\\
\Delta V^{\textrm{tot}}(\mathbf{r})=\int d\mathbf{r'} \epsilon^{-1}(\mathbf{r},\mathbf{r'})\Delta V^{\textrm{ext}}(\mathbf{r'}),
\label{eq:respdef}
\end{align}
where $\Delta V^{\textrm{ext}}$ is an external perturbing potential, $\Delta V^{\textrm{ind}}$ is the change of potential in the crystal induced by the external perturbation, $\Delta V^{\textrm{tot,H}} \coloneqq \Delta V^{\textrm{ind,H}}+\Delta V^{\textrm{ext}}$ and $\Delta V^{\textrm{tot}} \coloneqq \Delta V^{\textrm{ind}}+\Delta V^{\textrm{ext}}$ where the difference between $\Delta V^{\textrm{ind,H}}$ and $\Delta V^{\textrm{ind}}$ is respectively the neglect or the inclusion of the exchange correlation terms in the response (we use the same superscript meaning for the charge density $\rho$) \footnote{In other notations in the context of DFT or TDDFT $\chi^{0}$ is referred to as $\chi^{\textrm{KS}}$ or $\chi^{\textrm{ir}}$.}. Notice that $\chi^0$ is the IPP of the Khon-Sham system and is expressed in Eq. \ref{eq:3} with wavefunctions that have been computed with the inclusion of the $\textit{xc}$ potential \cite{giuliani2005quantum}, while $\Pi$ would not include the \textit{xc} contributions in the self consistent field calculation; in our work we use the RPA approximation in which the total response $\chi$ is approximated as
\begin{align}
\chi^{-1}=\chi^{0,-1}-v-K_{xc}\rightarrow\chi^{-1,RPA}=\chi^{0,-1}-v;
\end{align}
from this approximation and $\epsilon^{-1,RPA}=\mathcal{I}+v\chi^{RPA}$, it follows that $\epsilon^{-1,RPA}\chi^{-1,RPA}\chi^0=\mathcal{I}$ which means that $\epsilon^{RPA}=\mathcal{I}-v\chi^0$, i.e. Eq. \ref{eq:5}---see App. \ref{app:xc} for considerations beyond RPA. The definitions from Eq. \ref{eq:firsteq} to Eq. \ref{eq:respdef} are given for a general real space perturbation, and therefore in reciprocal space they can be used with the substitution $\Delta \rightarrow \delta$ where $\delta$ assumes the meaning given in App. \ref{app:A2}. In this case, to connect the external perturbation to the external charge change, we use the electrostatic relation
\begin{align}
\delta V^{\textrm{ext}}_{s,\alpha}(\mathbf{ q}+\mathbf{ G},z)=2\pi \int dz' \frac{e^{-|\mathbf{ q}+\mathbf{ G}||z-z'|}}{|\mathbf{ q}+\mathbf{ G}|}\delta \rho^{\textrm{ext}}_{s,\alpha}(\mathbf{ q}+\mathbf{ G},z').
\label{eq:A11}
\end{align}
Also, to our aims it is useful to define the averaging of a generic function $f(\star,z,z')$ over the layer width $t$ (defined in Sec. \ref{sec:IIsubA}) as 
\begin{align}
\tilde f(\star)=\int_{-\frac{t}{2}}^{\frac{t}{2}}dz \frac{dz'}{t} f(\star,z,z'), \label{eq:layeraveraging}
\end{align}
which we use to get rid of the $z,z'$ dependence of, e.g., the response functions. With the above definition, the layer-averaged response functions satisfy
\begin{align}
\int dz'' \sum_{\mathbf{ G''}} & \tilde f(\mathbf{ q}+\mathbf{ G},\mathbf{ q}+\mathbf{ G''}) \times \label{eq:invavdefbis}\\
& \tilde {f}^{-1}(\mathbf{ q}+\mathbf{ G''},\mathbf{ q}+\mathbf{ G'})=\delta_{\mathbf{ G} \, \mathbf{ G'}}. \nonumber
\end{align}
Notice that, from a more formal perspective, the layer-averaging procedure may be interpreted as keeping the leading order of the hyperbolic cosine in the appropriate expressions of Ref. \cite{PhysRevX.11.041027}.
\subsection{Definition of $\Delta_{\mathbf{ q} \nu}$}
\label{app:A4}
$\Delta_{\mathbf{ q} \nu}$ is defined as in Ref. \cite{RevModPhys.89.015003}, but with phase convention coherent with Eq. \ref{eq:phaseconv}, i.e. as
\begin{align}
\Delta_{\mathbf{ q} \nu} V=e^{i\mathbf{ q}\cdot\mathbf{ r}} \Delta_{\mathbf{ q} \nu} v,\\
\Delta_{\mathbf{ q} \nu} v=l_{\mathbf{ q}\nu}\sum_{s\alpha} \left(\frac{M_0}{M_s}\right)^{\frac{1}{2}}e^{\nu}_{s\alpha}(\mathbf{ q})\partial_{s\alpha,\mathbf{ q}}v,\\
\partial_{s\alpha,\mathbf{ q}}v=\sum_{p}e^{-i\mathbf{ q}\cdot (\mathbf{ r}-\mathbf{ R}_p-\boldsymbol{\tau}_s)}\frac{\partial V}{\partial  \tau_{s\alpha}}\Big|_{\mathbf{ r}-\mathbf{ R}_p-\boldsymbol{\tau}_s},
\end{align}
where $\Delta_{\mathbf{ q} \nu} v$ is a cell-periodic function and again we have restricted the derivation to be performed wit respect to in-plane displacements of the atoms. We also define
\begin{align}
l_{\mathbf{ q}\nu}=[\frac{\hbar}{2M_0\omega_{\mathbf{ q},\nu}}]^{1/2}.
\end{align}
Notice that 
\subsection{Independent particle polarizability and dielectric matrix}
\label{app:A5}
Starting from the approximation of Eq. \ref{eq:6}, Eq. \ref{eq:3} becomes Eq. \ref{eq:7}:
\begin{align}
& \chi^0(\mathbf{ q}+\mathbf{ G},\mathbf{ q}+\mathbf{ G'} ,z,z')=\frac{2e^2}{At^2}\sum_{mm'\mathbf{ k}}\frac{f_{m\mathbf{ k}}-f_{m'\mathbf{ k}+\mathbf{ q}}}{\epsilon_{m\mathbf{ k}}-\epsilon_{m'\mathbf{ k}+\mathbf{ q}}} \times \nonumber \\
& \braket{u_{m'\mathbf{ k}+\mathbf{ q}+\mathbf{ G}} | u_{m\mathbf{ k}}} \braket{ u_{m\mathbf{ k}}|u_{m'\mathbf{ k}+\mathbf{ q}+\mathbf{ G'}}} \theta(\frac{t}{2}-|z|) \theta(\frac{t}{2}-|z'|) \nonumber \\
& \coloneqq \frac{1}{t^2}\theta(\frac{t}{2}-|z|) \theta(\frac{t}{2}-|z'|) \chi^0(\mathbf{ q}+\mathbf{ G},\mathbf{ q}+\mathbf{ G'}).
\label{eq:7bis}
\end{align}
The brakets stand for integration on the in-plane real space variables and the expression $\chi^0(\mathbf{ q}+\mathbf{ G},\mathbf{ q}+\mathbf{ G'})$, as already anticipated in Sec. \ref{sec:II}, is the IPP of a two dimensional Khon-Sham system \cite{PhysRevB.91.165428,PhysRevB.75.205418}. To obtain an expression for the dielectric tensor, we plug Eq. \ref{eq:7bis} inside Eq. \ref{eq:5} obtaining
\begin{align}
\epsilon(\mathbf{ q}+\mathbf{ G} & ,\mathbf{ q}+\mathbf{ G'},z,z')=\delta(z-z')\delta_{\mathbf{ G} \, \mathbf{ G'}}+ \label{eq:8} \\
& -\frac{2\pi}{t^2} \frac{P(\mathbf{ q}+\mathbf{ G},z)}{|\mathbf{ q}+\mathbf{ G}|}  \chi^{0}(\mathbf{ q}+\mathbf{ G},\mathbf{ q}+\mathbf{ G'}) \theta(\frac{t}{2}-|z'|), \nonumber
\end{align}
where we have defined
\begin{align}
&P(\mathbf{ q}+\mathbf{ G},z) = \frac{2}{|\mathbf{ q}+\mathbf{ G}|} \times \label{eq:9}\\
&\times
\begin{cases}
e^{-|\mathbf{ q}+\mathbf{ G}|z}\sinh(|\mathbf{ q}+\mathbf{ G}|\frac{t}{2}) \quad &z>\frac{t}{2} \\
1-e^{-|\mathbf{ q}+\mathbf{ G}|\frac{t}{2}}\cosh(|\mathbf{ q}+\mathbf{ G}|z) \quad &|z|\leq \frac{t}{2} \\
e^{|\mathbf{ q}+\mathbf{ G}|z}\sinh(|\mathbf{ q}+\mathbf{ G}|\frac{t}{2}) \quad &z<-\frac{t}{2}
\end{cases} .\nonumber
\end{align}
We concentrate on the domain $z,z' \in [-\frac{t}{2},\frac{t}{2}]$ because it is the one where we are interested to evaluate $\epsilon^{-1}(\mathbf{ q}+\mathbf{ G},\mathbf{ q}+\mathbf{ G'},z,z')$. In this region Eqs. \ref{eq:8} and \ref{eq:9} can be further simplified using the definition of layer-averaged dielectric function of Eq. \ref{eq:10}. Plugging Eqs. \ref{eq:8} and \ref{eq:9} in Eq. \ref{eq:10}, we finally obtain Eq. \ref{eq:13} of the main text.

The appropriate long wavelength expansions of Eq. \ref{eq:13} are obtained analyzing the asymptotic behaviours of $\chi^0$. \textit{For insulators and undoped semiconductors at zero temperature} we can write for the leading orders
\begin{align}
&\lim_{\mathbf{ q}\rightarrow\mathbf{0}}\chi^{0}(\mathbf{ q})=\frac{2 q_{\alpha}q_{\beta}}{A}\sum_{mm'\mathbf{ k}} \frac{\theta(\epsilon_{m'\mathbf{ k}})-\theta(\epsilon_{m\mathbf{ k}})}{\epsilon_{m\mathbf{ k}}-\epsilon_{m'\mathbf{ k}}} \times \nonumber \\
&\braket{u_{m\mathbf{ k}}|\partial_{\mathbf{ k}_{\alpha}}u_{m'\mathbf{ k}}}^*\braket{u_{m\mathbf{ k}}|\partial_{\mathbf{ k}_{\beta}}u_{m'\mathbf{ k}}} \coloneqq \mathbf{ q} B \mathbf{ q}, \label{eq:qBq}\\
&\lim_{\mathbf{ q}\rightarrow\mathbf{0}}\chi^{0}(\mathbf{ q},\mathbf{ q}+\mathbf{ G'})= \frac{2 q_{\alpha}}{A}\sum_{mm'\mathbf{ k}} \frac{\theta(\epsilon_{m'\mathbf{ k}})-\theta(\epsilon_{m\mathbf{ k}})}{\epsilon_{m\mathbf{ k}}-\epsilon_{m'\mathbf{ k}}} \times \nonumber \\
&\braket{u_{m\mathbf{ k}}|\partial_{\mathbf{k}_{\alpha}}u_{m'\mathbf{ k}}}^{\textrm{c.c.}} \braket{u_{m\mathbf{ k}}|u_{m'\mathbf{ k}+\mathbf{ G'}}}\coloneqq \mathbf{ q}\cdot\mathbf{A}(\mathbf{ G'}), \label{eq:qA}\\
&\lim_{\mathbf{ q}\rightarrow\mathbf{0}}\chi^{0}(\mathbf{ q}+\mathbf{ G},\mathbf{ q}+\mathbf{ G'})=
\frac{2}{A}\sum_{mm'\mathbf{ k}} \frac{\theta(\epsilon_{m'\mathbf{ k}})-\theta(\epsilon_{m\mathbf{ k}})}{\epsilon_{m\mathbf{ k}}-\epsilon_{m'\mathbf{ k}}} \times \nonumber \\
&\braket{u_{m\mathbf{ k}}|u_{m'\mathbf{ k}+\mathbf{ G}}}^{\textrm{c.c.}}\braket{u_{m\mathbf{ k}}|u_{m'\mathbf{ k}+\mathbf{ G'}}}\coloneqq C(\mathbf{ G},\mathbf{ G'}) \label{eq:C},
\end{align}
where it is intended that Eq. \ref{eq:qA} is valid for $\mathbf{G'}\neq 0$ and Eq. \ref{eq:C} for $\mathbf{G}\neq 0$ and $\mathbf{G'}\neq 0$. To obtain the above expressions we have used that $\braket{u_{m\mathbf{ k}}|u_{m'\mathbf{ k}}}=0$ if $m$ is a valence state and $m'$ a conduction state (or viceversa) and that there are no intraband transitions that can vanish the energy denominator, so that the $\mathbf{ q}$ dependence of the limits is coming only from the long-wavelength expansion of the periodic part of the Bloch functions. For the dielectric matrix we find
\begin{align}
&\epsilon(\mathbf{ q}+\mathbf{ G},\mathbf{ q}+\mathbf{ G'}) = \frac{1}{t}
\begin{cases}
1-v( q)\mathbf{ q}\cdot B \cdot \mathbf{ q} \quad  \mathbf{ G},\mathbf{ G'}=0\\
-v( q) \mathbf{ q} \cdot \mathbf{A}^{\textrm{c.c.}}(\mathbf{ G'}) \quad \mathbf{ G}=0 \\
-v( G)\mathbf{ q} \cdot \mathbf{A}(\mathbf{ G}) \quad \mathbf{ G'}=0 \\
\delta_{\mathbf{ G}\mathbf{ G'}}-v( G) C(\mathbf{ G},\mathbf{ G'})
\end{cases}\label{eq:14}, 
\end{align}
where $v$ is defined in Eq. \ref{eq:12}. \textit{For metals or doped semiconductors} the asymptotic limits for the IPP instead read
\begin{align}
\lim_{\mathbf{ q}\rightarrow\mathbf{0}}\chi^{0}(\mathbf{ q})=\frac{2}{A}\sum_{m\mathbf{ k}} \frac{\delta f_{m\mathbf{ k}}-\delta  f_{m\mathbf{ k+  q}}}{\epsilon_{m\mathbf{ k}}-\epsilon_{m\mathbf{ k +  q}}}+\mathbf{ q}B\mathbf{ q}, \label{eq:C1}\\
\lim_{\mathbf{ q}\rightarrow\mathbf{0}}\chi^{0}(\mathbf{ q},\mathbf{ q}+\mathbf{ G'})= \frac{2}{A}\sum_{m\mathbf{ k}} \frac{\delta f_{m\mathbf{ k}}-\delta  f_{m\mathbf{ k+  q}}}{\epsilon_{m\mathbf{ k}}-\epsilon_{m\mathbf{ k +  q}}} \times \nonumber \\ 
\braket{u_{m\mathbf{ k}}|u_{m\mathbf{ k}+\mathbf{ G'}}}+ \mathbf{ q}\cdot\mathbf{A}(\mathbf{ G'}),\label{eq:C2}\\
\lim_{\mathbf{ q}\rightarrow\mathbf{0}}\chi^{0}(\mathbf{ q}+\mathbf{ G},\mathbf{ q}+\mathbf{ G'})=
\frac{2}{A}\sum_{mm'\mathbf{ k}} \frac{\delta  f_{m\mathbf{ k}}-\delta  f_{m'\mathbf{ k+  q}}}{\epsilon_{m\mathbf{ k}}-\epsilon_{m'\mathbf{ k+ q}}} \times \nonumber \\
\braket{u_{m\mathbf{ k}}|u_{m'\mathbf{ k}+\mathbf{ G}}}^{\textrm{c.c.}}\braket{u_{m\mathbf{ k}}|u_{m'\mathbf{ k}+\mathbf{ G'}}}+C(\mathbf{ G},\mathbf{ G'}), \label{eq:C3}
\end{align}
where we have used the approximation that $\braket{u_{m\mathbf{k}}|u_{m'\mathbf{k+q}}}=\delta_{mm'}$ and therefore the sum over $m$ is restricted to the states whose occupation is significantly different from 1 or 0. In this case the response functions depend on the carrier concentration $n$ and temperature $T$ through the occupation factors, i.e. $\epsilon(\mathbf{ q} +\mathbf{ G},\mathbf{ q} +\mathbf{ G'})\rightarrow \epsilon(\mathbf{ q} +\mathbf{ G},\mathbf{ q} +\mathbf{ G'},n,T)$. Using a purely qualitative argument, the expression of the dielectric matrix as a function of $n$ can be obtained in the degenerate limit performing the substitution $\mathbf{ q}\cdot B \cdot \mathbf{ q} \rightarrow k^2_{TF}(n)+\mathbf{ q}\cdot B \cdot \mathbf{ q}$ in Eq. \ref{eq:14}, where $k^2_{TF}(n)$ is the Thomas-Fermi screening wavevector of the material, or using $\mathbf{ q}\cdot B \cdot \mathbf{ q} \rightarrow k^2_{D}(n,T)+\mathbf{ q}\cdot B \cdot \mathbf{ q}$ in the non-degenerate limit, where $k_D(n,T)$ is the Debye screening wavevector \cite{EHRENREICH1959130}; we are here supposing that the wings of the dielectric response matrix are not affected by doping---as shown in Sec. \ref{sec:III} this is not rigorously true, but as a matter of fact wings are less sensitive to doping than the head. The above substitution cures the non-analiticity of the LRCs of the dynamical matrix and of the EPI typical of the insulator case (see Secs. \ref{sec:IIsubB} and \ref{sec:IIsubC}), smoothing their singular behaviours in a small region around $\Gamma$. Quantitatively, the above description of the doping effect on the screening is too rough since this is strongly dependent on the microscopic details of the material and temperature, and therefore more refined strategies, as the one proposed in Eq. \ref{eq:6.1}, have to be implemented.

\subsection{Inversion of $\w^{-1}$}
\label{app:D}
We define the inverse screened Coulomb potential $\w^{-1}$ as the inverse of Eq. \ref{eq:xiepsvc}:
\begin{align}
\w^{-1}(\mathbf{ q}+\mathbf{ G} & ,\mathbf{ q}+\mathbf{ G'}) = 
\frac{\epsilon(\mathbf{ q}+\mathbf{ G},\mathbf{ q}+\mathbf{ G'})}{v(\mathbf{ q}+\mathbf{ G})}  \label{eq:24}.
\end{align}
The $\w^{-1}$ matrix is Hermitian because of the Hermiticity of $\chi^0$ and Eq. \ref{eq:13}. For a generic Hermitian matrix $\w^{-1}$, following \cite{PhysRevB.1.910,PhysRevB.88.174106}, we write:
\begin{align}
\w^{-1}=
\begin{pmatrix}
P & Q \\
Q^{\dagger} & S
\end{pmatrix},
\\
\w=
\begin{pmatrix}
W & X \\
X^{\dagger} & Z
\end{pmatrix} \label{eq:xiinv},
\end{align}
where $P=\w^{-1}(\mathbf{ q})$ is the head of the matrix, $Q=\w^{-1}(\mathbf{ q},\mathbf{ q}+\mathbf{ G'})$ is the wing and $S=S^{\dagger}=\w^{-1}(\mathbf{ q}+\mathbf{ G},\mathbf{ q}+\mathbf{ G'})$ is the body, and the same goes for $W,X,Z$. We have
\begin{align}
W=(P-QS^{-1}Q^{\dagger})^{-1},\\
X=-WQS^{-1},\label{eq:X}\\
Z=S^{-1}+X^{\dagger}W^{-1}X.
\end{align}
$\w$ may then be rewritten as:
\begin{equation}
\w=
\begin{pmatrix}
W & X \\
X^{\dagger} & S^{-1}+X^{\dagger}W^{-1}X
\end{pmatrix}.
\label{eq:factorization}
\end{equation}
We also define
\begin{equation}
\hat \w =
\begin{pmatrix}
W & X \\
X^{\dagger} & X^{\dagger}W^{-1}X
\end{pmatrix},
\end{equation}
which can be rewritten as
\begin{align}
\hat \w(\mathbf{ q}+\mathbf{ G},\mathbf{ q}+\mathbf{ G'})= \frac{\hat \w(\mathbf{ q},\mathbf{ q}+\mathbf{ G'})\hat \w(\mathbf{ q}+\mathbf{ G},\mathbf{ q})}{\hat \w(\mathbf{ q})}= \nonumber \\ 
\frac{\hat \w(\mathbf{ q},\mathbf{ q}+\mathbf{ G'})\hat \w^{\textrm{c.c.}}(\mathbf{ q},\mathbf{ q}+\mathbf{ G})}{\hat \w(\mathbf{ q})}.
\end{align}
For the case of undoped semiconductors, the $\hat \w$ tensor contains all the non-analytical terms that give rise to the LRCs of the EPI and of the dynamical matrix; notice that the head and the wings of the $\w$ and the $\hat \w$ tensors coincide.
\subsubsection{Asymptotic expansion of the $\w$ tensor}
\label{app:A6subI}
To evaluate Eq. \ref{eq:30}, we need to know the asymptotic expressions of the $\w$ tensor, given the inversion formulae of App. \ref{app:D}. \textit{For insulators and undoped semiconductors} these are
\begin{align}
w(\mathbf{ q}+\mathbf{ G},\mathbf{ q}+\mathbf{ G'}) = 
\begin{cases}
\w(\mathbf{ q}) \quad \mathbf{ G},\mathbf{ G'}=0\\
\w(\mathbf{ q}) \left[ \mathbf{ q}\cdot \mathbf{A'}^{\textrm{c.c.}}(\mathbf{ G'})\right]  \quad \mathbf{ G}=0\\
\w(\mathbf{ q}) \left[ \mathbf{ q}\cdot \mathbf{A'}(\mathbf{ G})\right]   \quad \mathbf{ G'}=0\\
L(\mathbf{ G},\mathbf{ G'})+S^{-1}(\mathbf{ G},\mathbf{ G'})
\end{cases}
\label{eq:32}\\
\w(\mathbf{ q})=
\begin{cases}
\frac{2\pi}{|\mathbf{ q}|+\mathbf{ q}\cdot B'\cdot \mathbf{ q}} \quad \textrm{thin} \nonumber \\
\frac{4\pi}{t}\frac{1}{\mathbf{ q}\cdot B'' \cdot \mathbf{ q}} \quad \textrm{thick} \nonumber 
\end{cases}
\end{align}
where $\textrm{c.c.}$ stands for complex conjugate and the expressions for $\mathbf{A}',L,H$, which are different for the thin and thick limits, can be easily worked out.
\textit{For metals or doped semiconductors}, the above expressions have to be generalized to include the terms coming from the expansions Eqs. \ref{eq:C1}, \ref{eq:C2} and \ref{eq:C3}. The generalization can be easily worked out and is not reported here. We though notice that if we are just interested in the ratio between the wing and the head of the $\w$ entering Eq. \ref{eq:27} and Eq. \ref{eq:30}, then we can can recast it as
\begin{align}
&\frac{\w(\mathbf{ q},\mathbf{ q}+\mathbf{ G'},n,T)}{\w(\mathbf{ q},\mathbf{ q},n,T)}= \nonumber \\ 
&\begin{cases}
1\quad \mathbf{ G'}=0 \\
\mathbf{ q}\cdot \mathbf{A'}^{\textrm{c.c.}}(\mathbf{ G'})+\mathcal{C}(\mathbf{ q},n,T)\quad \mathbf{ G'}\neq 0
\end{cases},
\label{eq:Minterintra}
\end{align}
where the term $\mathcal{C}(\mathbf{ q},n,T)$ stems formally from the change of $\chi^0$ due to (marginally) the modification of the interband terms and to (substantially) the appearence of an intraband contribution to $\chi^0$ in presence of doping at finite temperature (see Eqs. \ref{eq:C1} and \ref{eq:C2} for the limiting values); practically, we associate the presence of $\mathcal{C}(\mathbf{ q},n,T)$ exclusively to the existence of intraband contributions. Such term can be detected in the density response in the region where the dielectric response acquires strong metallic features (see Sec. \ref{sec:III}).
\subsection{Proof of Eq. \ref{eq:expansion}}
\label{app:expansion}
To prove the expansion of Eq. \ref{eq:expansion}, we consider the electrostatic problem imposing that the $\mathbf{ G}=0$ component of the change of the total potential, $\delta  V^{\textrm{tot}}_{s,\alpha}(\mathbf{ q})$, is null for any possible charge density perturbation---notice that we are considering the electrostatic problem where each term of the Maxwell's equation has been layer-averaged independently (see also App. \ref{app:E}). This request can be satisfied only if in Eq. \ref{eq:xiinv} we have $W=\eta$ and $X=\eta$ where $\eta$ is an infinitesimally small number. We notice that $\lim_{\eta\rightarrow 0} \hat \w=0$, so that we are left only with the short range components of the $\w$ matrix. The total potential may now be written, for $\mathbf{ G} \neq 0$ as
\begin{align}
\delta \bar V^{\textrm{tot}}_{s,\alpha}(\mathbf{ q}+\mathbf{ G})=\sum_{\mathbf{ G'}\neq 0}S^{-1}(\mathbf{ q}+\mathbf{ G},\mathbf{ q}+\mathbf{ G'})t\delta \rho^{\textrm{ext}}_{s,\alpha}(\mathbf{ q}+\mathbf{ G'});
\label{eq:vbartot}
\end{align}
the bar over a quantity is here used to indicate that it is computed imposing $\delta  V^{\textrm{tot}}_{s,\alpha}(\mathbf{q})=0$---the bar notation for $\bar Z_{s,\alpha}$ introduced in Sec. \ref{sec:II} is not casual, as we will see in a moment. By the definition of $X^0$ we obtain
\begin{align}
t \delta \bar \rho^{\textrm{ind}}_{s,\alpha}(\mathbf{ q})= \sum_{\substack{\mathbf{ G}\neq 0 \\ \mathbf{ G'} \neq 0}}\chi^0(\mathbf{ q},\mathbf{ q}+\mathbf{ G}) S^{-1}(\mathbf{ q}+\mathbf{ G},\mathbf{ q}+\mathbf{ G'}) \times \nonumber \\
t \delta \rho^{\textrm{ext}}_{s,\alpha}(\mathbf{ q}+\mathbf{ G'}) \label{eq:indcharge}.
\end{align}
To obtain the total charge density change in this particular setup, $\delta \bar { \rho}^{\textrm{tot}}_{s,\alpha}(\mathbf{ q})$, we sum $\delta \rho^{\textrm{ext}}_{s,\alpha}(\mathbf{ q})$ to $\delta\bar \rho^{\textrm{ind}}_{s,\alpha}(\mathbf{ q})$. Using the asymptotic expressions for $\w$, $\chi^0$ and $S^{-1}$ (for the first see App. \ref{app:A6subI}, for the second see App. \ref{app:A5}, while for the third we just get $S^{-1}(\mathbf{ G},\mathbf{ G'})$) both in the thin and thick limits one finds that $t \delta \bar \rho^{\textrm{tot}}_{s,\alpha}(\mathbf{ q})$ is exactly equivalent to the asymptotic expansion of the left hand side of Eq. \ref{eq:30}, i.e.
\begin{align}
\delta \bar \rho^{\textrm{tot}}_{s,\alpha}(\mathbf{ q})=-i\frac{e  q}{At} \bar Z_{s,\alpha}(\mathbf{ q}).
\label{eq:rhototid}
\end{align}
This means that $-i\frac{e  q}{A} \bar Z_{s,\alpha}(\mathbf{ q})$ corresponds to the Fourier transform of the total charge density change generated by $t\delta \rho^{\textrm{ext}}_{s,\alpha}(\mathbf{q})$ once that we have imposed the absence of macroscopic electric fields, i.e. $t\delta \bar {\rho}^{\textrm{tot}}_{s,\beta}(\mathbf{ q})$---exactly as in the three dimensional case studied in Ref. \cite{PhysRevLett.129.185902}; the reason for the use of the bar notation for the unscreened effective charges is now evident. Eq. \ref{eq:indcharge} also shows us a fundamental property: $t\delta \bar {\rho}^{\textrm{tot}}_{s,\beta}(\mathbf{ q})$ \textit{is manifestly analytic} and as such may be expanded in a Taylor series, therefore justifying Eq. \ref{eq:expansion}.

From Eqs. \ref{eq:rhototid},\ref{eq:30}, \ref{eq:0}, \ref{eq:ZZbar} and the RPA electrostatic relation between charges and potentials we
deduce Eqs. \ref{eq:2inline}, from which the definition of $Z_{s,\alpha}(\mathbf{ q})$ starting from $\bar Z_{s,\alpha}(\mathbf{ q})$ (Eq. \ref{eq:ZZbar}) is now naturally clear, and is coherent with Eq. \ref{eq:0}.
Notice that the relation between screened and unscreened charges, explicitly deduced here for the first order of the expansion, is valid at all orders, as shown by the alternative derivation given in App. \ref{app:xc}.

The above argument is fundamental to prove that Eq. \ref{eq:expansion} contains the in-plane dynamical effective charges. For example, the Born effective charge is defined as in Eq. \ref{eq:35}, which in our notation (see App. \ref{app:A2}) is equivalent to write
\begin{align}
Z^*_{s,\alpha\beta}=\frac{A}{e}\frac{\partial P_{\alpha}}{\partial \lambda_{\beta}}\Big|_{\boldsymbol\lambda=\mathbf{0},\mathbf{q=0},\mathbf{E=0}};
\end{align}
notice that the star notation here does not mean complex conjugate, but has always been historically used to identify the Born effective charges, that are properly real quantities. We now define the cell-periodic layer-averaged charge density change on a single atom (SA) exploiting the superposition of the charge density change in the linear regime for the response
\begin{equation}
\delta \bar {\rho}^{\textrm{SA}}_{s,\beta}(\mathbf{ r}-\mathbf{ R}_p-\boldsymbol{ \tau}_s)=\frac{A}{(2\pi)^2}\int d\mathbf{ q} \delta \bar {\rho}^{\mathbf{ q},\textrm{tot}}_{s,\beta}(\mathbf{ r}) e^{-i\mathbf{ q}\cdot(\mathbf{ R}_p-\boldsymbol{ \tau}_s-\mathbf{ r})},
\end{equation}
with the inverse transform given by
\begin{equation}
\delta \bar {\rho}^{\mathbf{ q},\textrm{tot}}_{s,\beta}(\mathbf{ r})=\sum_p \delta \bar {\rho}^{\textrm{SA}}_{s,\beta}(\mathbf{ r}-\mathbf{ R}_p-\boldsymbol{ \tau}_s) e^{i\mathbf{ q}\cdot(\mathbf{ R}_p+\boldsymbol{ \tau}_s-\mathbf{ r})}.
\end{equation}
Then, we write
\begin{align}
\frac{\partial P_{\alpha}}{\partial \lambda_{\beta}}\Big|_{\mathbf{\lambda=0},\mathbf{q=0},\mathbf{E=0}}=\frac{1}{NA}\sum_p\int d\mathbf{r}( r_{\alpha}- R_{p,\alpha}- \tau_{s,\alpha}) \times \nonumber \\
\delta \bar {\rho}^{\textrm{SA}}_{s,\beta}(\mathbf{ r}-\mathbf{ R}_p-\boldsymbol{ \tau}_s)= i\frac{\partial}{\partial  q_{\alpha}} \delta \bar {\rho}^{\textrm{tot}}_{s,\beta}(\mathbf{ q})\Big|_{\mathbf{ q}=0}\label{eq:identification},
\end{align}
where the integration is intended to run over the whole crystal. The last equality of Eq. \ref{eq:identification} proves the identification between effective charges and the expansion of the unscreened charge density change; the same can be done also for higher orders. We need to stress here that these identifications are exact only in the thin and thick limits, i.e. in the regimes in which the LRCs can be matched to phenomenological expressions that involve only effective charges and the head of the inverse screening matrix. In these cases, these identifications are crucial in order to have an operative simple method to extract from \textit{ab-initio} calculations the value of the dynamical effective charges without passing from the explicit calculation of all the structure of the $\w$ tensor, which may depend on the adopted model.

\subsection{Derivation of Eq. \ref{eq:27}}
\label{app:A6}
For a single layer we can write for the dynamical matrix in reciprocal space  
\begin{align}
C_{ss',\alpha\beta}(\mathbf{ q})=\frac{2\pi \mathcal{Z}_s\mathcal{Z}_{s'}e^2}{A} \sum_{\mathbf{ G} \mathbf{ G'}} e^{i\mathbf{ G}\cdot \boldsymbol{ \tau}_{s}-i\mathbf{ G'}\cdot\boldsymbol{ \tau}_{s'}} \int_{-\infty}^{\infty} dz \big( q_{\alpha}+\nonumber \\
+ G_{\alpha}\big) \left( q_{\beta}+  G'_{\beta}\right) \epsilon^{-1}(\mathbf{ q}+\mathbf{ G},\mathbf{ q}+\mathbf{ G'},0,z)\frac{e^{-|\mathbf{ q}+\mathbf{ G'}||z|}}{|\mathbf{ q}+\mathbf{ G'}|}.
\label{eq:22}
\end{align}
To obtain asymptotic formulae, we now wish to get rid of the $z$ dependence of the above equation. To do so, we replace $\epsilon^{-1}(\mathbf{ q}+\mathbf{ G},\mathbf{ q}+\mathbf{\bar G'},0,z)$ with its layer average contextually restricting $z\in [-\frac{t}{2},\frac{t}{2}]$. In fact, we can approximate $\epsilon^{-1}$ assuming that the induced density is uniform inside the slab of width $t$, and vanishes outside, obtaining an expression similar to Eq. \ref{eq:8}---see App. \ref{app:B} for a discussion on this approximation. We also define
\begin{align}
Q(\mathbf{ q}+\mathbf{ G})=\frac{4 \pi}{|\mathbf{ q}+\mathbf{ G}|^2 t }\left( 1-e^{-|\mathbf{ q}+\mathbf{ G}|\frac{t}{2}} \right),
\label{eq:23}
\end{align}
which let us rewrite Eq. \ref{eq:22} as
\begin{align}
C_{ss',\alpha\beta}(\mathbf{ q})=\frac{ \mathcal{Z}_s\mathcal{Z}_{s'}e^2}{A} \sum_{\mathbf{ G} \, \mathbf{ G'}} \left( q_{\alpha}+  G_{\alpha}\right)\left( q_{\beta}+  G'_{\beta}\right) \times \nonumber \\
\w(\mathbf{ q}+\mathbf{ G},\mathbf{ q}+\mathbf{ G'})\frac{Q(\mathbf{ q}+\mathbf{ G'})}{v(\mathbf{ q}+\mathbf{ G'})} e^{i\mathbf{ G}\cdot \boldsymbol{ \tau}_{s}-i\mathbf{ G'}\cdot\boldsymbol{ \tau}_{s'}}.
\label{eq:26}
\end{align}
In the thin and thick limits we have $\frac{Q(|\mathbf{ q}+\mathbf{ G'}|)}{v(\mathbf{ q}+\mathbf{ G'})}\rightarrow 1$; since we are interested in these limits we can set the ratio of the two functions to 1, obtaining eventually Eq. \ref{eq:27} when isolating the non analytical terms coming from the $\w$ tensor. For intermediate regimes the behaviour of the dynamical matrix (and EPI) will be, in general, strongly model dependent and therefore the division into short and long range components is not well-defined. The evaluation of Eq. \ref{eq:26} passes through the knowledge of the $\w$ matrix; the inversion of the $\w^{-1}$ matrix can be performed symbolically, in line with Refs. \cite{PhysRevB.88.174106,PhysRevB.1.910}, as already shown in App. \ref{app:D}. 

\section{Layer thickness and layer-averaging}
\label{app:B}
\begin{figure*}%
\includegraphics[width=\columnwidth,angle=270]{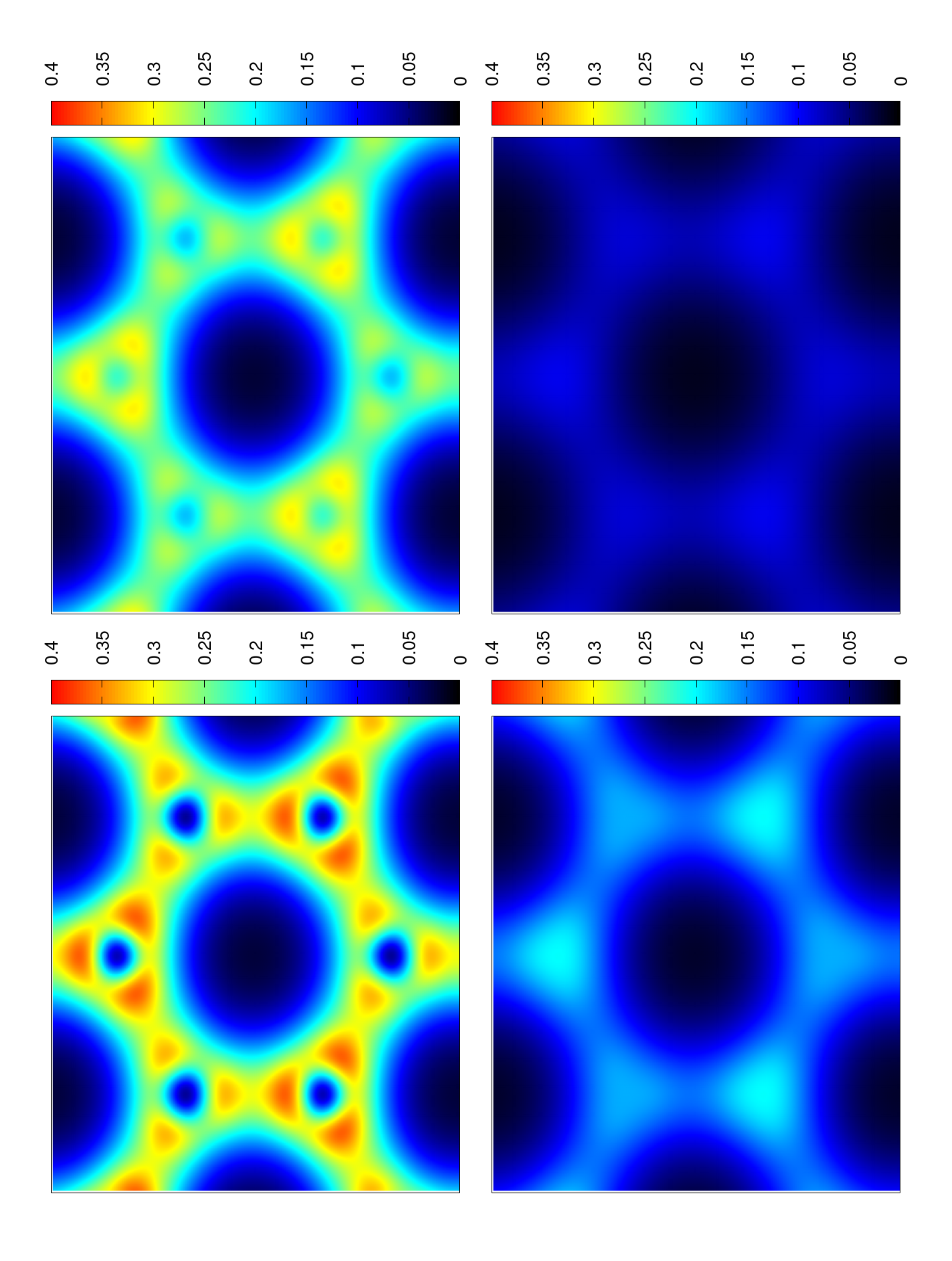}
\caption{\small Heat-maps of the in-plane electronic density of disproportionated graphene, expressed in atomic units, for several values of the $z$ coordinate. The graphene layer is situated at $z=0$. The heat-maps are taken at (upper left) $z=0$ Bohr, (upper right) $z=0.4648$ Bohr, (lower left) $z=0.9297$ Bohr and (lower right) $z=1.3946$ Bohr.}
\label{fig:appA1}
\end{figure*}

\begin{figure}%
\includegraphics[width=\columnwidth]{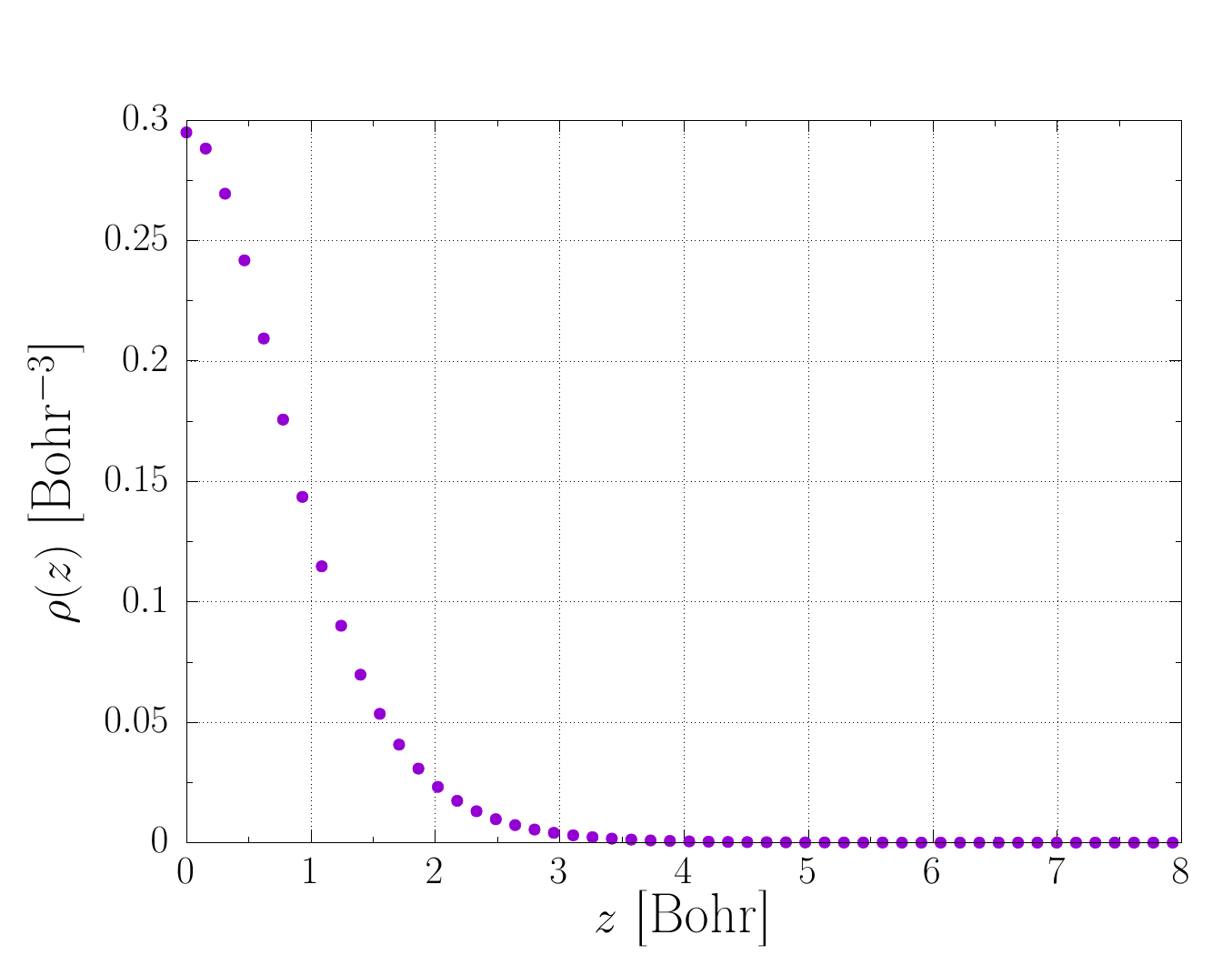}
\caption{ \small \it Slice of electronic density for a line along the $z$ direction whose projection on the graphene layer is situated at the point $\mathbf{ r}=(2.3244,0)$ Bohr. }
\label{fig:appA2}
\end{figure}

The physical working hypotesis assumed in Sec. \ref{sec:II} is that, in appropriate limits, we can approximate a quasi two-dimensional system as a electronically compact homogeneous layer of width $t$ along the $\hat z$ direction, following Eq. \ref{eq:6}. More specifically, we assume that whenever a $z$-dependent quantity enters an exact relation, we can substitute its value with its layer-average given by Eq. \ref{eq:layeraveraging}. The validity of this approximation is debatable, and indeed we show in Figs.  \ref{fig:appA1} and \ref{fig:appA2} that the unperturbed electronic density along the $\hat z$ direction is not a window function, but it decays exponentially on a certain characteristic length, in contradiction with what assumed in Eq. \ref{eq:6}. Nonetheless, we now argue that even in the exact treatment of the $z,z'$ dependence of our problem, we would obtain the same closed forms of the long-range components involving in-plane effective charges as if starting from Eq. \ref{eq:6}, with the only assumption that there exists a scale $\hat t$ which can be used to define the thin and thick limits.

We start from the expression of the dielectric tensor
\begin{align}
&\epsilon(\mathbf{ q}+\mathbf{ G},\mathbf{ q}+\mathbf{ G'},z,z',n,T)=\delta(z-z')\delta_{\mathbf{ G} \, \mathbf{ G'}}+ \label{eq:epsexact2} \\
& -2\pi \int dz'' \frac{e^{-|\mathbf{ q}+\mathbf{ G}||z-z''|}}{|\mathbf{ q}+\mathbf{ G}|}  \chi^0(\mathbf{ q}+\mathbf{ G},\mathbf{ q}+\mathbf{ G'},z'',z',n,T). \nonumber
\end{align}
In the thin limit, we can define a scale $\hat t$ corresponding to the typical size of the extension of the induced charge along the out-of-plane direction; with this definition, we can approximate $e^{-|\mathbf{ q}+\mathbf{ G}||z-z'|}\sim 1$ for $|\mathbf{ q}+\mathbf{ G}|\hat t \ll 1$, to obtain
\begin{align}
&\epsilon(\mathbf{ q}+\mathbf{ G},\mathbf{ q}+\mathbf{ G'},z,z',n,T)\sim \delta(z-z')\delta_{\mathbf{ G} \, \mathbf{ G'}}+ \label{eq:appB1} \\
&-\frac{2\pi}{|\mathbf{ q}+\mathbf{ G}|}  \int dz''\chi^{0}(\mathbf{ q}+\mathbf{ G},\mathbf{ q}+\mathbf{ G'},z'',z',n,T). \nonumber
\end{align}
A similar relation holds between $\epsilon^{-1}$ and $\chi^{-1}$. Now, since to get to the formulae of the LRC of the dynamical matrix all the relevant approximations involve products of $\epsilon^{-1}$ with exponentials in the $z,z'$ coordinates, that can be approximated to unity, the final form of the LRC depends only on the layer-averaged value of the inverse dielectric matrix that can be defined from Eq. \ref{eq:appB1} using Eq. \ref{eq:invavdefbis} over the length $\hat t$. The asymptotic formal expressions of the LRC of the dynamical matrix are therefore equal to the one of our model, since the layer-averaging does not introduce in general changes to the asymptotic limits of the inverse dielectric matrix; of course, this is true only if stick to the in-plane components of the expansions, otherwise further terms coming from the out-of-plane direction may be present as shown in Ref. \cite{PhysRevX.11.041027}. For the LRC of the EPI modifications of the asymptotic formulae may come from the out-of-plane integration of the product of the inverse dielectric matrix with the periodic parts of the Bloch functions at $\mathbf{ k}$ and $\mathbf{ k}+\mathbf{ q}$. Nonetheless, in the general case we expect such integration to tend to a constant non zero value in the limit $\mathbf{ q}\rightarrow 0$ and therefore to recover the form of the LRCs of our model even in this case. Even if the form of the asymptotic models are the same, one would need to obtain the correct prefactors coming from the different kinds of approximation: we overcome this obstacle in our work by taking the values of the effective charges and of the dielectric constants directly from \textit{ab-initio} calculations.

Analogously, for $|\mathbf{ q}+\mathbf{ G}|\hat t \gg 1$ we can use that the integral of Eq. \ref{eq:epsexact2} is relevantly different from zero only when in the IIP we have $z''\approx z$: 
\begin{align}
&\epsilon(\mathbf{ q}+\mathbf{ G} ,\mathbf{ q}+\mathbf{ G'},z,z')\sim \delta(z-z')\delta_{\mathbf{ G} \, \mathbf{ G'}}+ \label{eq:appB2} \\
& -2\pi \frac{P(\mathbf{ q}+\mathbf{ G},z)}{|\mathbf{ q}+\mathbf{ G}|}  \chi^0(\mathbf{ q}+\mathbf{ G},\mathbf{ q}+\mathbf{ G'},z,z') \sim \nonumber\\
& \delta(z-z')\delta_{\mathbf{ G} \, \mathbf{ G'}}-4\pi \frac{1}{|\mathbf{ q}+\mathbf{ G}|^2}  \chi^0(\mathbf{ q}+\mathbf{ G},\mathbf{ q}+\mathbf{ G'},z,z').
\end{align}
In the derivation of the LRCs for every occurrence of the inverse dielectric function under the integration sign together with $e^{-|\mathbf{ q}+\mathbf{ G}||z-z'|}$ we can use the same approximation of Eq. \ref{eq:appB2}. Again, the asymptotic in-plane limits are not modified and the LRCs have the same formal expression as for our model. 

From the above considerations it follows that the typical scale $\hat t$ shall be regarded as the scale where the charge response of the system is substantially different from zero. 
\section{Layer-averaged Maxwell's equations}
\label{app:E}
The Maxwell's equations are classical equations which relate macroscopic physical densities and currents to electromagnetic fields. In this appendix, we discuss the form of the first Maxwell's equation in the thin limit within the RPA approximation. We start from the RPA electrostatic relation between charge and potential
\begin{align}
\delta V^{\textrm{tot}}(\mathbf{ q}+\mathbf{ G},z)= 2\pi \int dz' \frac{e^{-|\mathbf{ q}+\mathbf{ G}||z-z'|}}{|\mathbf{ q}+\mathbf{ G}|}\delta \rho^{\textrm{tot}}(\mathbf{ q}+\mathbf{ G},z').
\label{eq:E2}
\end{align}
We now substitute each term of the above equation with its layer-average, as defined in App. \ref{app:A3} and discussed in App. \ref{app:B}; we have
\begin{align}
\delta V^{\textrm{tot}}(\mathbf{ q}+\mathbf{ G})= v(\mathbf{ q}+\mathbf{ G}) t \delta \rho^{\textrm{tot}}(\mathbf{ q}+\mathbf{ G});
\label{eq:E3}
\end{align}
if we now specify $\delta \rho^{\textrm{tot}}=\delta \rho^{\textrm{ind}}+\delta \rho^{\textrm{ext}}$ and use the relation of Eq. \ref{eq:A14} between $\delta \rho^{\textrm{ind}}$ and $\delta V^{\textrm{tot}}$, we obtain
\begin{align}
\sum_{\mathbf{ G'}} \epsilon(\mathbf{ q}+\mathbf{ G},\mathbf{ q}+\mathbf{ G'})\frac{\delta V^{\textrm{tot}}(\mathbf{ q}+\mathbf{ G'})}{ v(\mathbf{ q}+\mathbf{ G})}= \nonumber \\
t\delta \rho^{\textrm{ext}}(\mathbf{ q}+\mathbf{ G}),
\end{align}
which using Eq. \ref{eq:24} becomes
\begin{align}
\sum_{\mathbf{ G'}} \w^{-1}(\mathbf{ q}+\mathbf{ G},\mathbf{ q}+\mathbf{ G'}) \delta  V^{\textrm{tot}}(\mathbf{ q}+\mathbf{ G'})=t\delta \rho^{\textrm{ext}}(\mathbf{ q}+\mathbf{ G}).
\end{align}
For the thin limit in particular we can write
\begin{align}
\sum_{\mathbf{ G'}} \epsilon(\mathbf{ q}+\mathbf{ G},\mathbf{ q}+\mathbf{ G'}) \delta V^{\textrm{tot}}(\mathbf{ q}+\mathbf{ G'})=2\pi \frac{t\delta \rho^{\textrm{ext}}(\mathbf{ q}+\mathbf{ G})}{|\mathbf{ q}+\mathbf{ G}|},
\label{eq:Maxwell}
\end{align}
where $t\delta \rho^{\textrm{ext}}(\mathbf{ q}+\mathbf{ G})$ is dimensionally a 2d density and the left hand side is not explicitly dependent on $t$. The above relation, which follows from Eqs. \ref{eq:E2} and \ref{eq:chiir}, relates the total potential change to the external charge perturbation, and includes all the in-plane non-local-fields component of the response.

The form of the classical Maxwell's equation instead follows from the relation between the total potential change and the macroscopically unscreened total charge change through the macroscopic inverse dielectric tensor, as shown in Sec. \ref{sec:IIsubB} and in particular in Eq. \ref{eq:30}. Such equation may be rewritten for clarity as a function of the density as
\begin{align}
\delta V^{\textrm{tot}}_{s,\alpha}(\mathbf{ q})=
\w(\mathbf{q})t\delta \bar \rho^{\textrm{tot}}_{s,\alpha}(\mathbf{ q}). \label{eq:maxwellmacro}
\end{align}
Notice that if we compare the above expression with Eq. \ref{eq:E3} evaluated at $\mathbf{ G}=0$ we obtain the relation of Eq. \ref{eq:2inline} between macroscopically screened and unscreened charges. Physically, this means that $\delta \bar \rho^{\textrm{tot}}_{s,\alpha}(\mathbf{ q})$ acts electrostatically as an \textit{external macroscopic perturbing potential} that has been renormalized by the local fields on the system under consideration.

\section{\textit{xc} and response functions}
\label{app:xc}
The many body theoretical treatment of the response functions has been performed up to now directly in the RPA approximation, therefore neglecting any effect stemming from the presence of an exchange-correlation kernel $K_{xc}$ in the response. We will now try to extend the treatment beyond such an approximation. For the sake of clarity, we will restrict to the thick limit only and will take $t=1$. 

\textit{In absence of $K_{xc}$}, the screening function and its inverse are written as
\begin{align}
\epsilon^{-1,RPA}=\mathcal{I}+v\chi^{RPA},\\
\epsilon^{RPA}=\mathcal{I}-v\chi^{0};
\end{align}
We define the screened Coulomb interaction as
\begin{align}
w=\epsilon^{-1,RPA}v;
\end{align}
if we now take Eq. \ref{eq:vbartot} and \ref{eq:indcharge}, using Eq. \ref{eq:X} we obtain
\begin{align}
\delta \bar \rho^{\textrm{ind}}_{s,\alpha}(\mathbf{ q})+\delta \rho^{\textrm{ext}}_{s,\alpha}(\mathbf{ q})=\sum_{\mathbf{ G'}}\frac{\w(\mathbf{ q},\mathbf{ q}+\mathbf{ G'})}{\w(\mathbf{ q})} \delta \rho^{\textrm{ext}}_{s,\alpha}(\mathbf{ q}+\mathbf{ G'}).
\label{eq:rhototexact}
\end{align}
Now, one uses Eq. \ref{eq:30} (where it is only used that $V^{ext}=v\rho^{ext}$) and eventually it follows
\begin{align}
\delta V^{\textrm{tot}}_{s,\alpha}(\mathbf{ q})=
w(\mathbf{q})\delta \bar \rho^{\textrm{tot}}_{s,\alpha}(\mathbf{ q}),\\
\delta \rho^{\textrm{tot}}_{s,\alpha}(\mathbf{ q})=\epsilon^{-1}(\mathbf{ q})\delta \bar \rho^{\textrm{tot}}_{s,\alpha}(\mathbf{ q}),
\\
Z_{s,\alpha}(\mathbf{ q})= \epsilon^{-1}(\mathbf{ q})\bar Z_{s,\alpha}(\mathbf{ q}).
\end{align}
A first crucial observation is that, numerically, in the undoped case the value of the unscreened Born effective charges is the same when computed in the RPA or in the RPA+\textit{xc} with an LDA kernel, while this is not true for the value of $\epsilon^{-1}(\mathbf{q})$ \footnote{In two dimensions the difference between the dielectric screenin in RPA and RPA+\textit{xc} for undoped semiconductors is anyway small. The same conclusions are reached though also for the three dimensional case of Ref. \cite{PhysRevLett.129.185902}, where the difference is more marked.}; with Eq. \ref{eq:indcharge} in mind, we use this finding to state that at small $\mathbf{q}$ the values of all $S^{-1}$ elements are fairly independent of \textit{xc} effects in the undoped case \footnote{The effects of the exchange-correlation terms in $\chi^0$ are present through the periodic part of the Bloch function and the electronic energies; nonetheless, such effects are considered also in the RPA approximation.}.

\textit{In presence of $K_{xc}$}, the screening function and its inverse (which cannot be called `dielectric' anymore) are written as
\begin{align}
\epsilon^{-1}=\mathcal{I}+(v+K_{xc})\chi,\\
\epsilon=\mathcal{I}-(v+K_{xc})\chi^0;
\end{align}
properly speaking $K_{xc}$ should present a long-wavelength behaviour proportional to $1/q^2$, in the most common approximations $K_{xc}$ is independent of $\mathbf{q}$ (in LDA) \cite{PhysRevB.35.5585} or dependent on positive powers of $\mathbf{q}$ (in GGA) \cite{PhysRevB.56.12811}. We still define the screened Coulomb interaction as
\begin{align}
w=\epsilon^{-1}v,
\end{align}
even though the matrix is now not Hermitian and the rules for inversion change. For a generic non Hermitian matrix $\w^{-1}$, we write \cite{PhysRevB.1.910}:
\begin{align}
\w^{-1}=
\begin{pmatrix}
P & Q \\
R & S
\end{pmatrix},
\\
\w=
\begin{pmatrix}
W & X \\
Y & Z
\end{pmatrix},
\end{align}
where the inversion rules are now
\begin{align}
W=(P-QS^{-1}R)^{-1},\\
X=-WQS^{-1},\label{eq:X2}\\
Y=-S^{-1}RW,\\
Z=S^{-1}+YW^{-1}X,\label{eq:factorization2}
\end{align}
where we notice that Eq. \ref{eq:X2} is equal to Eq. \ref{eq:X}. The wing of $\w^{-1}$ is now (assuming local $K_{xc}$ with no local field dependence for the sake of simplicity of the argument)
\begin{align}
Q(\mathbf{q},\mathbf{q+G})=-\frac{v(\mathbf{q})+K_{xc}(\mathbf{q})}{v(\mathbf{q})}\chi^0(\mathbf{q},\mathbf{q+G}).
\end{align}
As already discussed before, in the undoped case the presence of $K_{\textrm{xc}}$ in the response affects mainly the terms of $\w$ which depend on $W$. Also, \textit{xc} terms are not expected to cure the non analytical behaviours typical of the RPA case, so that we are still induced to split the $\w$ matrix in a short range component involving only the body $S^{-1}$, which is mainly RPA-like, and a remainder which is non analytical. To isolate the short-range component, we are therefore induced to adopt the same prescription of the RPA case to nullify the total macroscopic potential, comprised now also of the \textit{xc} component; this explain the procedure described in Sec. \ref{sec:compeffcharges}. Also, within our interpolating procedure explained in Sec. \ref{sec:IIIsubBsub3}, this splitting means that in the doped case we are approximating $S^{-1}$ as the one of the undoped case; the goodness of the approximation is therefore connected to the magnitude of \textit{xc} contributions to the local fields in presence of doping.
\\
With the above prescription, Eq. \ref{eq:indcharge} can now be written as 
\begin{align}
\delta \bar \rho^{\textrm{ind}}_{s,\alpha}(\mathbf{ q})= -\sum_{\substack{\mathbf{ G}\neq 0 \\ \mathbf{ G'} \neq 0}}\frac{v(\mathbf{q})}{v(\mathbf{q})+K_{xc}(\mathbf{q})}Q(\mathbf{ q},\mathbf{ q}+\mathbf{ G})\times \nonumber \\
S^{-1}(\mathbf{ q}+\mathbf{ G},\mathbf{ q}+\mathbf{ G'})
\delta \rho^{\textrm{ext}}_{s,\alpha}(\mathbf{ q}+\mathbf{ G'}) .
\end{align}
We use Eq. \ref{eq:X} obtaining
\begin{align}
\delta \bar \rho^{\textrm{ind}}_{s,\alpha}(\mathbf{ q})=\sum_{\mathbf{ G'}\neq 0}\frac{\w(\mathbf{ q},\mathbf{ q}+\mathbf{ G'})}{\epsilon^{-1}(\mathbf{q})[v(\mathbf{ q})+K_{xc}(\mathbf{q})]} \delta \rho^{\textrm{ext}}_{s,\alpha}(\mathbf{ q}+\mathbf{ G'})\\
\delta \bar \rho^{\textrm{tot}}_{s,\alpha}(\mathbf{ q})=\delta \bar \rho^{\textrm{ind}}_{s,\alpha}(\mathbf{ q})+\delta \rho^{\textrm{ext}}_{s,\alpha}(\mathbf{ q})
\end{align}
As evident, in principle we lose the correspondence between the above equation and Eq. \ref{eq:30}, and therefore barred quantities are no more related to the unbarred ones simply via the screening function. In practice, for the undoped case or small dopings and for the ordinary approximations to $K_{xc}$, we find that the correspondence is respected to a satisfactory precision. Notice that our considerations imply the well-known consequence that in the undoped case the Born-Huang formulae works even when their ingredients are computed in a RPA+\textit{xc} approach; in fact, $\epsilon^{-1}$ in the above equations now contains the effects due to the the presence of $K_{\textrm{xc}}$ in the response, while as already mentioned the effective charge functions are fairly insensitive to \textit{xc} effects. Notice also that, formally, with the current prescription the definition of, e.g., the Born effective charge tensors of Eq. \ref{eq:35} has to be intended as evaluated at total macroscopic field equal to zero, and not only the electrostatic one. 

In a more general treatment, we find it interesting to rewrite the responses as
\begin{align}
\epsilon^{-1}=\mathcal{I}+v\tilde \chi \quad 
\epsilon=\mathcal{I}-v\tilde \chi^0,\\
\tilde \chi = v^{-1} [v+K_{xc}] \chi \quad
\tilde \chi^0 = v^{-1} [v+K_{xc}] \chi^0.
\end{align}
In this case, we can interpret $\tilde \chi^0$ as the polarizability response that enters in the Maxwell's equations. We now define a new induced density, namely
\begin{align}
\delta \tilde {\bar \rho}^{\textrm{ind}}_{s,\alpha}(\mathbf{ q})= \sum_{\substack{\mathbf{ G}\neq 0 \\ \mathbf{ G'} \neq 0}}\tilde \chi^0(\mathbf{ q},\mathbf{ q}+\mathbf{ G}) S^{-1}(\mathbf{ q}+\mathbf{ G},\mathbf{ q}+\mathbf{ G'}) \times \nonumber \\
\delta \rho^{\textrm{ext}}_{s,\alpha}(\mathbf{ q}+\mathbf{ G'});
\end{align}
summing the above expression to the external density, we obtain
\begin{align}
\delta \tilde{\bar \rho}^{\textrm{ind}}_{s,\alpha}(\mathbf{ q})+\delta \rho^{\textrm{ext}}_{s,\alpha}(\mathbf{ q})=\sum_{\mathbf{ G'}}\frac{\w(\mathbf{ q},\mathbf{ q}+\mathbf{ G'})}{\w(\mathbf{ q})} \delta \rho^{\textrm{ext}}_{s,\alpha}(\mathbf{ q}+\mathbf{ G'})
\end{align}
and therefore we can identify it with Eq. \ref{eq:30}; we therefore have
\begin{align}
\delta V^{\textrm{tot}}_{s,\alpha}(\mathbf{ q})=
w(\mathbf{q})\delta \tilde{\bar \rho}^{\textrm{tot}}_{s,\alpha}(\mathbf{ q})\coloneqq -i\frac{eq}{A}\w(\mathbf{q}) \tilde{\bar Z}_{s,\alpha}(\mathbf{ q}).
\end{align}
Interestingly, the expression of $g^{\textrm{L}}$ is the same as Eq. \ref{eq:gasymp} with the replacement $Z\rightarrow \tilde Z$, while in the expression for $C^{\textrm{L}}$ we can approximate $\bar Z$ at its undoped value and send $Z\rightarrow \tilde Z$.
\\
We end this section noting that the definition of effective charges given in presence of \textit{xc} in the context of this work, i.e. in the context of interpolation of short and long range components of the dynamical matrix and of the EPI, is not physical for the case of doped semiconductors or metals. The physical value would be in fact obtained defining the effective charges from the total density change obtained at null macroscopic electrostatic potential, leaving the \textit{xc} component of the total potential unaltered; such density would be automatically analytical and would respect translational invariance. We leave the investigation of such effective charges to future works.
\bibliography{biblio}

\end{document}